\documentclass[trackchanges, twocolumn]{aastex701}
\hypersetup{linkcolor=red,citecolor=blue,filecolor=cyan,urlcolor=magenta}
\usepackage{hyperref}
\usepackage{placeins}
\usepackage{xurl}
\begin{document}

%\title{X-ray Spectral Properties of Four Classical TeV High Energy Peaked Blazars using NICER and NuSTAR} 
\title{X-ray Spectral Properties of Four Classical TeV Blazars using Simultaneous Observations from NICER and NuSTAR}

\author[orcid=0000-0002-7658-0350]{Riya Bhowmick}
%\altaffiliation{}
\affiliation{Aryabhatta Research Institute of Observational Sciences (ARIES), Manora Peak, Nainital 263001, India}
\email[show]{riyabhowmickmalda@gmail.com}  

\author[orcid=0000-0002-9331-4388]{Alok C. Gupta} 
%\altaffiliation{}
\affiliation{Aryabhatta Research Institute of Observational Sciences (ARIES), Manora Peak, Nainital 263001, India}
\email[show]{acgupta30@gmail.com}

% \author[gname=Savannah,sname=Africa]{S. Africa}
% \affiliation{South African Astronomical Observatory}
% \affiliation{University of Cape Town, Department of Astronomy}
% \email{fakeemail3@google.com}

% \author{River Europe}
% \affiliation{University of Heidelberg}
% \email{fakeemail4@google.com}

% \author[0000-0000-0000-0003,sname=Asia,gname=Mountain]{Asia Mountain}
% \altaffiliation{Astrosat Post-Doctoral Fellow}
% \affiliation{Tata Institute of Fundamental Research, Department of Astronomy}
% \email{fakeemail5@google.com}

% \author[0000-0000-0000-0004]{Coral Australia}
% \affiliation{James Cook University, Department of Physics}
% \email{fakeemail6@google.com}

% \author[gname=IceSheet]{Penguin Antarctica}
% \affiliation{Amundsen–Scott South Pole Station}
% \email{fakeemail7@google.com}

%\collaboration{all}{The Terra Mater collaboration}

%% Use the \collaboration command to identify collaborations. This command
%% takes an optional argument that is either a number or the word "all"
%% which tells the compiler how many of the authors above the command to
%% show. For example "\collaboration[all]{(DELVE Collaboration)}" wil include
%% all the authors above this command.
%%
%% Mark off the abstract in the ``abstract'' environment. 
\begin{abstract}
\noindent
We present a detailed study of X-ray spectral properties observed in 4 classical TeV (tera-electron volt) photon-emitting high synchrotron-peaked BL Lacertae objects using the simultaneous data of NICER and NuSTAR satellites. We analyzed 13 spectra in total from four BL Lacertae objects: Mrk 421, Mrk 501, PG 1553+113, and PKS 2155-304. We fitted all the spectra using the absorbed Log-Parabolic (LP) model first. While 7 spectra were fitted well using the absorbed LP model, we observed that 4 spectra of Mrk 421 and 2 spectra of Mrk 501 were not fitted satisfactorily using the absorbed LP model. 
%\sout{While studying the flux state, we observed that during these observations the sources were in a} \sout{low to moderate flux state} 
The investigation of the flux states of the sources revealed that Mrk~421 was in a moderate to low-flux state during the 4 epochs and Mrk~501 was in a low-flux state during the 2 epochs. We concluded that there was a contribution from the disk in these 6 spectra. The moderate to low-flux state can justify the contribution of disk emission in the X-ray spectra. In the case of 4 spectra of Mrk 421, we observed a Gaussian feature between 1.42 and 1.70 keV.
\end{abstract}

%% Keywords should appear after the \end{abstract} command. 
%% The AAS Journals now uses Unified Astronomy Thesaurus (UAT) concepts:
%% https://astrothesaurus.org
%% You will be asked to selected these concepts during the submission process
%% but this old "keyword" functionality is maintained in case authors want
%% to include these concepts in their preprints.
%%
%% You can use the \uat command to link your UAT concepts back its source.
%\keywords{\uat{Blazars} --- \uat{BL Lacs} --- \uat{X-ray} --- \uat{HSP} ---  }

%% From the front matter, we move on to the body of the paper.
%% Sections are demarcated by \section and \subsection, respectively.
%% Observe the use of the LaTeX \label
%% command after the \subsection to give a symbolic KEY to the
%% subsection for cross-referencing in a \ref command.
%% You can use LaTeX's \ref and \label commands to keep track of
%% cross-references to sections, equations, tables, and figures.
%% That way, if you change the order of any elements, LaTeX will
%% automatically renumber them.

\section{Introduction} 
\noindent
%Blazars are a rare and extreme type of jetted active galactic nucleus (AGN), which have very bright centers of some galaxies powered by a supermassive black hole (SMBH) (Mass = 10$^{6} - \rm{10}^{10} \rm{M}_{\odot}$).
Blazars are a rare and extreme type of jetted active galactic nucleus (AGN), %\sout{which have very bright centers of some galaxies} 
which are very bright sources located at the centers of some distant galaxies, and are powered by a supermassive black hole (SMBH) (Mass = 10$^{6} - \rm{10}^{10} \rm{M}_{\odot}$). 
%\citep{1984ARA&A..22..471R}. 
Blazars are divided into two subclasses: BL Lacertae objects (BL Lacs) and flat spectrum radio quasars (FSRQs). BL Lacs display featureless or very weak emission lines (equivalent width, EW $\leq$ 5\AA), whereas FSRQs show strong emission lines in their composite UV/optical spectra. 
%\citep{1991ApJS...76..813S,1996MNRAS.281..425M}. 
Blazars emit radiation in the whole electromagnetic (EM) spectrum, which is predominantly nonthermal. TeV blazars are a tiny subclass of blazar that emits very-high-energy (VHE) gamma-rays in the tera-electronvolt (TeV, $10^{12}$ eV) range. Blazars emit relativistic charged particle jets towards almost the observer's line of sight, with viewing angle generally $\lesssim 10^\circ$ \citep{1995PASP..107..803U}. The emission from blazars is Doppler boosted and highly variable in flux, spectra, and polarization in all possible time scales, ranging from a few minutes to days to months and even several years \citep[see for review][]{2025A&ARv..33....8R}. Blazars' radio-to-gamma-ray multi-wavelength (MW) spectral energy distributions (SEDs) are double-humped, 
%\sout{and in TeV blazars} 
the first hump peaks in IR to X-rays, whereas the second hump peaks in 
 $\gamma-$ray energies \citep{1998MNRAS.299..433F}. Synchrotron emission from radio to UV/soft X-ray frequencies and inverse Compton (IC) scattering for hard X-ray and gamma-ray energies are probably the dominant emission mechanisms in blazars. Blazars are classified into three subclasses based on their low-energy hump (synchrotron peak $\nu_{syn}$) of SED: low synchrotron peak (LSP) if $\nu_{syn} \lesssim$ $10^{14}$ Hz, intermediate synchrotron peak (ISP) if $10^{14}$ Hz $\lesssim \nu_{syn} \lesssim$ $10^{15}$ Hz, and high synchrotron peak (HSP) if $\nu_{syn} \gtrsim$ $10^{15}$ Hz \citep{2010ApJ...716...30A}. To date, most of the TeV-emitting blazars belong to the HSP class (59 out of 88)\footnote{https://www.tevcat.org/}. 
The principal emitters of blazars are thought to be ultra-relativistic electrons or positrons (leptonic scenario), and occasionally protons (hadronic scenario) \citep{2013ApJ...768...54B}. \\
\\
VHE gamma-ray astronomy started in the 1980s; the first detection of TeV emission from the blazar Mrk~421 was reported by \citet{1992Natur.358..477P}. Till 2006, TeV emission was detected from only seven blazars, namely Mrk~421 \citep{1992Natur.358..477P},
%,1996A&A...311L..13P}, 
Mrk~501 \citep{1996ApJ...456L..83Q,1997A&A...320L...5B}, 1ES 2344+514 \citep{1998ApJ...501..616C}, PKS~2155-304 \citep{1999ApJ...513..161C}, H 1426+428 \citep{2002ApJ...571..753H,2002A&A...384L..23A}, 1ES 1959+650 \citep{2003ApJ...583L...9H,2003A&A...406L...9A}, and PG 1553+113 \citep{2006A&A...448L..19A}.
%,2007ApJ...654L.119A}. 
These can be called classical TeV-emitting blazars, and except for PG 1553+113, the other six of these 
TeV-emitting blazars' gamma-ray properties are given in \citet{2004NewAR..48..367K}. Thanks to the development of new facilities, e.g., HESS (High Energy Stereoscopic System), MAGIC (Major Atmospheric Gamma-ray Imaging Cerenko), and VERITAS (Very Energetic Radiation Imaging Telescope Array System) in the 2000s \citep[e.g.][and references therein]{2004APh....22..285F,2004NIMPA.518..188B,2006APh....25..391H}, which have made a complete revolution in TeV gamma-ray astronomy. These discoveries motivated scientists to develop several more facilities for TeV gamma-ray sciences around the globe, e.g., HAWC (High-Altitude Water Cherenkov Gamma-ray Observatory), MACE (Major Atmospheric Cerenkov Experiment Telescope), LHAASO (Large High Altitude Air Shower Observatory), etc. \citep[e.g.][and references therein]{2012APh....35..641A,2013APh....50...26A,2024APh...15902960B,2021ChPhC..45b5002A}. With the help of TeV gamma-ray observing facilities around the globe, the recent catalogue shows that to date 322 astronomical sources have been detected as TeV gamma-ray emitters\footnote{https://www.tevcat.org/}. \\
\\
The enlarged sample of TeV blazars will be very useful for our understanding of the emission mechanism of these extreme blazars through the study of their variability properties across the whole range of EM bands. Since 2009, we have been working on a pilot project to search for blazar flux and/or spectral variabilities and also searching for quasi-periodic oscillations (QPOs) using various X-ray mission data, e.g., XMM-Newton \citep{2009A&A...506L..17L,2010ApJ...718..279G,2014MNRAS.444.3647B,2016NewA...44...21B,2015MNRAS.451.1356K,2017MNRAS.469.3824K,2016MNRAS.462.1508G,2021MNRAS.506.1198D,2022MNRAS.511.3101P,2022ApJS..262....4N,2022ApJ...939...80D,2025ApJS..278...20D,2026ApJ...998..107H}, RXTE \citep{2009ApJ...696.2170R}, 	NuSTAR \citep{2017ApJ...841..123P,2018ApJ...859...49P}, Chandra \citep{2018MNRAS.480.4873A}, Swift and XMM-Newton \citep{2019ApJ...880...19K}, and Suzaku \citep{2019ApJ...884..125Z,2021ApJ...909..103Z,2024MNRAS.532.3285Z}. Here, we report our first search for X-ray spectral variability properties of four classical TeV blazars 
%\sout{from the simultaneous observations by the NICER} \sout{and NuSTAR satellites' public archive.} 
using simultaneous NICER and NuSTAR observations available in the public archives.\\
\\
Section 2 provides a brief description about NICER and	NuSTAR satellites, their data, and the reduction procedure.  Spectral analysis and results are reported in Section 3 and Section 4, respectively. A discussion and conclusion are provided in Section 5.
\begin{table*}
\centering
\small
\caption{Observation log of blazar sources with NICER and NuSTAR \label{Table:log}}
\resizebox{\textwidth}{!}{%
\begin{tabular}{l c c | c c c | c c c}
\hline\hline
Blazar Name & MJD & Date of Obs. & & NICER & & & NuSTAR \\
 &  & YYYY-MM-DD & OBSID & Exposure (s) & Energy range (keV) & OBSID & Exposure (s) & Energy range (keV)\\
 ~~~~~~~(1)    &   (2)  &  (3)   &   (4)   &   (5)   &   (6)  &   (7)   & (8) & (9) \\
\hline\hline 
Mrk~421 & 60077 & 2023-05-13 & 6100110102 & 7195 & 0.5 -- 10.0 & 80801643002 & 44571 & 4.0 -- 78.0 \\
              & 60289 & 2023-12-11 & 6100110129 & 5250 & 0.5 -- 10.0 & 60902024004 & 21235 & 4.0 -- 78.0 \\
              & 60296 & 2023-12-18 & 6100110136 & 5960 & 0.5 -- 6.0 & 60902024006 & 19990 & 4.0 -- 78.0 \\
              & 60298 & 2023-12-20 & 6100110138 & 913  & 0.5 -- 10.0 & 60902024008 & 21099 & 4.0 -- 78.0 \\
              & 60429 & 2024-04-29 & 7566016001 & 598  & 0.5 -- 8.0 & 60902024010 & 20967 & 4.0 -- 78.0 \\
              & 60433 & 2024-05-03 & 7566016401 & 404  & 0.5 -- 8.0 & 60902024012 & 21025 & 4.0 -- 78.0 \\
              & 60439 & 2024-05-09 & 7566017001 & 428  & 0.5 -- 10.0 & 60902024014 & 19502 & 4.0 -- 78.0 \\
              & 60443 & 2024-05-13 & 7566017401 & 321  & 0.5 -- 10.0 & 60902024016 & 18122 & 4.0 -- 78.0 \\ [3.5ex]
Mrk~501 & 59660 & 2022-03-22 & 5202660102 & 5886 & 0.5 -- 10.0 & 60502009002 & 21303 & 4.0 -- 78.0 \\
              & 59662 & 2022-03-24 & 4700040101 & 7153 & 0.5 -- 10.0 & 60502009004 & 29724 & 4.0 -- 78.0 \\
              & 59665 & 2022-03-27 & 4700040102 & 2915 & 0.5 -- 10.0 & 60702062004 & 20297 & 4.0 -- 78.0 \\[3.5ex]
PG~1553+113   & 60048 & 2023-04-14 & 6202110101 & 6809 & 0.5 -- 6.0 & 90901615002 & 20699 & 4.0 -- 78.0 \\ [3.5ex]
PKS~2155-304  & 60285 & 2023-12-07 & 6010070114 & 1734 & 0.5 -- 6.0 & 11001611002 & 42190 & 4.0 -- 25.0 \\
\hline\hline
\end{tabular}
}
\end{table*}
\section{Observation and Data Analysis}
\noindent
We have utilized simultaneous data from the Neutron Star Interior Composition Explorer (NICER) and the Nuclear Spectroscopic Telescope Array Mission (NuSTAR) for the spectral analysis of the four classical TeV blazars, Mrk~421, Mrk~501, PG~1553+113 and PKS~2155-304. The MJDs, the observation IDs (OBSIDs), the corresponding exposure time and the energy range used for the NICER and NuSTAR data for this study are listed in \autoref{Table:log}. A brief introduction to the satellites used, along with the data reduction process, is provided in this section.
\subsection{NICER}
\noindent
The Neutron Star Interior Composition Explorer (NICER) is a scientific payload installed on the International Space Station (ISS). NICER is designed to investigate X-ray sources using soft X-ray timing techniques. It was launched on June 3, 2017, aboard a SpaceX Falcon 9 and installed on the ISS on June 16, 2017. The NICER's X-ray Timing Instrument (XTI), consists of 56 pairs of X-ray concentrator optics (XRCs) and silicon drift detectors (SDDs) \citep{10.1117/12.2231304}. Each XRC gathers X-rays from a sky region of about 30 $arcmin^2$ and focuses them onto a small corresponding SDD. Individual photons are detected by the SDD, with their energies recorded to a few‑percent spectral resolution and their arrival times measured with an unprecedented precision of 100 nanoseconds RMS relative to Universal Time. NICER provides high signal-to-noise photon-counting capabilities across the 0.2–12.0~keV energy band, ideally suited for studying neutron star spectra as well as a wide variety of other astrophysical sources. NICER employs a star-tracker-guided pointing system that enables the X-ray Timing Instrument (XTI) to aim at and follow celestial sources across almost an entire hemisphere of the sky. \\
\\
In this current study, we analysed the NICER spectra of the four TeV blazars in between 0.5 to 10.0 keV energy range, where the instrument calibration is well established and the source signal dominates over instrumental and background uncertainties. Although, for some OBSIDs we had to restrict the energy range to 0.5--8.0~keV or 0.5--6.0~keV depending on the source and background spectra. 
\subsubsection{Data Reduction}
\noindent
We have used SciServer \citep{2020A&C....3300412T} using HEASARCv6.36 for the data reduction of NICER. Using heasoftpy we ran the nicerl2 pipeline to process and clean the data which is required to produce the spectra. For spectral analysis, the standard task nicerl3-spect was used to produce the spectral products, including the source spectrum and the associated response files (ARF and RMF). The SCORPEON background model was employed to generate background spectra for the NICER observations. This model was selected over alternatives such as 3C50 and Space Weather for several reasons. The SCORPEON model explicitly handles astrophysical X-ray sky backgrounds, including the Cosmic X-ray Background (CXB), the Galactic X-ray Halo (Halo) and the Local Hot Bubble (LHB). The other background models (3C50 and Space Weather) are library-type models, where as the SCORPEON is a parameterized model and it provides smooth template spectra, continuously varying with its parameters. A background model produced by SCORPEON, can be fitted jointly with the source model, thereby yielding a closer match between the background and the observational data. \\
\\
The source spectra were grouped using the optimal binning method \citep {2016A&A...587A.151K}. We chose the `grouptype' as `optmin' with a minimum number of 25 counts using `groupscale'. 
\subsection{NuSTAR}
\noindent
The Nuclear Spectroscopic Telescope Array (NuSTAR) is the first focusing hard X-ray satellite placed in orbit, delivering an improvement in sensitivity of more than two orders of magnitude compared to earlier high-energy missions operating in similar energy ranges. NuSTAR was launched on June 13, 2012 as part of NASA’s Small Explorer program. NuSTAR is the first mission in which focusing telescopes were employed to image the sky in the high‑energy X‑ray band (3.0--79.0~keV) of the electromagnetic spectrum. The observatory is equipped with two co-aligned, identical grazing-incidence X-ray telescopes, designated by their focal plane modules, A and B (FPMA and FPMB) \citep{2013ApJ...770..103H}. In this study we analysed the NuSTAR spectra of three TeV blazars (Mrk~421, Mrk~501 and PG~1553+113) in the energy range of 4.0 to 78.0 keV. For the blazar PKS~2155-304, after careful examination of the source and background spectra, we could analyse the spectra in the energy range of 4.0 to 25.0 keV only.
\subsubsection{Data Reduction}
\noindent
We analysed the NuSTAR data using HEASoft version 6.34. The spectra were extracted using the NuSTAR data analysis software \textit{NuSTARDAS}. The \textit{nupipeline} command was executed to produce stage-II data for the focal plane module FPMA and FPMB. Using the SAOImageDS9, circular region of 40 arcseconds centered on the source position was defined to generate the source region file, while a background region of equal size was selected away from the source. Spectra, along with the corresponding RMF and ARF response files, were produced using \textit{nuproduct}. The extracted spectra were re-binned with \textit{grppha} to ensure a minimum of 25 counts per bin.
\section{Spectral Analysis} 
\begin{figure*}
\centering
  \begin{minipage}[b]{0.49\textwidth}
    \includegraphics[width=\textwidth, angle=0]{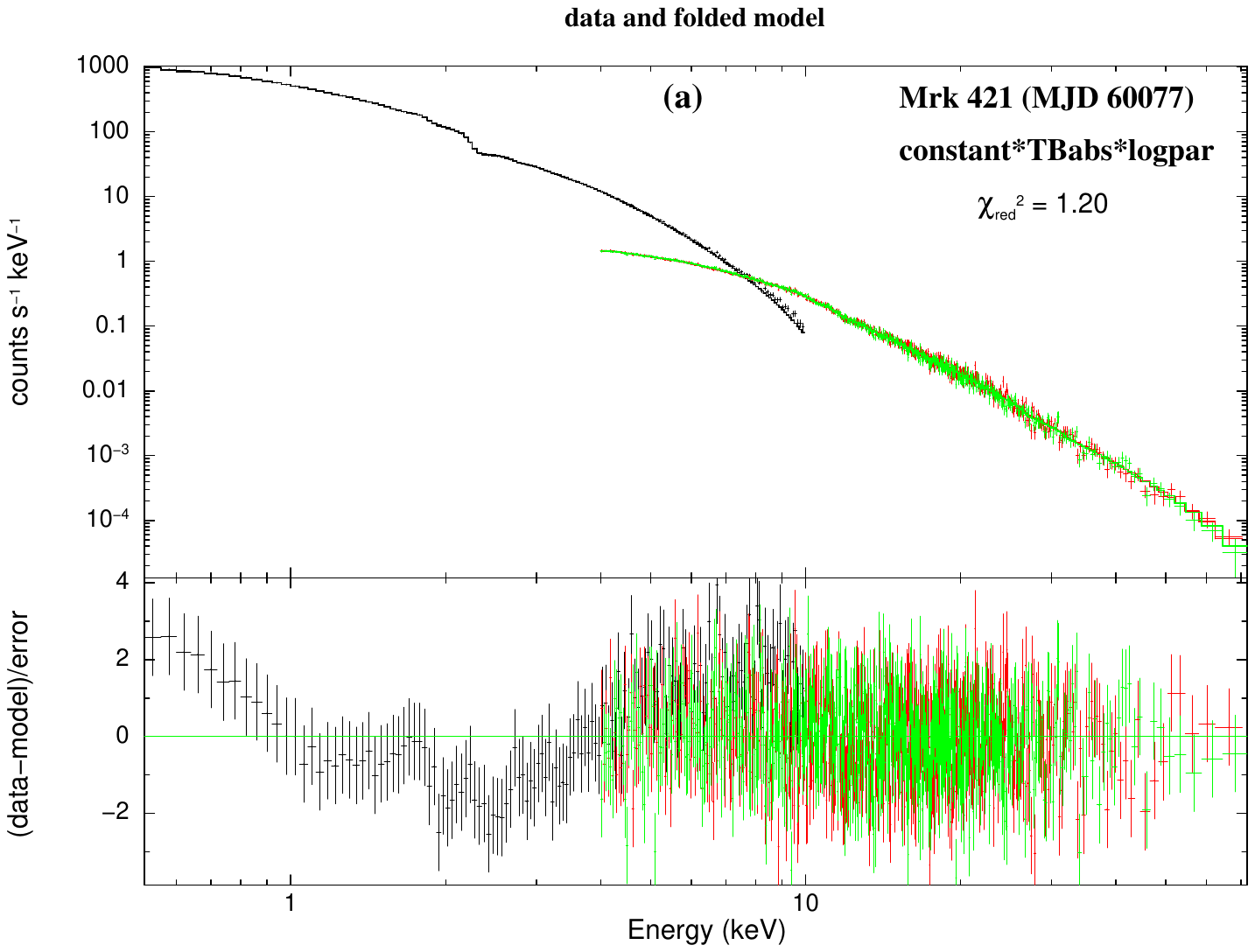}
\end{minipage}
\hfill
  \begin{minipage}[b]{0.49\textwidth}
    \includegraphics[width=\textwidth, angle=0]{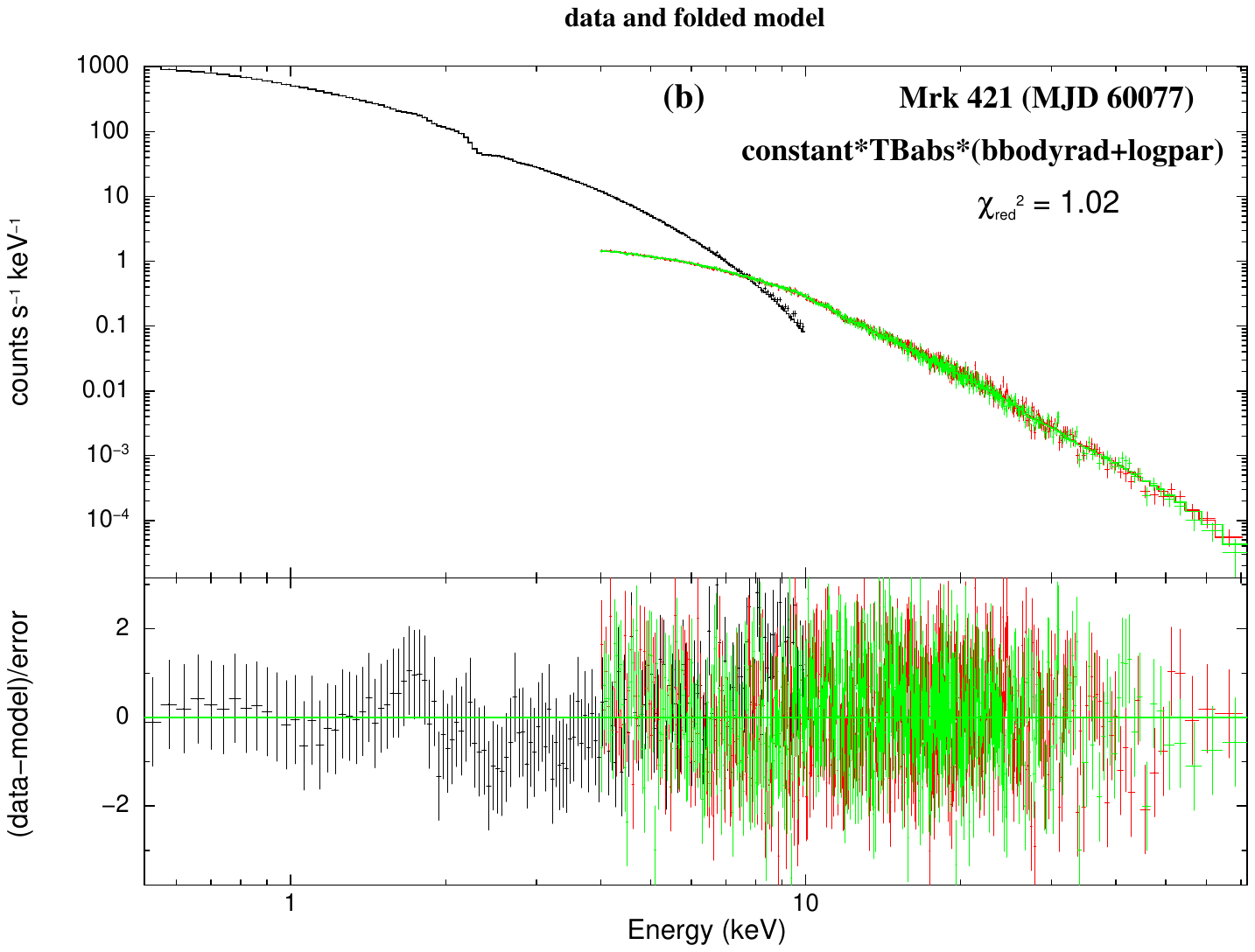}
\end{minipage}
\begin{minipage}[b]{0.49\textwidth}
    \includegraphics[width=\textwidth, angle=0]{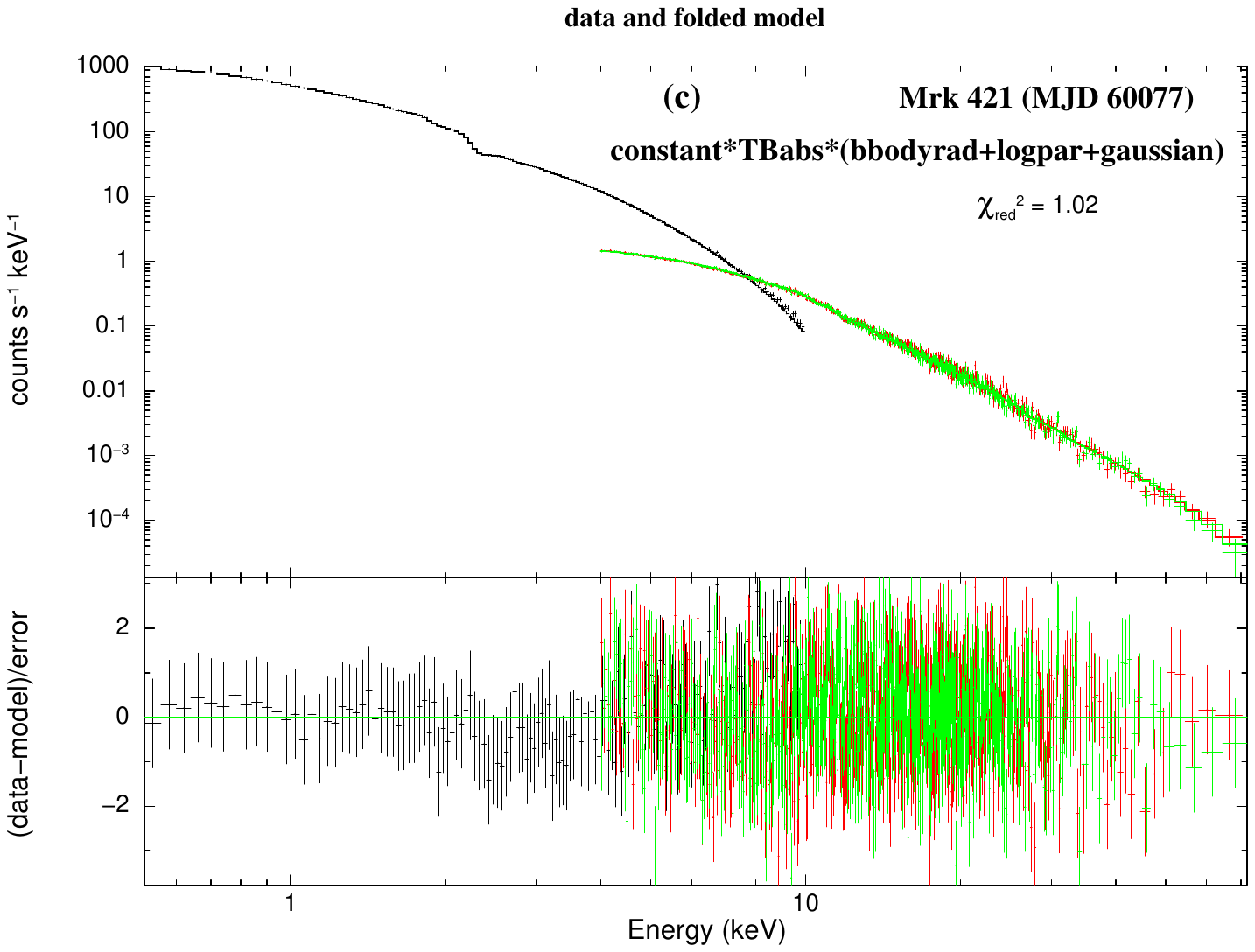}
\end{minipage}
\caption{The combined NICER (black) plus NuSTAR (red points denote FPMA and green points denote FPMB) fitted spectrum of Mrk~421 during MJD 60077. Panel (a) shows fitted spectrum using constant*TBabs*logpar model. Panel (b) shows fitted spectrum using constant*TBabs*(bbodyrad+logpar) model. Panel (c) shows fitted spectrum using constant*TBabs*(bbodyrad+logpar+gaussian) model. The source name, MJD value, fitting models and $\chi_{red}^2$ are mentioned in the panels. \label{Mrk_421_60077}}
\end{figure*}

\noindent
We performed the X-ray spectral analysis with the XSPEC software package version: 12.14.1 \citep {1996ASPC..101...17A}. At first all the spectra were fitted using the Log-Parabolic (LP) model \citep{2004A&A...413..489M}. The LP model is commonly applied in blazar spectral studies due to its effectiveness in representing curved spectral profiles. The LP model (\textit {logpar}) is defined as,
\begin{equation}
%\[
F(E) = K \left(\frac{E}{E_1}\right)^{-\;(\alpha \;+\; \beta \log (E/{E_1}))}
%\]
\end{equation}
in unit of $ph/(cm^2~s~keV)$. Here, $K$ denotes the normalization constant, while $E_1$ represents the pivot energy fixed at 1~keV during spectral fitting. The parameter $\alpha$ corresponds to the photon index at $E_1$, and $\beta$ is the curvature term that characterizes the extent of spectral curvature. \\
\\
In case of four OBSIDs of the BL Lac Mrk~421, we observed that the spectra were not fitted well with the LP model. To fit the spectra properly, we added the \textit{bbodyrad} model along with the LP model. While fitting two spectra of Mrk~501, we needed to add the \textit{bbodyrad} model as well. The \textit {bbodyrad} model is defined as,
\begin{equation}
%\[
F(E) = \frac{K \times 1.0344 \times 10^{-3} \, E^{2} \, dE}{\exp(E/kT) - 1}
%\]
\end{equation}
The model has two parameters. Parameter 1, kT gives the temperature in keV and parameter 2, K is the normalization. $K= r_{km}^2/d_{10}^2$, where $r_{km}$ is the radius of the source in km and $d_{10}$ is the distance of the source in units of 10 kpc. \\
\\
%\sout{We observed a Gaussian nature in spectra of four} 
%\sout{OBSIDs of Mrk~421. We added the \textit{gaussian} model}
%\sout{to account for the Gaussian nature in the spectra.} 
Four spectral fits of Mrk~421 showed significant improvement after the inclusion of a Gaussian line component in the model (see \autoref{Mrk_421_60077}, \autoref{Mrk421_bbrad_lp_ga_rest} and \autoref{Table:results}). The \textit{gaussian} model is defined as,
\begin{equation}
%\[
F(E) = K \frac{2}{\sigma \sqrt{2\pi} \left(1 - {erf}\!\left(-\frac{E_l}{\sqrt{2}\,\sigma}\right)\right)} 
\exp\!\left(-\frac{(E - E_l)^{2}}{2\sigma^{2}}\right)
%\]
\end{equation}
The line profile has three parameters. The parameter 1, $E_l$ is the line energy in keV. Parameter 2, $\sigma$ is the line width in keV. Parameter 3, K is the normalization i.e. the total $photons/cm^2/s$ in the line. \\
\\
In case of PG~1553+113, we had to add an \textit{edge} model along with LP model to fit the spectrum properly. The \textit{edge} model is defined as,
\begin{equation}
%\[
F(E) =
\left\{
\begin{array}{ll}
1 & \text{for } E \leq E_t \\[0.2cm]
\exp\left[-D\left(\frac{E}{E_t}\right)^{-3}\right] & \text{for } E \geq E_t
\end{array}
\right.
%\]
\end{equation}
where, parameter 1, $E_t$ is the threshold energy and parameter 2, D is the absorption depth at the threshold energy. \\
\\
Galactic absorption was addressed by applying the \textit{TBabs} model (Tubingen–Boulder interstellar medium absorption model; \citet {2000ApJ...542..914W}), which was convolved with the previously described models. The $N_H$ values for the blazars were obtained from the HI4PI Collaboration \citep{2016A&A...594A.116H} and for the spectral fitting the $N_H$ values were fixed to the average value. 
%%
% \begin{figure*}
% \centering
%   \begin{minipage}[b]{0.49\textwidth}
% \includegraphics[width=0.99\textwidth]{logpar_ld_60429_421_A.pdf}
% \end{minipage}
% \hfill
%   \begin{minipage}[b]{0.49\textwidth}
% \includegraphics[width=0.99\textwidth]{logpar_ld_59660_501_A.pdf}
% \end{minipage}
% \hfill
%   \begin{minipage}[b]{0.49\textwidth}
% \includegraphics[width=0.99\textwidth]{logpar_ld_60285_PKS_A.pdf}
% \end{minipage}
% \caption{The combined NICER (black) plus NuSTAR (red) fitted spectra of Mrk~421, Mrk~501 and PKS~2155-304 using absorbed LP (constant*TBabs*logpar) model. The source names, MJD values, fitting model and $\chi_{red}^2$ are mentioned in the panels. \label{logpar}}
% \end{figure*} 
\begin{figure*}
\centering
\begin{minipage}[b]{0.49\textwidth}
\includegraphics[width=0.99\textwidth]{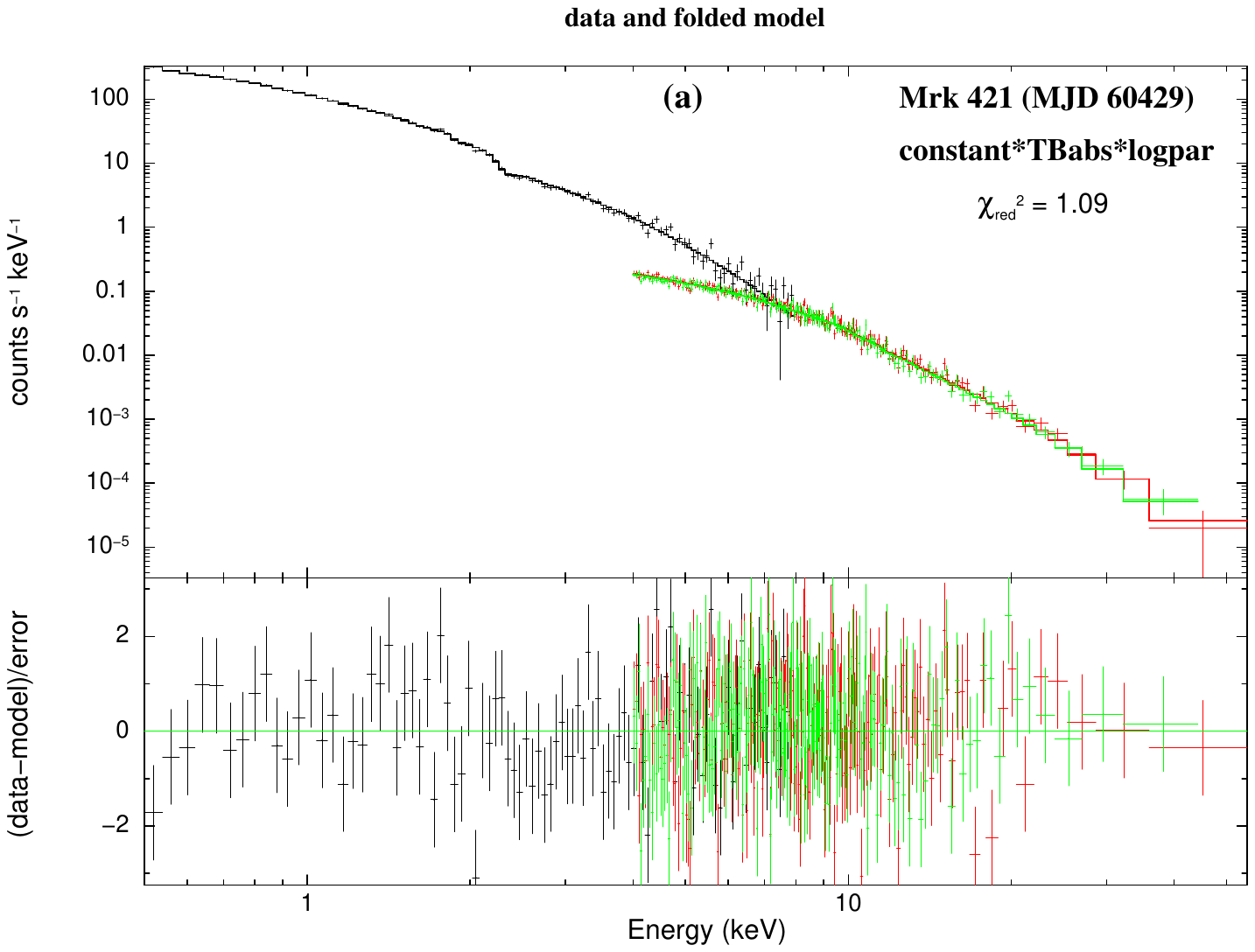}
\end{minipage}
\hfill
  \begin{minipage}[b]{0.49\textwidth}
\includegraphics[width=0.99\textwidth]{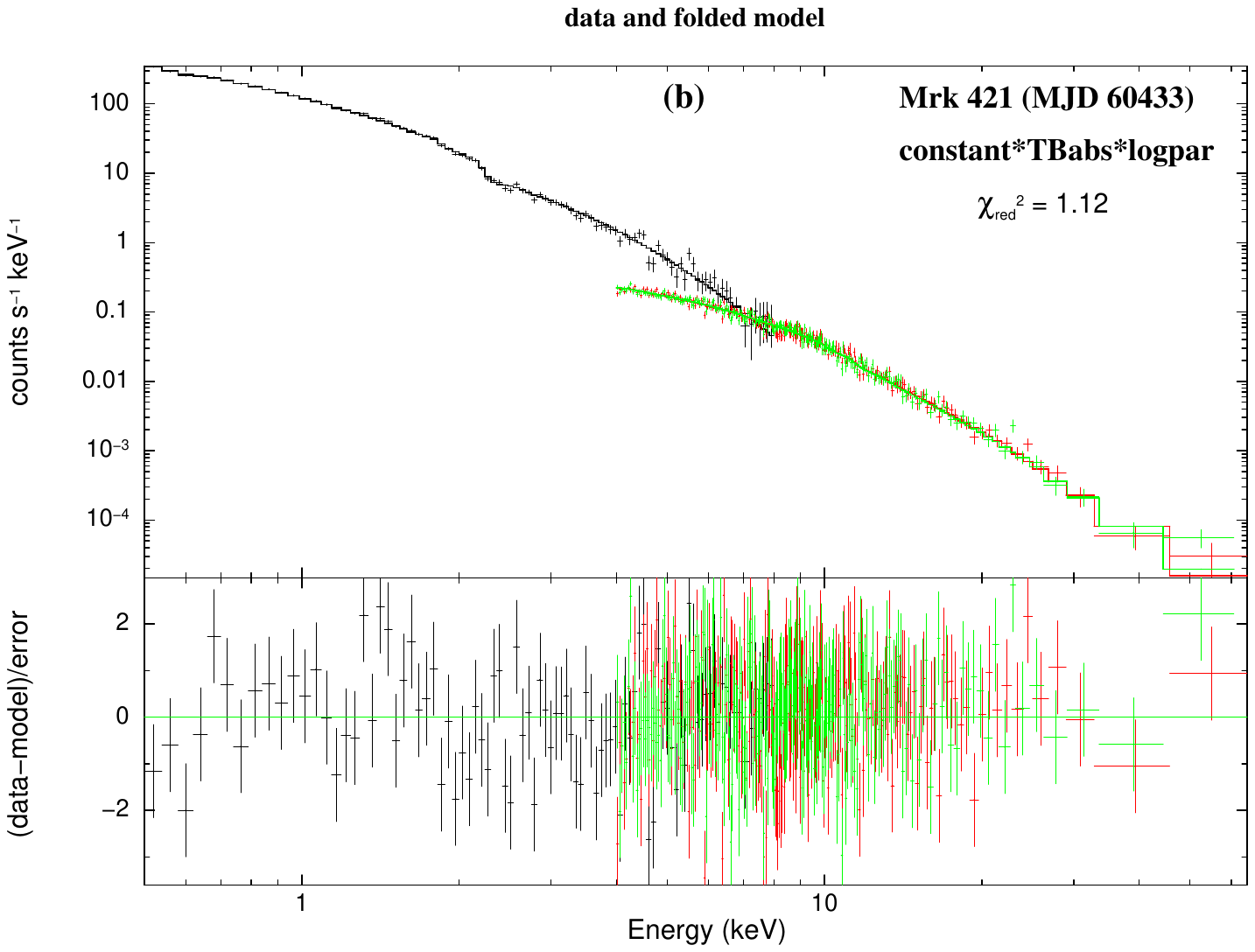}
\end{minipage}
\hfill
  \begin{minipage}[b]{0.49\textwidth}
\includegraphics[width=0.99\textwidth]{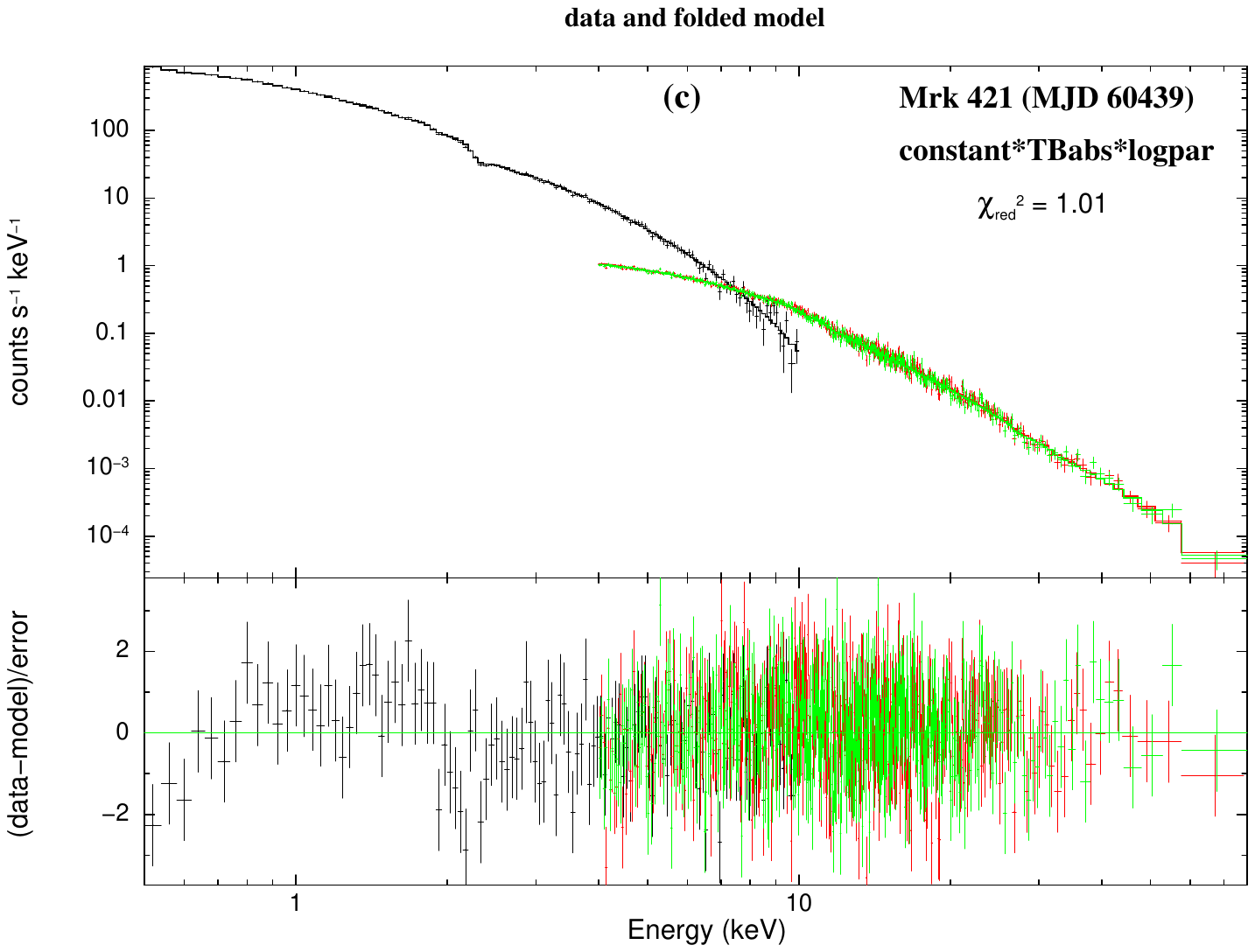}
\end{minipage}
\hfill
  \begin{minipage}[b]{0.49\textwidth}
\includegraphics[width=0.99\textwidth]{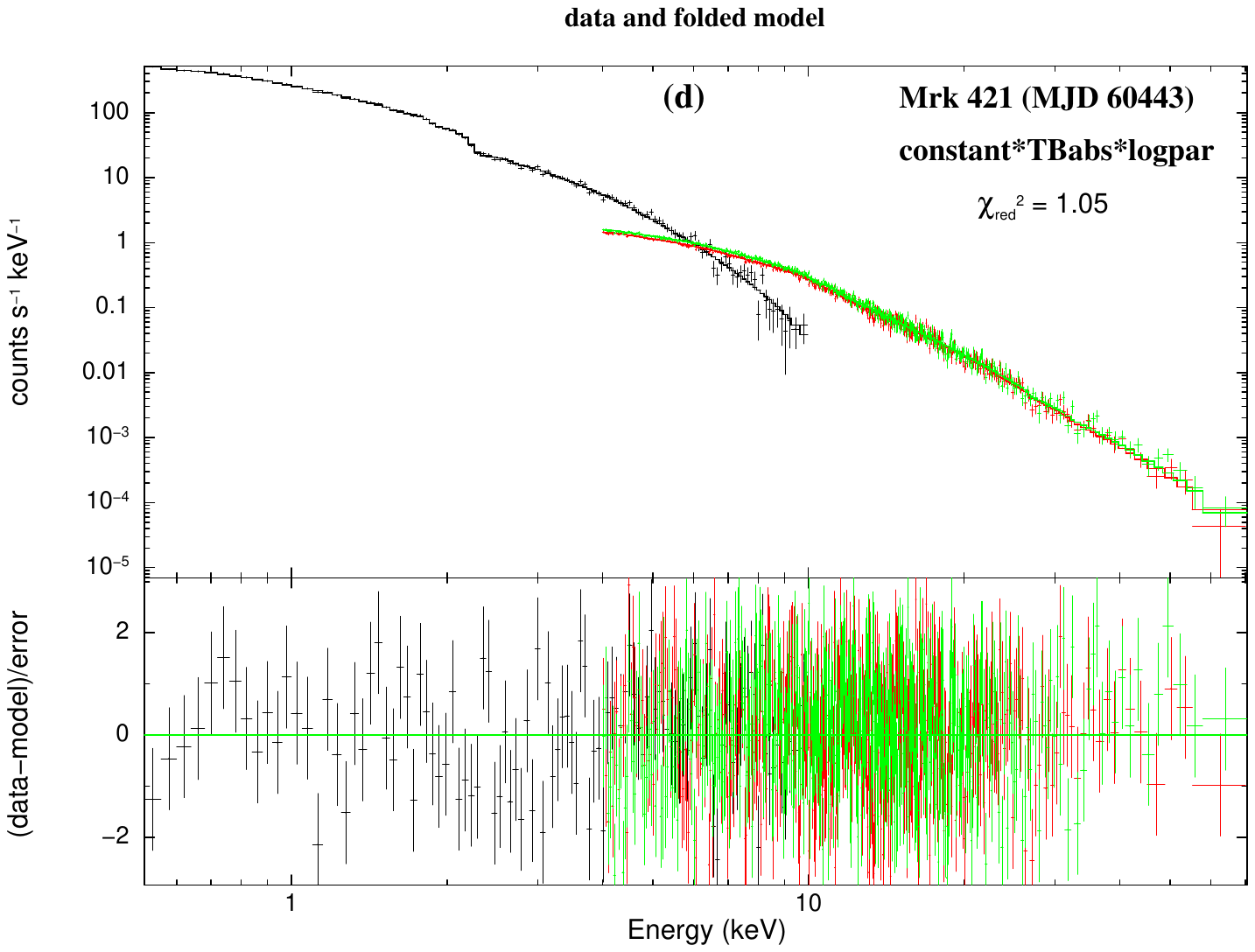}
\end{minipage}
\caption{The combined NICER (black) plus NuSTAR (red points denote FPMA and green points denote FPMB) fitted spectra of Mrk~421 during MJD 60429, 60433, 60439 and 60443 using absorbed LP (constant*TBabs*logpar) model. The source name, MJD values, fitting model and $\chi_{red}^2$ are mentioned in the panels.\label{Mrk_421_logpar}}
\end{figure*} 
%%%%%%%%%%%%
%%%%%%%%%%%%%
\begin{figure*}
\centering
\includegraphics[width=0.90\textwidth]{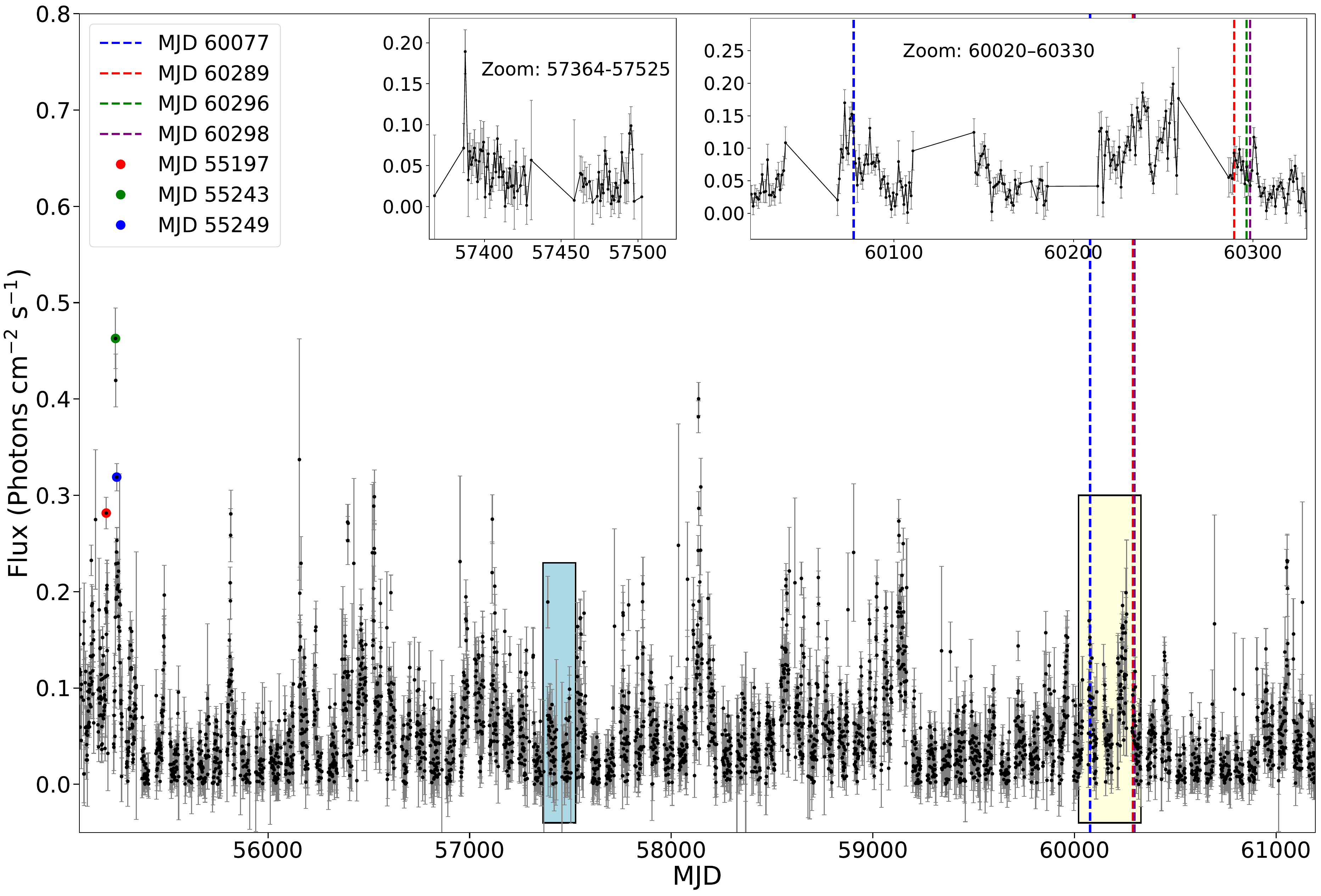}
\caption{The 2--20~keV MAXI/GSC light-curve of Mrk~421 from MJD 55065 to 61195. The interval enclosed by the light-yellow rectangular box, spanning MJD 60020--60330, is shown in the zoomed-in panel on the upper right. The epochs corresponding to MJD 60077, 60289, 60296, and 60298 are marked by blue, red, green, and purple dashed vertical lines, respectively. The light-blue shaded box marks a low-flux state region (MJD 57364--57525), which is displayed in the zoomed-in panel at the upper center. The red, green and blue dots, corresponding to MJD 55197, 55243 and 55249 respectively, mark the high-flux states of the source.
\label{lc_Mrk_421}}
\end{figure*} 
%%%%%%%%%%%%%
\section{Results}
\noindent
In the following, we discuss the spectral analysis results for each source individually. To conduct the spectral analysis, the NICER spectrum and the spectra from the two NuSTAR detector units, FPMA and FPMB, were fitted simultaneously for each observational epoch. %\sout{In case of NuSTAR data, we used both the FPMA} \sout{and FPMB spectra and observed that the value of} \sout{the fitted parameters did not vary much. In the} \sout{\autoref{Table:results} we have noted the value obtained using} \sout{spectra of FPMA module of NuSTAR along with} \sout{NICER spectra. In all the Figures we have shown} \sout{the fitted spectra of simultaneous NuSTAR (FPMA)} \sout{data (red) and NICER data (black)}. 
The value of the best fitted parameters obtained from the spectral fits are listed in \autoref{Table:results}. The spectral-fitting figures present the fitted spectra of simultaneous NICER and NuSTAR observations, where the black points represent the NICER data, and the red and green points correspond to the NuSTAR FPMA and FPMB data respectively.
\subsection{Mrk~421}
\noindent
Mrk~421 is the closest known (redshift z = 0.031) and the first VHE TeV $\gamma-$ray emitting blazar \citep{1992Natur.358..477P}. The optical spectrum of its host galaxy is being used to estimate its central SMBH mass, which was evaluated to be (2 -- 9) $\times \ \rm{10}^{9} \rm{M}_{\odot}$ \citep{2002ApJ...569L..35F,2002A&A...389..742W,2003ApJ...583..134B}. The source is one of the prime targets of variability studies on diverse timescales in MW covering the whole EM spectrum \citep[e.g.][and references therein]{2000ApJ...542L.105T,2005ApJ...630..130B,2008ApJ...677..906F,2009ApJ...695..596H,2011ApJ...738...25A,2021MNRAS.504.1427A,2015A&A...576A.126A,2016ApJ...819..156B}. Since Mrk~421 is a strong X-ray and $\gamma-$ray emitter, it has been extensively studied in these energy ranges using various satellites (in X-ray and $\gamma-$rays) and ground-based $\gamma-$ray telescopes \citep[e.g.][and references therein]{2005A&A...437...95A,2007ApJ...663..125A,2007A&A...466..521T,2009A&A...501..879T,2011ApJ...736..131A,2016ApJ...819..156B,2022ApJS..262....4N,2023NatAs...7.1245D}. X-ray timing, polarization and spectral properties of Mrk~421 using the data from various satellites have been extensively carried out \citep[e.g.][and references therein]{2007A&A...466..521T,2013MNRAS.434.2684M,2017ApJ...842..129C,2017ApJ...848..103K,2020MNRAS.499.2094G,2021MNRAS.508.5921H,2022ApJS..262....4N,2022MNRAS.513.1662M,2022A&A...663A.178M,2023NatAs...7.1245D}. X-ray spectral analysis using all Swift/XRT observations of Mrk 421 between April 2006 and July 2006 was carried out, in which the X-ray spectrum is described well by an LP distribution \citep{2009A&A...501..879T}. Coordinated multi-wavelength observations of Mrk 421 taken in 2013 January–March, involving Swift and NuSTAR, resulted in the spectrum softening when the source is dimmer until the X-ray spectral shape saturates into a steep power law \citep{2016ApJ...819..156B}. The simultaneous Swift/XRT and NuSTAR observations of Mrk 421 show significant spectral curvature that can be reproduced by an LP function \citep{2020MNRAS.499.2094G}. The first evidence of a clear detection of a hard X-ray excess in Mrk 421 above 20 keV was reported by \citet{2016ApJ...827...55K}, which was later explained as being produced from the spine/layer jet structure \citep{2017ApJ...842..129C}. \\
\\
We used eight simultaneous archived data of NICER and NuSTAR for the spectral analysis of the blazar Mrk~421. We used the NuSTAR spectra in the energy range 4.0--78.0~keV for all the OBSIDs. Depending on the source and the background spectra we had to use different energy ranges for the NICER spectra of different OBSIDs. The energy ranges used for the OBSIDs in this study are noted in the \autoref{Table:log}. We fitted all of the eight spectra using the absorbed LP model ($constant*TBabs*logpar$) first. The spectra of the four MJDs: 60429, 60433, 60439 and 60443, were fitted satisfactorily using the absorbed LP model. The fitted spectra (using absorbed LP model) of MJD 60429, %is shown in \autoref{logpar}a and the fitted spectra (using absorbed LP model) of MJDs 
60433, 60439 and 60443 can be seen in \autoref{Mrk_421_logpar}. The value of the fitted parameters are mentioned in the \autoref{Table:results}. \\
\\
In case of the four MJDs (60077, 60289, 60296 and 60298) it can be seen from the residual (\autoref{Mrk_421_60077}a and \autoref{Mrk421_bbrad_lp_ga_rest}), that fitting the spectra with the absorbed LP model did not yield satisfactory results. The spectra were not fitted well in the low energy range (below 4.0 keV). In \citet {2022A&A...663A.178M}, the authors showed that an accretion disk–driven model can be employed for Mrk~421 during periods of low to moderate flux states. We added the \textit{bbodyrad} model to fit the soft energy part of the spectra along with the absorbed LP model and the fitting model became: $constant*TBabs*(bbodyrad+logpar)$. The fitted spectra using the $constant*TBabs*(bbodyrad+logpar)$ model can be seen in \autoref{Mrk_421_60077}b and \autoref{Mrk421_bbrad_lp_ga_rest}. It can be clearly seen that the fitting improved in the soft energy range after adding the \textit{bbodyrad} model. 
%\sout{and the value of the reduced chi-square ($\chi_{red}^2$) improved} \sout{from 1.30 to 1.02 in case of MJD 60077.} 
While fitting the spectra using $constant*TBabs*(bbodyrad+logpar)$ model we could see the presence of a Gaussian feature in the residual of all the four MJDs. We added an additional Gaussian line model (\textit{gaussian}) to fit the spectra properly and the fitting model became: $constant*TBabs*(bbodyrad+logpar+gaussian)$. The fitted spectra using the $constant*TBabs*(bbodyrad+logpar+gaussian)$ model can be seen in the \autoref{Mrk_421_60077}c and \autoref{Mrk421_bbrad_lp_ga_rest} and from the residual we can observe that the spectra are fitted satisfactorily. 
%\sout{The unfolded spectra of these four MJDs using the same} \sout{model are shown in the \autoref{MJD_60077_421}d and \autoref{Mrk421_A}.}
\\
\\
In case of Mrk~421, the value of photon index ($\alpha$) varied in between $2.03_{-0.01}^{+0.01}$ to $2.50_{-0.02}^{+0.02}$ while curvature parameter ($\beta$) ranged from $0.25_{-0.02}^{+0.02}$ to $0.46_{-0.03}^{+0.04}$. In case of MJD 60077, the value of the disk temperature was found to be $1.51_{-0.06}^{+0.06}$~keV and we detected a Gaussian line at $1.70_{-0.23}^{+0.10}$~keV having a width of $0.10_{-0.09}^{+0.08}$~keV. For the MJDs 60289, 60296 and 60298 the obtained disk temperature was $\sim$ 0.15~keV. We observed a Gaussian line at $1.47_{-0.42}^{+0.15}$~keV having a width of $0.28_{-0.08}^{+0.07}$~keV for the MJD 60289. For the MJD 60296, the Gaussian line was found at $1.42_{-0.55}^{+0.17}$~keV having a width of $0.31_{-0.05}^{+0.05}$~keV. For MJD 60298, the Gaussian line was detected at $1.56_{-0.12}^{+0.11}$~keV having a width of $0.18_{-0.08}^{+0.09}$~keV. We have shown the value of the fitted parameters using the absorbed LP model and the $constant*TBabs*(bbodyrad+logpar+gaussian)$ model in the \autoref{Table:results}. The origin of the observed spectral lines could not be conclusively identified. In NICER spectra, instrumental features such as the Al-K$\alpha$ line at $\sim$1.56~keV and the Si-K$\alpha$ line at $\sim$1.84~keV are known to arise from the detector and other instruments of the satellite\footnote{\url{https://heasarc.gsfc.nasa.gov/docs/nicer/data_analysis/workshops/NICER-CalStatus-Markwardt-2021.pdf}}. The Gaussian features detected in the present study can be associated with instrumental effects. The lines can also be originated from background contamination. \\
\\
We calculated flux from the fitted spectra of Mrk~421 using $constant*TBabs*(bbodyrad+logpar+gaussian)$ model for the OBSIDs having MJDs 60077, 60289, 60296 and 60298, in which cases we observed the disk contribution. In 0.5--10.0~keV energy range the flux values obtained for the MJDs 60077, 60289, 60296 and 60298 are 13.60~$\times~10^{-10}$, 9.89~$\times~10^{-10}$, 11.48~$\times~10^{-10}$ and 9.14~$\times~10^{-10}$ in units of $ergs~cm^{-2}~sec^{-1}$ respectively. In 10.0--78.0~keV energy range the flux values obtained for the MJDs 60077, 60289, 60296 and 60298 are 2.11~$\times~10^{-10}$, 0.72~$\times~10^{-10}$, 0.90~$\times~10^{-10}$ and 0.79~$\times~10^{-10}$ in units of $ergs~cm^{-2}~sec^{-1}$ respectively.\\  
\\
The \autoref{lc_Mrk_421} shows 2--20~keV MAXI/GSC light-curve\footnote{\url{https://maxi.riken.jp/star_data/J1104+382/J1104+382.html}} of the blazar Mrk~421 over the period from MJD 55065--61195. 
The four epochs corresponding to MJD 60077, 60289, 60296, and 60298 are marked by dashed vertical lines (blue, red, green, and purple respectively). A zoomed-in view of the interval spanning MJD 60020--60330 (marked by light-yellow shaded box) is shown in the upper-right panel of the figure to highlight the flux variations around the four observational epochs in greater detail. During MJDs 60077, 60289, 60296 and 60298 the 2--20~keV flux of the source was found to be $0.13\pm0.02$, $0.09\pm0.02$, $0.07\pm0.02$, and $0.04\pm0.02$ in units of $photons~cm^{-2}~sec^{-1}$ respectively.  Considering the entire light-curve over the period MJD 55065--61195, the average 2--20~keV flux of the source is $\sim0.05~photons~cm^{-2}~s^{-1}$.
The source was reported to be in high-flux state during MJD 55197, 55243 and 55249 \citep{2010PASJ...62L..55I}. From the 2--20~keV MAXI/GSC light-curve, during MJDs 55197 (indicated by red dot), 55243 (indicated by green dot) and 55249 (indicated by blue dot), the flux of the source was found to be $0.28\pm0.02$, $0.46\pm0.03$ and $0.32\pm0.01$ in units of $photons~cm^{-2}~sec^{-1}$ respectively. The source was reported to be in low-flux state during MJD 57364--57525 \citep{2021MNRAS.504.1427A}. This interval (MJD 57364--57525) is marked by a light-blue shaded box and is shown in the zoomed-in panel at the upper center of \autoref{lc_Mrk_421}. It can be observed that during this period the 2--20~keV flux of the source remained $\lesssim$0.10~$photons~cm^{-2}~s^{-1}$. Therefore, in comparison with the high and low-flux states, the source can be considered to be in a moderate to low-flux state during the four epochs, MJD 60077, 60289, 60296, and 60298.

%%%%%%%%%%%%%%%%%%%%%%%%%%%%%%%%%%%%%%%%%%%%%%%%%%%%%%%%%%%%%%%%%%%%%%%%%%%%%%%%%%
\begin{figure*}
\centering
\begin{minipage}[b]{0.52\textwidth}
\includegraphics[width=0.99\textwidth]{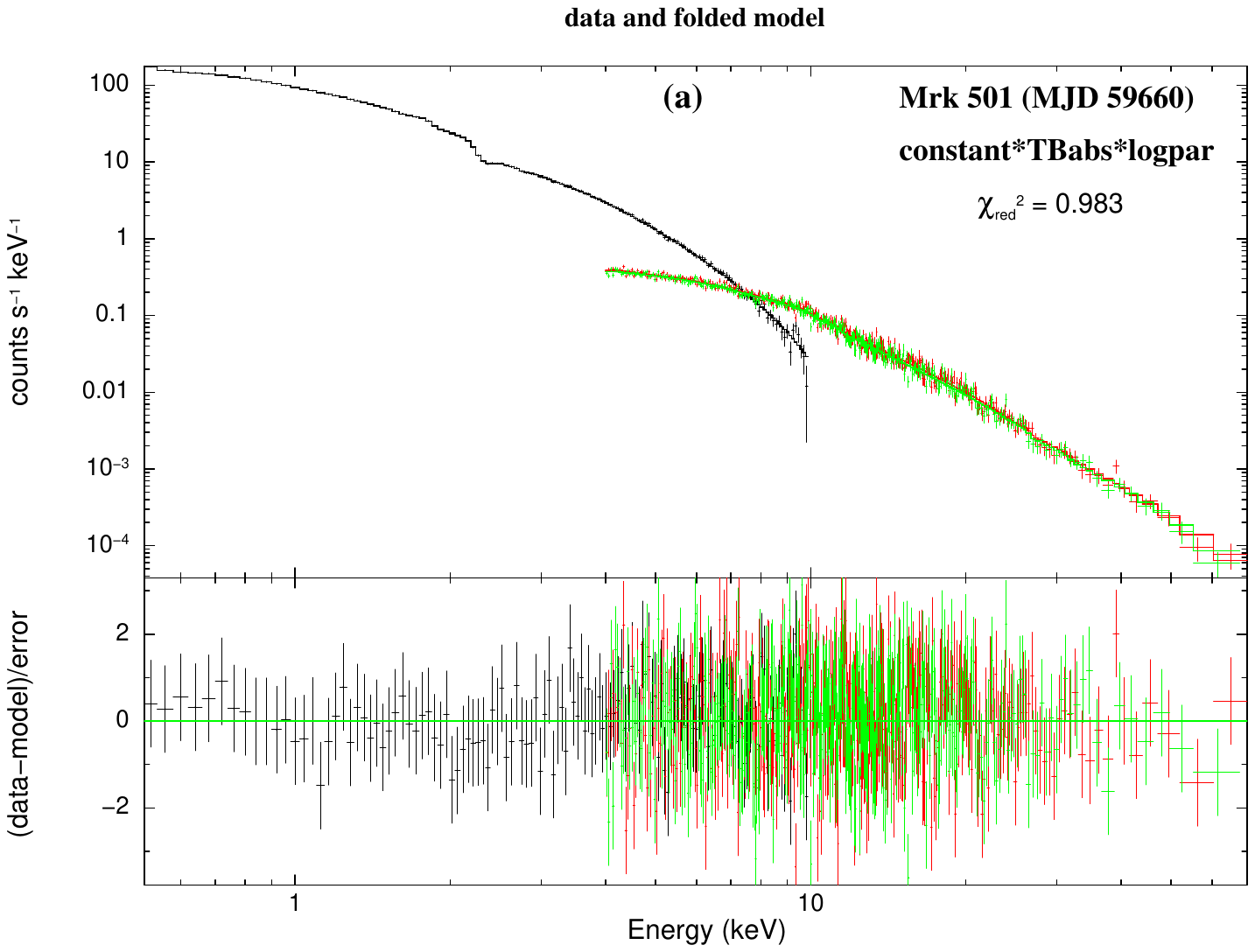}
\end{minipage}
\hfill
\begin{minipage}[b]{0.49\textwidth}
    \includegraphics[width=\textwidth, angle=0]{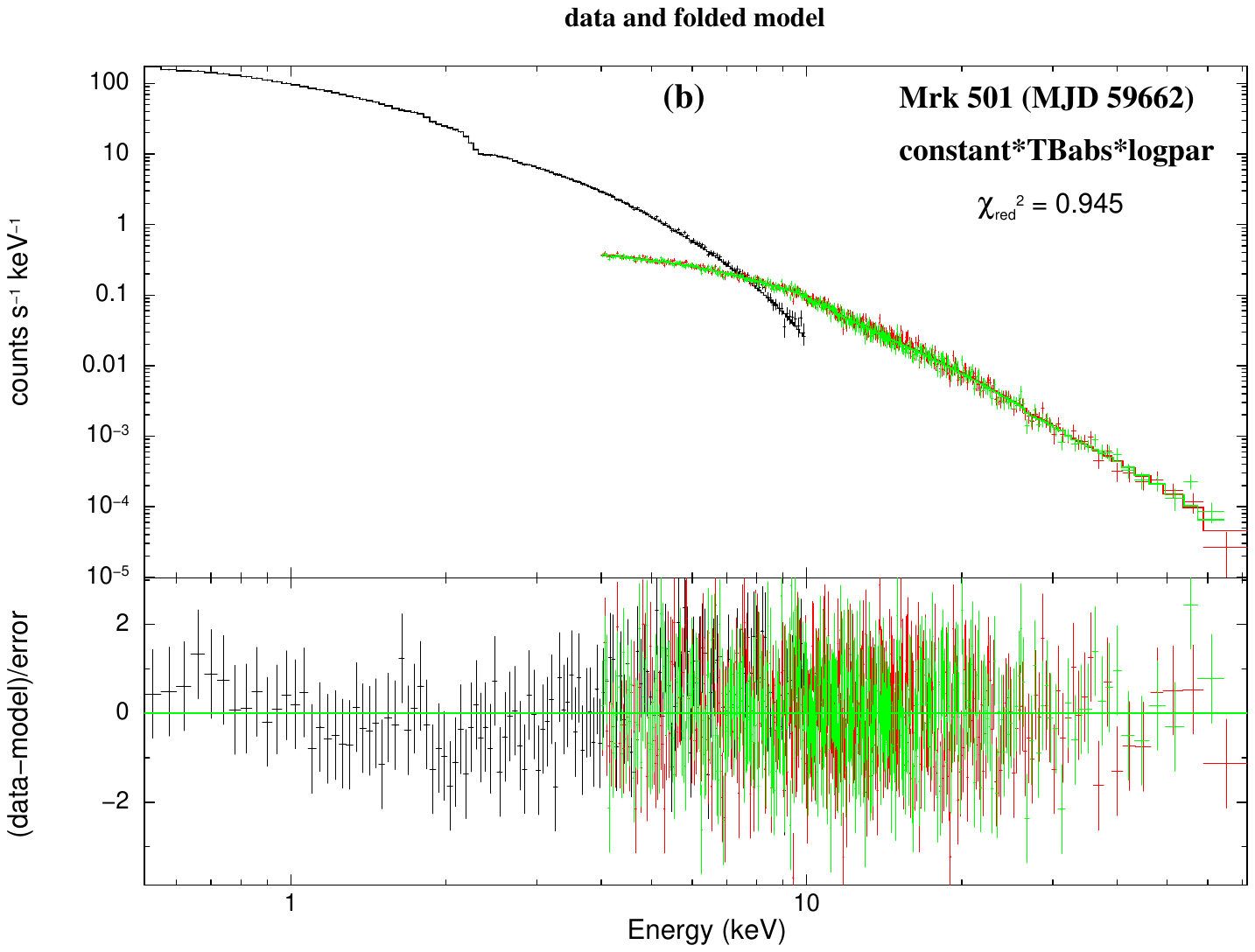}
\end{minipage}
\hfill
\begin{minipage}[b]{0.49\textwidth}
    \includegraphics[width=\textwidth, angle=0]{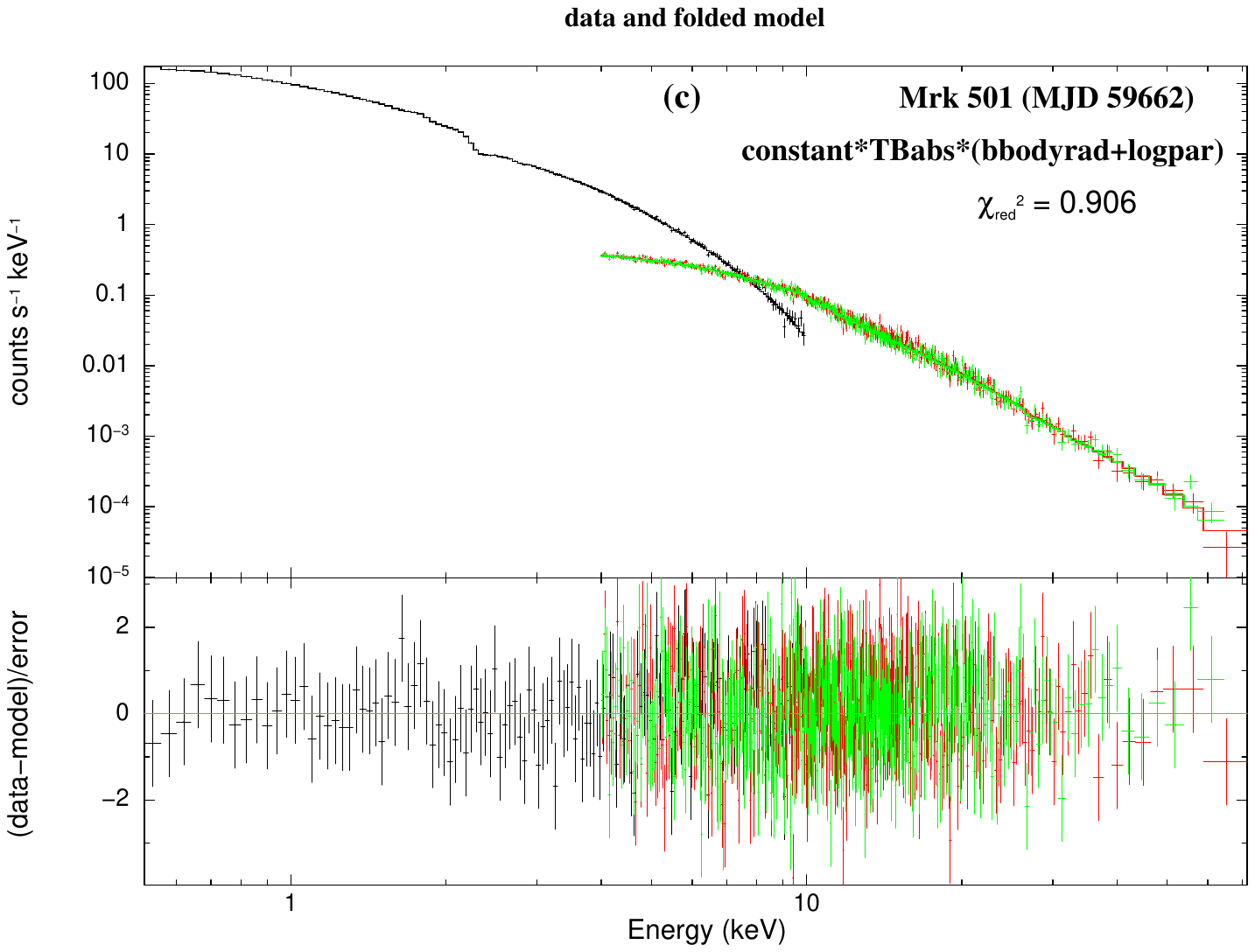}
\end{minipage}
\hfill
  \begin{minipage}[b]{0.49\textwidth}
    \includegraphics[width=\textwidth, angle=0] {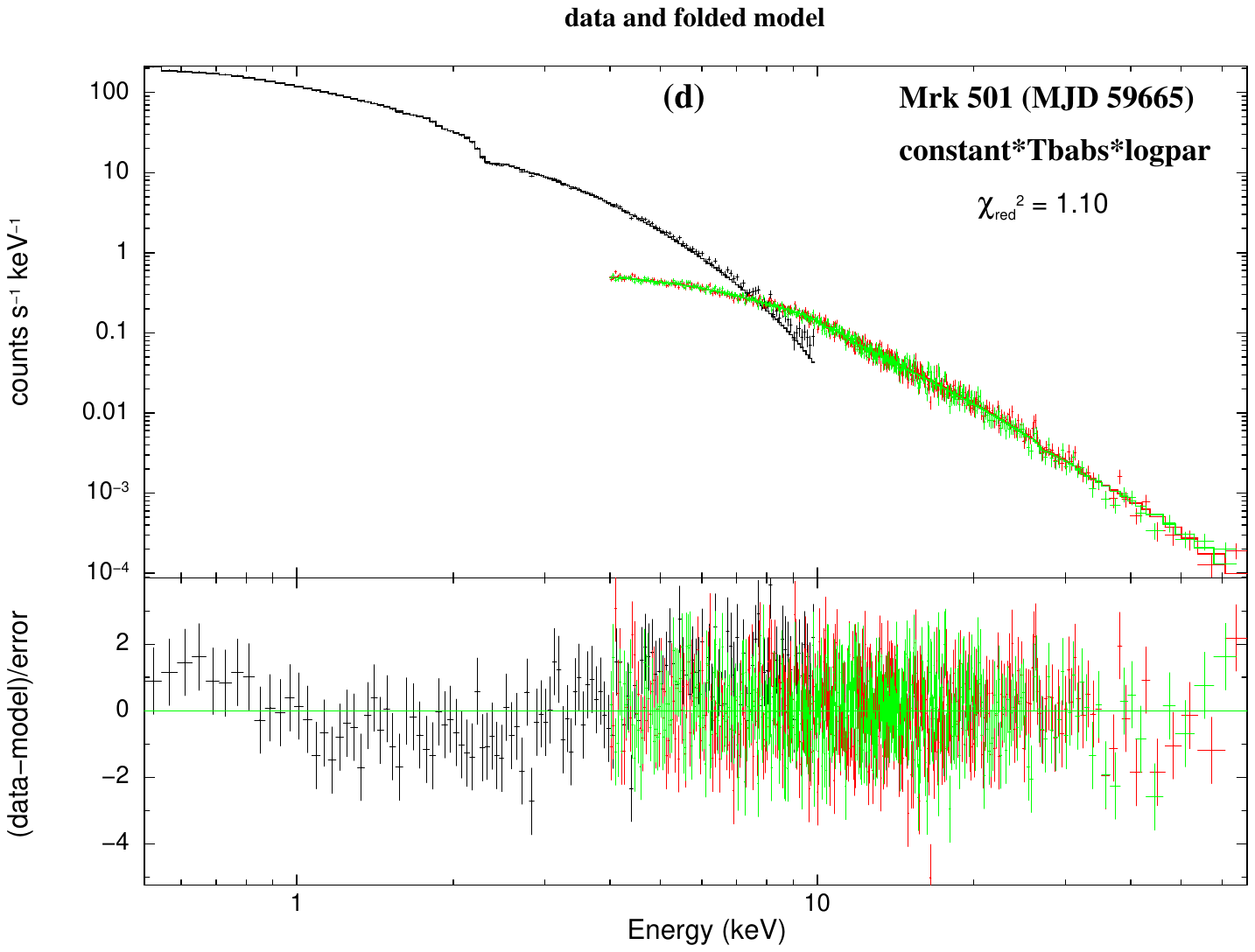}
\end{minipage}
\hfill
\begin{minipage}[b]{0.49\textwidth}
    \includegraphics[width=\textwidth, angle=0]{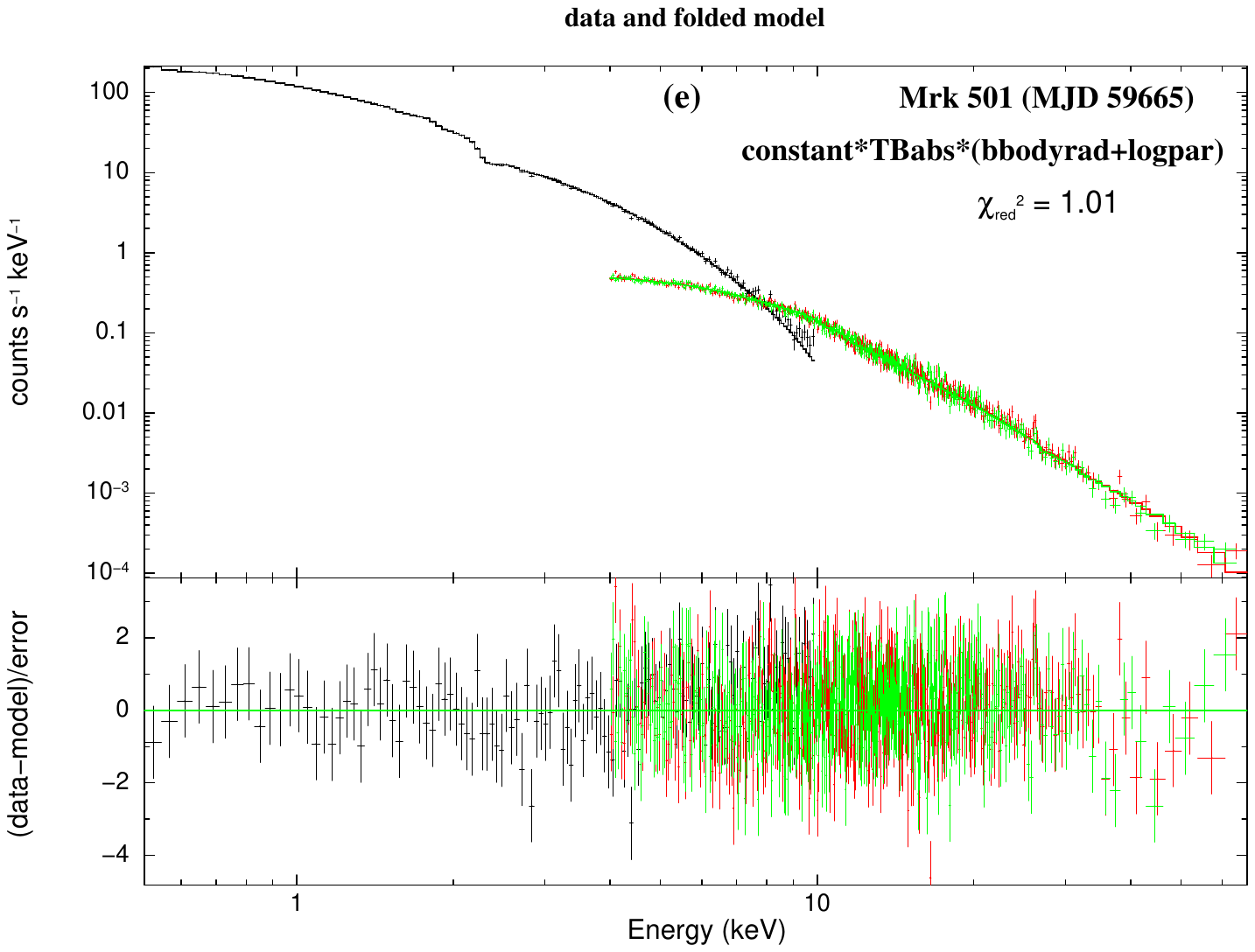}
\end{minipage}
\caption{The combined NICER (black) plus NuSTAR (red points denote FPMA and green points denote FPMB) fitted spectra of Mrk~501 during different MJDs. Panel (a),(b) and (d) show fitted spectra using absorbed LP (constant*TBabs*logpar) model for MJD 59660, 59662 and 59665 respectively. Panel (c) and (e) show fitted spectra using constant*TBabs*(bbodyrad+logpar) model for MJD 59662 and 59665 respectively. The source name, MJD values, fitting models and $\chi_{red}^2$ are mentioned in the panels. \label{Mrk_501}}
\end{figure*}
\begin{figure*}
\centering
\includegraphics[width=0.90\textwidth]{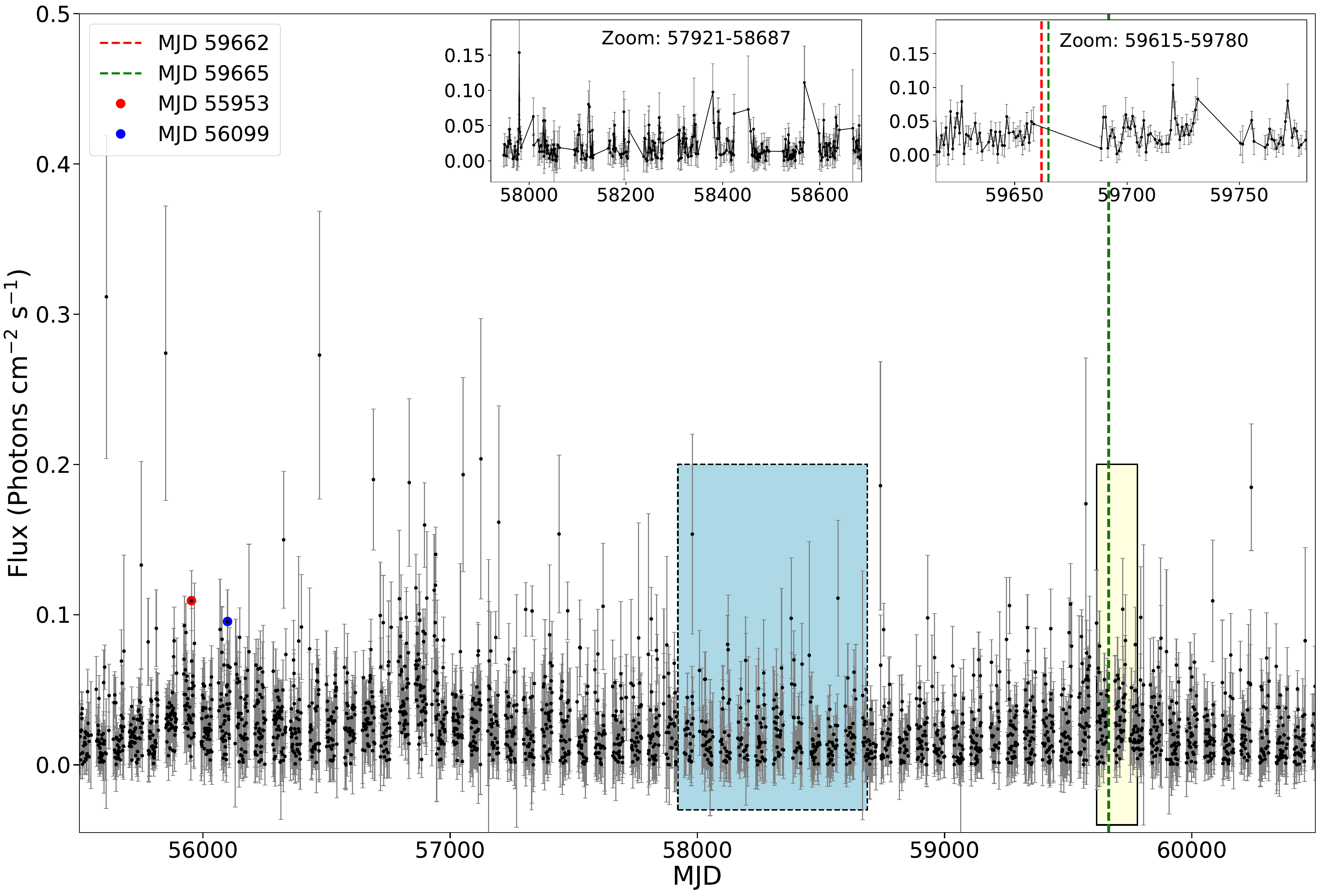}
\caption{The 2--20~keV MAXI/GSC light-curve of Mrk~501 from MJD 55500 to 60500. The interval enclosed by the light-yellow rectangular box, spanning MJD 59615--59780, is shown in the zoomed-in panel at the upper right side. The epochs corresponding to MJD 59662 and 59665 are marked by red and green dashed vertical lines, respectively. The light-blue shaded box marks a low-flux state region (MJD 57921--58687), which is displayed in the zoomed-in panel at the upper center. The red and blue dots, corresponding to MJD 55953 and 56099 respectively, mark the high-flux state of the source during the X-ray flaring activity from MJD 55840--56140. \label{lc_Mrk_501}}
\end{figure*} 
%%%%%%%%%%%%%%%
\subsection{Mrk~501}
\noindent 
Mrk 501 is one of the closest known (redshift z = 0.0337) \citep{1975ApJ...198..261U} and HSP blazars defined as having a $\nu_{syn} > \rm{10}^{15}$ Hz \citep{2020ApJ...892..105A} and among the first VHE TeV gamma-ray-emitting blazars \citep{1996ApJ...456L..83Q,1997A&A...320L...5B}. By using the optical spectrum of Mrk 501, the stellar velocity dispersion of its host galaxy, and assuming that if the source follows the M$_{BH} - \sigma_{*}$ correlation defined for local galaxies, then its central SMBH has a mass of (0.9 -- 3.4) $\times \ \rm{10}^{9} \ \rm{M}_{\odot}$ \citep{2002ApJ...566L..13B}. 
%\sout{Mrk 501 is one of the prime TeV emitting blazars of} \sout{variability studies on diverse timescales in MW covering} \sout{the entire EM spectrum} 
Mrk 501 is one of the prime TeV-emitting blazars and is extensively studied for its variability across diverse timescales through MW observations covering the entire EM spectrum \citep[e.g.][and references therein]{2000ApJ...536..742P,2001A&A...367..809K,2009ApJ...705.1624A,2015A&A...573A..50A,2017A&A...603A..31A}. Since Mrk 501 is a powerful X-ray and $\gamma-$ray emitter, it has been thoroughly investigated in these energy ranges using a variety of X-ray and $\gamma-$ray satellites as well as ground-based $\gamma-$ray telescopes \citep[e.g.][and references therein]{1996ApJ...456L..83Q,1998ApJ...492L..17P,1999ApJ...514..138K,2000A&A...353...97K,2002MNRAS.336..721K,2007ApJ...669..862A,2011ApJ...727..129A}. Numerous studies have been conducted on the X-ray timing, polarization, and spectral characteristics of Mrk 501 using data from different satellites \citep[e.g.][and references therein]{1998ApJ...492L..17P,2001ApJ...554..725T,2004A&A...422..103M,2015ApJ...812...65F,2020A&A...637A..86M,2024A&A...685A.117M}. In a high TeV flux state, Mrk 501 was observed by BeppoSAX in April 1997 over the whole energy range 0.1–200~keV and found that the spectrum was exceptionally hard, indicating that the X-ray power output peaked at (or above) $\sim$100~keV \citep{1998ApJ...492L..17P}. X-ray spectral analysis of Mrk 501 using BeppoSAX observations from 1996 to 2001 was carried out during diverse flux states, e.g., from spring 1997 to early 1999 in a very active and bright state, whereas the source went into the low flux state in late 2001, and it was noticed that the X-ray energy distribution of Mkn 501 is well described by a log-parabolic law in all flux states \citep{2004A&A...413..489M}. In simultaneous broadband observations of Mrk 501 between April and August 2013, the characterization of the synchrotron peak of SED with Swift and NuSTAR data was obtained, and it was found that the NuSTAR and combined Swift and NuSTAR spectra of the source were well fitted by the LP model \citep{2015ApJ...812...65F}. Using six pointed observations of Mrk 501 with XMM-Newton from 2002 to 2011, \citet{2022MNRAS.510.5280M} performed X-ray spectral analysis and found different spectra are well fitted with PL (power law), LP, and BPL (broken power law) models. \\
\\
We used three available simultaneous archived data of NICER and NuSTAR for the spectral analysis of the blazar Mrk~501. All three spectra were analyzed using NICER data from 0.5--10.0~keV and NuSTAR data from 4.0--78.0~keV. Firstly, we fitted all of the three simultaneous spectra of MJDs 59660, 59662 and 59665 using absorbed LP model. The absorbed LP model ($constant*TBabs*logpar$) was able to fit the spectrum of the MJD 59660 satisfactorily. \autoref{Mrk_501}a shows the fitted spectrum and residual of the MJD 59660 using the absorbed LP model. However, in case of the MJDs 59662 and 59665 (\autoref{Mrk_501}b and \autoref{Mrk_501}d), we can see that the spectra are not fitted properly below the 10~keV energy range. %\sout{We can see a concave nature in the residual of} \sout{the fitted spectra.}
We added a \textit{bbodyrad} model along with the absorbed LP model to fit the spectra properly, and the fitting model became: $constant*TBabs*(bbodyrad+logpar$). From the residual of \autoref{Mrk_501}c and \autoref{Mrk_501}e it can be seen that using the $constant*TBabs*(bbodyrad+logpar$) model, the fitting improved. The $\chi_{red}^2$ improved from 1.10 to 1.01 in case of MJD 59665. The $\chi_{red}^2$ changed from 0.95 to 0.91 in case of MJD 59662. In case of Mrk~501, the photon index ($\alpha$) varied between $1.83_{-0.01}^{+0.01}$ and $1.90_{-0.01}^{+0.01}$, while the curvature parameter ($\beta$) ranged from $0.20^{+0.01}_{-0.01}$ to $0.24_{-0.01}^{+0.01}$. The disk temperature was found to be $1.75_{-0.19}^{+0.22}$ and $1.71_{-0.13}^{+0.14}$~keV for MJDs 59662 and 59665 respectively. The value of all the best fitted parameters are given in the \autoref{Table:results}. \\
\\
We derived the flux from the fitted spectra of Mrk~501 using the model $constant*TBabs*(bbodyrad+logpar)$ for the OBSIDs corresponding to MJDs 59662 and 59665. In 0.5--10.0~keV energy range the flux values obtained for the MJDs 59662 and 59665 are 2.73~$\times~10^{-10}$ and 3.67~$\times~10^{-10}$ in units of $ergs~cm^{-2}~sec^{-1}$ respectively. In 10.0--78.0~keV energy range the flux values obtained for the MJDs 59662 and 59665 are 0.97~$\times~10^{-10}$ and 1.65~$\times~10^{-10}$ in units of $ergs~cm^{-2}~sec^{-1}$ respectively. \\
\\
The \autoref{lc_Mrk_501} shows 2--20~keV MAXI/GSC light-curve\footnote{\url{https://maxi.riken.jp/star_data/J1653+397/J1653+397.html}} of the blazar Mrk~501 over the period from MJD 55500--60500. The two epochs corresponding to MJD 59662 and 59665 are indicated by dashed vertical lines (red and green respectively). A zoomed-in view of the interval spanning MJD 59615--59780 (indicated by light-yellow shaded box) is shown in the upper-right side of the figure for a more detailed examination of the flux variations across these epochs. 
During the particular MJD 59662 and 59665, the MAXI/GSC data was not available. However, the flux values of the neighboring observations indicate that the flux of the source remained $\sim0.05~photons~cm^{-2}~sec^{-1}$ during these two epochs. Mrk~501 exhibited a prolonged X-ray flaring activity during 2012 (MJD 55840--56140; \citealt{2017MNRAS.469.1655K}). 
%During the high-flux state, the 2--20~keV flux remained $\gtrsim$0.10~$photons~cm^{-2}~sec^{-1}$, as indicated by the red and blue dots in \autoref{lc_Mrk_501}. 
During the high-flux states corresponding to MJD 55953 (indicated by red dot) and 56099 (indicated by blue dot), the 2--20~keV MAXI/GSC flux was found to be 0.11$\pm$0.02 and 0.10$\pm$0.02 in units of $photons~cm^{-2}~sec^{-1}$ respectively.  
The light-blue shaded region in \autoref{lc_Mrk_501} marks the interval from MJD 57921 to 58687 (displayed in the zoomed-in panel at the upper center), during which the source was reported to be in very low flux state \citep{2023ApJS..266...37A}. Throughout this low-flux state, the 2--20~keV average flux value was observed to be $\lesssim0.05~photons~cm^{-2}~sec^{-1}$ (\autoref{lc_Mrk_501}). Therefore, from the observed flux variations, the blazar Mrk~501 can be considered to be in a low-flux state when the 2--20~keV flux is $\lesssim0.05~photons~cm^{-2}~sec^{-1}$. The flux values during the high and low-flux states suggest that the source remained in a low-flux state during the two epochs, MJD 59662 and 59665. This interpretation is consistent with the results of \citet{2024JHEAp..44..393T}, where the authors reported that Mrk~501 remained in a low-flux state across the EM spectrum between August 15, 2016 and March 27, 2022 (MJD 57615--59665).
%%%%%%%%%%%%%%%%%%%%%%%%%%%%%%%%%%%%%%%%%%%%%%%%%%%%%%%%%%%%%%%%%%%%%%%%%%%%%%%%%%
\subsection{PG~1553+113}
\begin{figure*}
\centering
  \begin{minipage}[b]{0.49\textwidth}
    \includegraphics[width=\textwidth, angle=0]{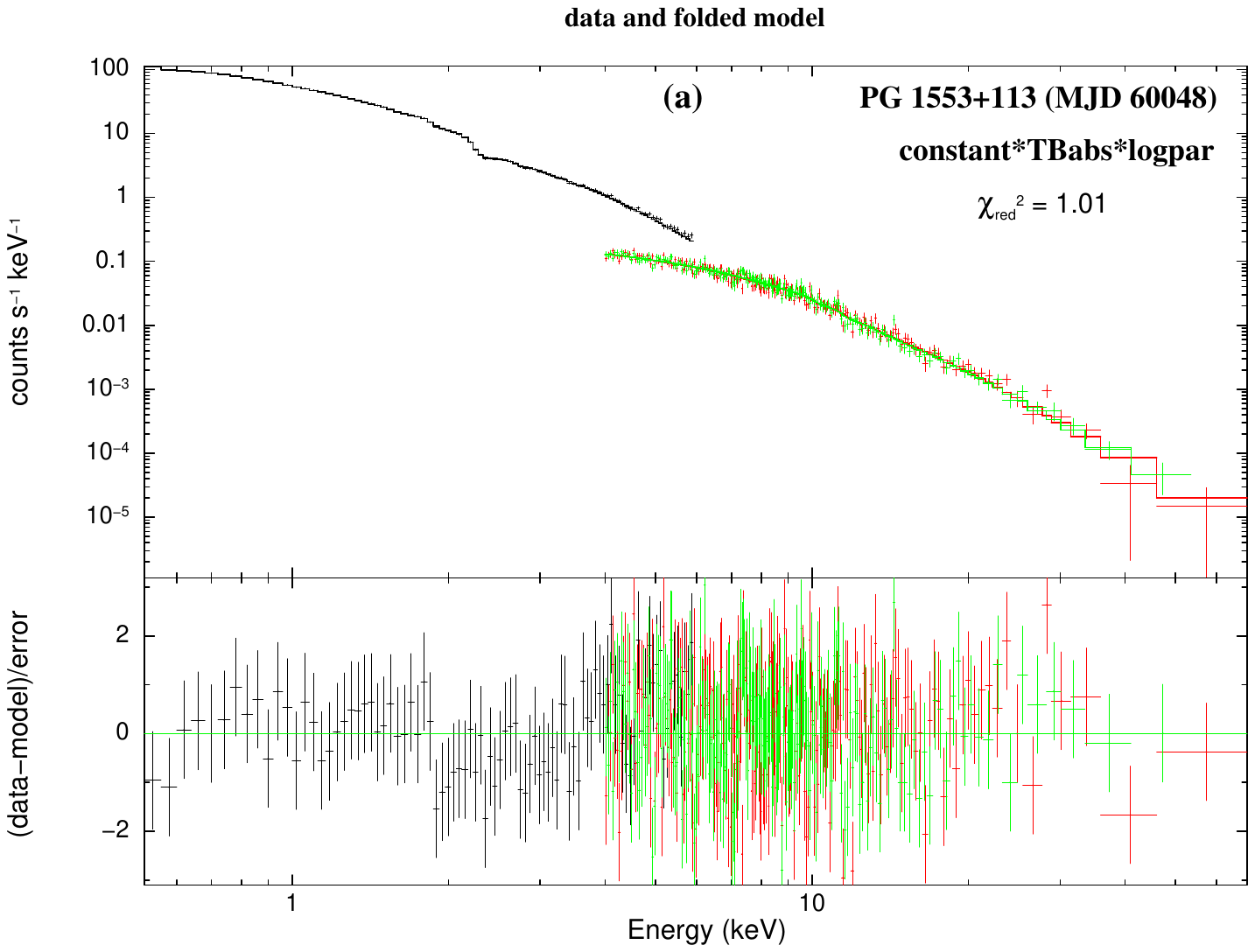}
\end{minipage}
\hfill
  \begin{minipage}[b]{0.49\textwidth}
    \includegraphics[width=\textwidth, angle=0]{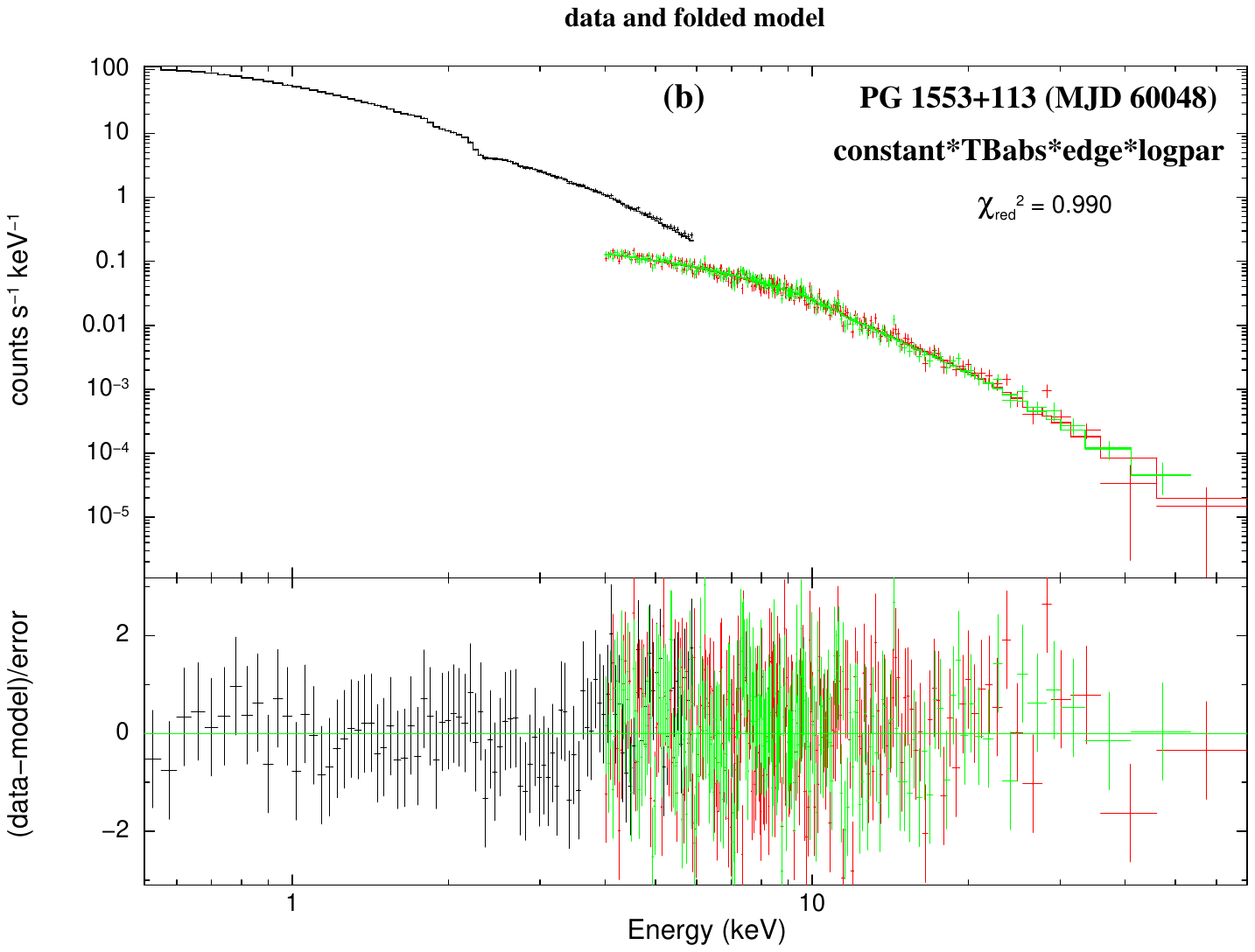}
\end{minipage}

\caption{The combined NICER (black) plus NuSTAR (red points denote FPMA and green points denote FPMB) fitted spectrum of PG~1553+113 during MJD 60048. Panel (a) shows fitted spectrum using absorbed LP (constant*TBabs*logpar) model. Panel (b) shows fitted spectrum using constant*TBabs*edge*logpar model. The source name, MJD value, fitting models and $\chi_{red}^2$ are mentioned in the panels. \label{PG_60048}}
\end{figure*}

\noindent
PG 1553+113 is classified as a HSP BL Lac \citep{2019A&A...632A..77C}. Since its spectra appear as a featureless continuum, the estimation of its redshift is under debate. Recently, Ly$\alpha$-forest-based redshift estimation reported its redshift z = 0.433 \citep{2022MNRAS.509.4330D}, whereas using two different photohadronic models, its redshift was found to be 0.491 $< \rm{z} <$ 0.537 \citep{2025MNRAS.540.3483S}. It is among the first VHE TeV gamma-ray-emitting blazars \citep{2006A&A...448L..19A,2007ApJ...654L.119A}. The potential relationship between inflow accretion (disk luminosity) and outflow/jet (jet power) in blazars is used to determine the SMBH mass of PG 1553+113 and found to be 1.6 $\times \rm{10}^{8} \rm{M}_{\odot}$ \citep{2014Natur.515..376G}. PG 1553+113 has witnessed several simultaneous MW monitoring of the blazar on diverse time scales \citep[e.g.][and references therein]{2006AJ....132..873O,2010A&A...515A..76A,2015MNRAS.454..353R,2024MNRAS.529.3894M}. It has also been observed in several studies conducted on the X-ray timing, polarization, and spectral characteristics using data from different satellites \citep[e.g.][and references therein]{2008ApJ...682..775R,2013ApJ...769...90N,2021MNRAS.506.1198D,2023ApJ...949...26S,2023ApJ...953L..28M}. In PG 1553+113, QPOs have been reported in different EM bands, and it has also been reported as a binary SMBH blazar \citep[e.g.][and references therein]{2015ApJ...813L..41A,2017ApJ...851L..39C,2018ApJ...854...11T,2023ApJ...949...26S,2024ApJ...976..203A}. {\it Suzaku} observations of the blazar in May 2006 revealed that it was in the low-flux state. Its XIS and HXD PIN data clearly show spectral curvature up to the highest hard X-ray data point (30~keV), which manifests as softening with increasing energy. This spectral shape was described by either a broken power law or a log-parabolic fit with equal statistical goodness of fits \citep{2008ApJ...682..775R}. \citet{2015MNRAS.454..353R} carried out X-ray spectral analysis of observations taken using Swift/XRT from 2009 to 2013 and an observation from the European Photon Imaging Camera (EPIC) onboard XMM–Newton taken on 2013 July 24–25. The Swift/XRT were fitted using PL and LP models whereas XMM-Newton spectra was fitted using PL, LP, and BPL models. Recently, complete XMM-Newton of 29 pointed observations were taken from 2001 to 2025. The blazar spectral analysis was carried out in the energy range 0.6--7.0~keV, and it was found nearly 50\% of the spectra (14/29) fitted well with absorbed PL, and the rest nearly 50\% of the spectra (15/29) fitted well with the absorbed LP model \citep{devanand2026}. \\
\\
We had only one simultaneous NICER+NuSTAR spectrum for the blazar PG~1553+113. The spectrum was analyzed using NICER data from 0.5--6.0~keV and NuSTAR data from 4.0--78.0~keV. When we fitted the spectrum using the absorbed LP model ($constant*TBabs*logpar$), we observed a silicon edge at around 1.87~keV in the NICER spectrum (\autoref{PG_60048}a). We added an \textit{edge} along with the absorbed LP model and the fitting model became: $constant*TBabs*edge*logpar$. The fitted spectrum using the $constant*TBabs*edge*logpar$ model can be seen in the \autoref{PG_60048}b. The value of the photon index ($\alpha$) and curvature parameter ($\beta$) obtained from the absorbed LP model were $2.26^{+0.01}_{-0.01}$ and $0.19^{+0.02}_{-0.02}$ respectively. When we used the $constant*TBabs*edge*logpar$ model, the value of the $\alpha$ and $\beta$ became $2.25^{+0.01}_{-0.01}$ and $0.20^{+0.02}_{-0.02}$ respectively (\autoref{Table:results}). 

%%%%%%%%%%%%%%%%%%%%%%%%%%%%%%%%%%%%%%%%%%%%%%%%%%%%%%%%%%%%%%%%%%%%%%%%%%
\begin{figure*}
\centering
\includegraphics[width=0.53\textwidth]{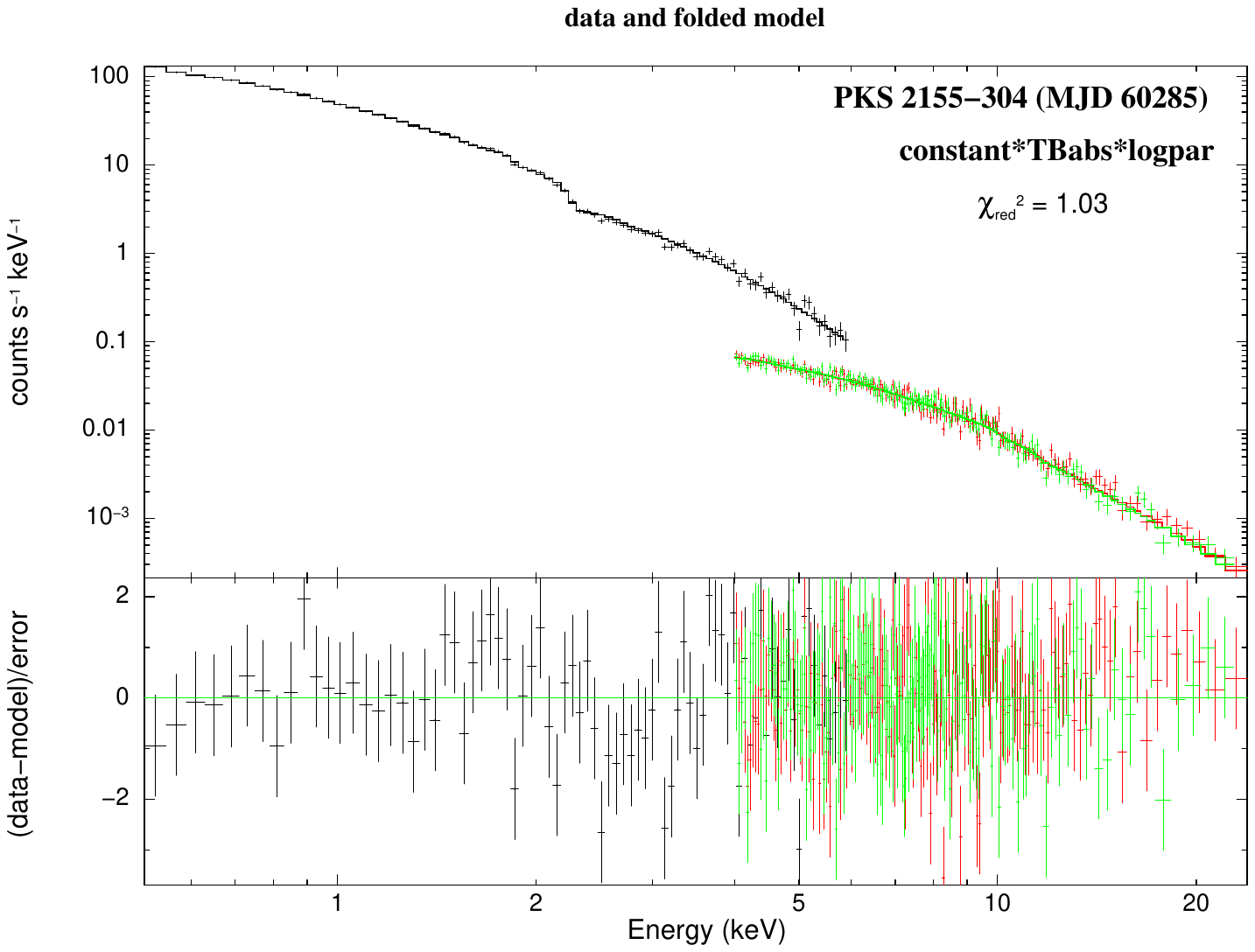}
\caption{The combined NICER (black) plus NuSTAR (red points denote FPMA and green points denote FPMB) fitted spectra of PKS~2155-304 during MJD 60285 using absorbed LP (constant*TBabs*logpar) model. The source name, MJD value, fitting model and $\chi_{red}^2$ are mentioned in the panels. \label{PKS_60285}}
\end{figure*} 

\subsection{PKS~2155-304}
\noindent
PKS 2155-304 is an HSP blazar at z = 0.116 $\pm$ 0.002 \citep{1993ApJ...411L..63F} and the brightest TeV $\gamma-$ray emitting astronomical aggregate in the Southern Hemisphere \citep{2007ApJ...664L..71A}. It is among the first TeV gamma-ray-emitting blazars \citep{1999ApJ...513..161C} and witnessed an exceptional VHE gamma-ray strong flare, which detected five bursts within two hours of observation \citep{2007ApJ...664L..71A}. By using rapid optical flux variability of the blazar, the SMBH mass was estimated to be 5.5 $\pm \rm{0.4} \times \rm{10}^{8} \rm{M}_{\odot} \leq \rm{M} \leq 8.1 \pm \rm{1.0} \times \rm{10}^{9}$ \citep{1992ApJ...385..146C}, and by using the normalized excess variance of XMM-Newton observations of the blazar, its SMBH mass was estimated to be 1.45 $\times \ \rm{10}^{8} \ \rm{M}_{\odot}$ \citep{2005ApJ...629..686Z}. On another occasion, using the X-ray QPO period of $\sim$4.6 hours from one of its pointed observations by XMM-Newton in 2006, the SMBH mass of the blazar was estimated in the range of 3.29 $\times \ \rm{10}^{7} \ \rm{M}_{\odot}$ to 2.09 $\times \ \rm{10}^{8} \ \rm{M}_{\odot}$ \citep{2009A&A...506L..17L}. PKS 2155-304 has been observed in several simultaneous multi-wavelength observational campaigns \citep[e.g.][and references therein]{1989ApJ...341..733T,1994A&A...288..433B,1995ApJ...438..120E,1997ApJ...486..799U,2005A&A...442..895A,2009ApJ...696L.150A,2009A&A...502..749A, 2020A&A...639A..42A,2026MNRAS.546f2282H}. Using various X-ray missions, the source has been extensively studied for temporal \citep[e.g.][and references therein]{2000ApJ...528..243K,2001ApJ...554..274E,2002ApJ...572..762Z,2005ApJ...629..686Z,2010ApJ...718..279G,2021ApJ...909..103Z}, spectral \citep[e.g.][and references therein]{1993ApJ...404..112S,1999ApJ...521..552C,2014MNRAS.444.3647B,2016ApJ...831..142M,2017ApJ...850..209G}, and polarization variability \citep{2024A&A...689A.119K}. In 1988 and 1989 observations of the blazar PKS 2155-304 by Ginga, it was found that in the energy range of 2--6~keV, the spectrum is characterized by a break at $\sim$4~keV \citep{1993ApJ...404..112S}. In a $\sim$100 ks observation in November 1996 of the source with BeppoSAX in the energy range 0.1--100~keV, the source shows diverse features; e.g., between 0.1 and 10 keV, the spectrum can be well described by a convex spectrum with an (energy) slope gradually steepening from 1.1 to 1.6 keV, an emission line at 6.4~keV in the source rest frame is seen, and evidence for a sharp spectral hardening was found at higher energies \citep{1998A&A...333L...5G}. In three observational campaigns in 1996, 1997, and 1999 of the source by BeppoSAX, the time-resolved X-ray spectral fits with a curved model show that peak energies of the synchrotron component are located in the very soft X-ray range or even below the BeppoSAX lower energy limit of 0.1~keV \citep{2002ApJ...572..762Z}. Simultaneous optical, UV, and X-ray 20-point observations taken from 2000 to 2012 with XMM-Newton show the spectrum was well fitted with the PL+LP model \citep{2014MNRAS.444.3647B}. The X-ray spectrum of the blazar during different intensity states in 2009–2014 using the XMM-Newton observation in 0.6--10.0~keV was fitted well using the LP model \citep{2017ApJ...850..209G}. \\
\\
For the blazar PKS~2155-304 we had a single simultaneous NICER+NuSTAR spectrum. We examined the source and background spectra carefully and we could use NICER data from 0.5--6.0~keV and NuSTAR data from 4.0--25.0~keV only.  We fitted the spectrum using the absorbed LP model: $constant*TBabs*logpar$ (\autoref{PKS_60285}). The spectrum was fitted properly using the absorbed LP model with $\chi_{red}^2 =$ 1.03. The obtained value of the photon index ($\alpha$) was $2.45^{+0.02}_{-0.02}$ and curvature parameter ($\beta$) was $0.31^{+0.02}_{-0.02}$ (\autoref{Table:results}).

\setlength{\tabcolsep}{3.5pt}
\begin{longrotatetable}
\begin{deluxetable*}{c|c|cccc|ccccccccc}
\tabletypesize{\scriptsize}
\tablewidth{0pt}
\tablecaption{Spectral Properties of four TeV Blazars using NICER and NuSTAR\label{Table:results}}
\tablehead{
\multicolumn{1}{c|}{Source} & \multicolumn{1}{c|}{MJD} &  \multicolumn{4}{c|}{Model 1} & \multicolumn{9}{c}{Model 2} \\ 
\multicolumn{1}{c|}{(1)} & \multicolumn{1}{c|}{(2)} & \colhead{(3)} & \colhead{(4)} & \colhead{(5)} & \multicolumn{1}{c|}{(6)} & \colhead{(7)} & \colhead{(8)} & \colhead{(9)} & \colhead{(10)} & \colhead{(11)} & \colhead{(12)} & \colhead{(13)} & \colhead{(14)} & \colhead{(15)} 
}
\startdata
\multicolumn{1}{c|}{\bf{Mrk~421}}  & \multicolumn{1}{c|}{} & \multicolumn{4}{c|}{\bf{absorbed LP model (\textit{constant*TBabs*logpar})}} & \multicolumn{9}{c}{\bf{constant*TBabs (bbodyrad+logpar+gaussian) model}}\\ [1.5ex]
\multicolumn{1}{c|}{}  & \multicolumn{1}{c|}{} & $\alpha$ & $\beta$ & LP~Norm & $\chi^2$/DOF & kT & bNorm & $\alpha$ & $\beta$ & LP~Norm & LineE & $\sigma$ & ga~Norm & $\chi^2$/DOF\\
\multicolumn{1}{c|}{}  & \multicolumn{1}{c|}{} &  &  & ($\times10^{-2}$) & & (keV) & & & & ($\times10^{-2}$) & (keV) & (keV) & ($\times10^{-3}$) & \\ [1.5ex]
& 60077 & $1.97_{-0.01}^{+0.01}$ & $0.39_{-0.01}^{+0.01}$ & $34.70_{-0.15}^{+0.15}$ & 1453/1227 & $1.51_{-0.06}^{+0.06}$ & $0.71_{-0.15}^{+0.17}$ & $2.03_{-0.01}^{+0.01}$ & $0.34_{-0.01}^{+0.01}$ & $34.60_{-0.20}^{+0.20}$ & $1.70_{-0.23}^{+0.10}$ & $0.10_{-0.09}^{+0.08}$ & $1.00_{-0.43}^{+0.53}$ & 1243/1222 \\ [1.0ex]
& 60289 & $2.31_{-0.01}^{+0.01}$ & $0.36_{-0.01}^{+0.01}$ & $30.40_{-0.13}^{+0.13}$ & 787/772 & $0.15_{-0.02}^{+0.01}$ & $(1.96_{-0.99}^{+1.93})\times10^{4}$ & $2.19_{-0.06}^{+0.05}$ & $0.43_{-0.03}^{+0.03}$ & $27.30_{-1.30}^{+1.00}$ & $1.47_{-0.42}^{+0.15}$ & $0.28_{-0.08}^{+0.07}$ & 3.66$_{-1.65}^{+1.82}$ & 750/767 \\ [1.0ex]
& 60296 & $2.39_{-0.01}^{+0.01}$ & $0.33_{-0.01}^{+0.01}$ & $36.10_{-0.20}^{+0.20}$ & 735/779 & $0.15_{-0.02}^{+0.02}$ & $(2.96_{-1.57}^{+3.27})\times10^{4}$ & $2.24_{-0.07}^{+0.06}$ & $0.42_{-0.03}^{+0.04}$ & $31.70_{-1.72}^{+1.46}$ & $1.42_{-0.55}^{+0.17}$ & $0.31_{-0.05}^{+0.05}$ & 6.51$_{-2.18}^{+2.31}$ & 682/774 \\ [1.0ex]
& 60298 & $2.36_{-0.01}^{+0.01}$ & $0.38_{-0.01}^{+0.01}$ & $28.80_{-0.14}^{+0.14}$ & 799/788 & $0.15_{-0.01}^{+0.01}$ & $(2.39_{-1.04}^{+2.36})\times10^{4}$ & $2.23_{-0.07}^{+0.05}$ & $0.46_{-0.03}^{+0.04}$ & $26.10_{-1.47}^{+1.05}$ & $1.56_{-0.12}^{+0.11}$ & $0.18_{-0.08}^{+0.09}$ & 1.63$_{-1.00}^{+1.26}$ & 774/783 \\ [1.0ex]
& 60429 & $2.48_{-0.02}^{+0.02}$ & $0.33_{-0.02}^{+0.02}$ & $7.49_{-0.05}^{+0.05}$ & 526/482 & --- & --- & --- & --- & --- & --- & --- & --- & --- \\ [1.0ex]
& 60433 & $2.50_{-0.02}^{+0.02}$ & $0.25_{-0.02}^{+0.02}$ & $8.80_{-0.07}^{+0.07}$ & 594/532 & --- & --- & --- & --- & --- & --- & --- & --- & --- \\ [1.0ex]
& 60439 & $2.11_{-0.01}^{+0.01}$ & $0.28_{-0.01}^{+0.01}$ & $25.50_{-0.14}^{+0.14}$ & 964/951 & --- & --- & --- & --- & --- & --- & --- & --- & ---  \\ [1.0ex]
& 60443 & $2.06_{-0.01}^{+0.01}$ & $0.36_{-0.01}^{+0.01}$ & $26.30_{-0.17}^{+0.17}$ & 1026/975 & --- & --- & --- & --- & --- & --- & --- & --- & ---  \\ [1.5ex]
\hline
%%%
\multicolumn{1}{c|}{\bf{Mrk~501}}  & \multicolumn{1}{c|}{} & \multicolumn{4}{c|}{\bf{absorbed LP model (\textit{constant*TBabs*logpar})}} & \multicolumn{6}{c}{\bf{constant*TBabs (bbodyrad+logpar) model}} & & & \\ [1.5ex]
\multicolumn{1}{c|}{}  & \multicolumn{1}{c|}{} & $\alpha$ & $\beta$ & LP~Norm & $\chi^2$/DOF & kT & bNorm & $\alpha$ & $\beta$ & LP~Norm & $\chi^2$/DOF &  &  & \\
\multicolumn{1}{c|}{}  & \multicolumn{1}{c|}{} &  &  & ($\times10^{-2}$) & & (keV) &  &  &  & ($\times10^{-2}$) &  &  &  & \\
& 59660 & $1.85_{-0.01}^{+0.01}$ & $0.23_{-0.01}^{+0.01}$ & $5.87_{-0.03}^{+0.03}$ & 848/863 & --- & --- & --- & --- & --- & --- & & & \\ [1.0ex]
& 59662 & $1.87_{-0.01}^{+0.01}$  & $0.26_{-0.01}^{+0.01}$ & $6.09_{-0.03}^{+0.03}$  & 853/903 & $1.75_{-0.19}^{+0.22}$ & $0.07_{-0.03}^{+0.04}$ & $1.90_{-0.01}^{+0.01}$ & $0.24_{-0.01}^{+0.01}$ & $6.07_{-0.03}^{+0.03}$ & 816/901 & & & \\ [1.0ex]
& 59665 & $1.78_{-0.01}^{+0.01}$ & $0.24_{-0.01}^{+0.01}$ & $7.45_{-0.04}^{+0.04}$ & 1019/934 & $1.71_{-0.13}^{+0.14}$  & $0.18_{-0.06}^{+0.07}$ & $1.83_{-0.01}^{+0.01}$ & $0.20_{-0.01}^{+0.01}$ & $7.40_{-0.04}^{+0.04}$ & 944/932 & & & \\ [1.5ex]
\hline
%%%
\multicolumn{1}{c|}{\bf{PG~1553+113}} & \multicolumn{1}{c|}{} & \multicolumn{4}{c|}{\bf{absorbed LP model (\textit{constant*TBabs*logpar})}} & \multicolumn{6}{c}{\bf{constant*TBabs*edge*logpar model}} & & & \\ [1.5ex]
\multicolumn{1}{c|}{} & \multicolumn{1}{c|}{} & $\alpha$ & $\beta$ & LP~Norm & $\chi^2$/DOF & edgeE & MaxTau & $\alpha$ & $\beta$ & LP~Norm & $\chi^2$/DOF &  &  & \\
\multicolumn{1}{c|}{} & \multicolumn{1}{c|}{} & & & ($\times10^{-2}$) & & (keV) & & & & ($\times10^{-2}$) & &  &  & \\
& 60048 & $2.26_{-0.01}^{+0.01}$ & $0.19_{-0.02}^{+0.02}$ & $3.74_{-0.02}^{+0.02}$ & 508/501 & $1.87_{-0.06}^{+0.07}$ & $(5.12_{-2.22}^{+2.22})\times10^{-2}$ & $2.25_{-0.01}^{+0.01}$ & $0.20_{-0.02}^{+0.02}$ & $3.75_{-0.02}^{+0.02}$ & 494/499 & & & \\ [1.5ex]
\hline
%%%
\multicolumn{1}{c|}{\bf{PKS~2155-304}} & \multicolumn{1}{c|}{} & \multicolumn{4}{c|}{\bf{absorbed LP model (\textit{constant*TBabs*logpar})}} & & & & \\ [1.5ex]
\multicolumn{1}{c|}{} & \multicolumn{1}{c|}{} & $\alpha$ & $\beta$ & LP~Norm & $\chi^2$/DOF & & & & & & &  &  & \\
\multicolumn{1}{c|}{} & \multicolumn{1}{c|}{} &  &  & ($\times10^{-2}$) & & & & & & & &  &  & \\
& 60285 & $2.45_{-0.02}^{+0.02}$ & $0.31_{-0.02}^{+0.02}$ & $3.17_{-0.02}^{+0.02}$ & 436/423 & --- & --- & --- & --- & --- & --- & & & \\ [1.5ex]
\hline \hline
\enddata
\tablecomments{ The $\alpha$ and $\beta$ and $LP~Norm$ represent the photon index, curvature parameter and the normalization parameter of the \textit{Log-Parabolic (LP)} model, respectively. The $kT$ and $bNorm$ correspond to the temperature and normalization parameter of the \textit{bbodyrad} model. The $LineE$, $\sigma$ and $ga~Norm$ denote the line energy, line width and normalization parameter of the Gaussian line, respectively. The $edgeE$ and $MaxTau$ represent the threshold energy and the absorption depth at the threshold energy of the \textit{edge} model.}
\end{deluxetable*}
\end{longrotatetable}
%%%%%%%%%%%%%%%%%%%%%%%%%%%%%%%%%%%%%%%%%%%%%%%%%%%%%%%%%%%%%%%%%%%%%%%%%%%%%%%%%%%
\section{Discussion and Conclusion}
\noindent
In the present work we studied the X-ray spectra of 4 classical TeV blazars that were observed simultaneously by NICER and NuSTAR. These sources were chosen from TeVCat\href{https://www.tevcat.org}{$^1$}. We had 8 simultaneous data for the Mrk~421, 3 simultaneous data for Mrk~501, 1 simultaneous data for PG~1553+113 and 1 simultaneous data for PKS~2155-304. Out of the 8 NICER spectra of Mrk~421, 2 were studied by \citet{2025arXiv251208531K}. The spectra of all the other NICER observations are analyzed for the first time in this study. Out of the 8 NuSTAR spectra of Mrk~421, 3 spectra (having OBSIDs: 60902024004, 60902024006 and 60902024008) were studied by \citet{2025ApJ...986..230M} to examine the variability of X-ray polarisation of the source. The rest of the NuSTAR spectra (5 out of 8) of Mrk~421 are studied for the first time in this work. The NuSTAR spectra for the Mrk~501 were studied separately by \citet{2025JHEAp..4800417A}. %\sout{All of the NuSTAR spectra for the Mrk~421,}
The NuSTAR spectra of the PG~1553+113 and PKS~2155-304 are analyzed for the first time in this present work. We found 7 out of the 13 X-ray spectra were fitted satisfactorily with the absorbed LP model. \\
\\
FSRQs have radiatively efficient, optically thick and geometrically thin accretion disk while most of the BL Lacs have radiatively inefficient and geometrically thick accretion disk \citep{2019MmSAI..90..154G, 1997ApJ...478L..79N, 2000ApJ...539..798N}. In case of BL Lacs, the ultraviolet (UV) emission is significantly reduced and disk radiation is unable to photo-ionize the broad line region (BLR), resulting in weak or absent broad emission lines  in the spectra \citep{2009MNRAS.396L.105G}. During high jet activity states of the BL Lacs, the continuum is amplified due to Doppler boosting and becomes dominant. This phenomena can also explain the weakness or absence of the emission lines in the spectra of the BL Lacs. The faint emission lines observed in BL Lacs can be observed due to the combined action of weak accretion disk and relativistic jet \citep{2012RAA....12..359F}. Removing the Doppler boosting effect of the jet continuum, the intrinsic FWHM of the emission lines falls within the typical range of other AGNs. This suggests that during low jet activity or faint states of BL Lacs, the observed X-ray emission may include a significant contribution from the accretion disk \citep{2022A&A...663A.178M}. In \citet{2018ApJ...859L..21C}, the authors observed a break in the power spectral density (PSD) while studying the X-ray light-curve of Mrk~421 during 2017, using Indian space telescope AstroSat. Galactic black hole X-ray binaries and Seyfert galaxies show similar characteristic in their X-ray variability, while in these systems X-rays are produced in the accretion disk or corona. Therefore, similar type of break in the PSD of the BL Lac Mrk~421 can also be related to the properties of the accretion flow. Spectral analysis of Mrk~421 shows that the epochs of 2017 are well described by the accretion disk based Two Conponent Advective Flow (TCAF) model. The estimated relations between disk and jet flux with radio flux indicate that the accretion disk can contribute to the observed X-ray emission \citep{2022A&A...663A.178M}.\\
\\Mrk~421 is a HSP blazar that generally shows no emission lines in its optical spectra. In such cases, the observed X-ray flux variations are generally attributed to jet activity \citep{2011ApJ...736..131A}. However, during less active states, the accretion disk may also contribute to the emission, though it is unlikely to dominate over the jet. In \citet {2022A&A...663A.178M}, the authors observed contribution from the disk in case of Mrk~421 during the epochs of 2017 while the source was in a low flux state. The authors explained that accretion disk based models such as TCAF can fit the spectra without requiring additional components. This suggests that, although Mrk~421 is a HSP blazar, during its low to moderate activity states the accretion disk may contribute significantly to the observed X-ray emission, although the jet remains dominant. \citet{2018ApJ...859L..21C} reported a break in the X-ray PSD of Mrk~421. Their AstroSat observations were taken during a moderate brightness state. Such breaks, commonly seen in Seyfert galaxies, are often linked to X-rays produced in the accretion disk. This motivates the hypothesis that during low or moderate X-ray activity in Mrk~421, disk emission could contribute to the observed X-rays. In this present study of Mrk~421, we also observed disk contribution in case of spectra of four OBSIDs. For this HSP BL Lac, we analysed eight simultaneous archived spectra of NICER and NuSTAR. The spectra of the four MJDs, 60429, 60433, 60439 and 60443 were fitted well using the absorbed LP model. However, in case of the spectra of MJDs 60077, 60289, 60296 and 60298, it can be observed from the residual of the fitted spectra that the absorbed LP model could not fit the four spectra satisfactorily. While fitting the spectra, we observed that the spectra were not fitting in the lower energy regime, below 4~keV. When we added \textit{bbodyrad} model, the lower energy range seemed to be fitted better. During these four spectra the source was not in a high flux state. In \citet {2022A&A...663A.178M}, the authors calculated the flux during the historic bright state on April 14, 2013 using the Swift/XRT (0.3--10.0~keV) as 27.57~$\times~10^{-10}$ $ergs~cm^{-2}~sec^{-1}$. In this present study, in the energy range of 0.5--10.0~keV, the flux values obtained for the MJDs 60077, 60289, 60296 and 60298 are 13.60~$\times~10^{-10}$, 9.89~$\times~10^{-10}$, 11.48~$\times~10^{-10}$ and 9.14~$\times~10^{-10}$ in units of $ergs~cm^{-2}~sec^{-1}$ respectively. By comparing the X-ray flux during these four MJDs (60077, 60289, 60296 and 60298) with the bright state (MJD 56397), we can comment that during these observations X-ray flux values are much weaker, therefore the contribution of the jet emission during these four MJDs is relatively lower. \\
\\
If we consider the 2--20~keV MAXI/GSC light-curve (\autoref{lc_Mrk_421}), it can be observed during the high-flux state, the source flux remained around $\gtrsim0.30~photons~cm^{-2}~s^{-1}$. The source can be considered to be in low-flux state when the flux value remains $\lesssim0.10~photons~cm^{-2}~s^{-1}$. During MJD 60077, the value of 2--20~keV MAXI/GSC flux was $0.13\pm0.02$ and during the MJDs 60289, 60296 and 60298 the flux was $<0.10~photons~cm^{-2}~s^{-1}$. In comparison with the previously reported high and low-flux states, the source can be considered to be in a moderate to low-flux state during the four epochs, MJD 60077, 60289, 60296, and 60298. Therefore, the contribution of the disk in the X-ray band can be justified during these observations. The improvement in the spectral fitting, while adding the \textit{bbodyrad} model, can also explain the contribution from the disk. Thus we can conclude that during these four epochs having MJD 60077, 60289, 60296, and 60298, there were emission from the disk along with the synchrotron emission from the jet. We also observed a Gaussian line feature around the 1.42--1.70~keV energy range in these X-ray spectra. We added a Gaussian line model to fit this feature. The origin of the observed lines could not be determined. 
%\sout{In case of NICER spectra we can observe a Al-K$\alpha$ at} \sout{1.49~keV and Si-K$\alpha$ line at 1.74~keV.} 
The Gaussian lines observed in the spectra of this current study can be instrumental features or can be originated from the background.\\
\\In case of Mrk~501, we fitted all three simultaneous NICER+NuSTAR spectra using absorbed LP model firstly. 
Although the spectrum corresponding to MJD 59660 was fitted satisfactorily, the spectra corresponding to MJDs 59662 and 59665 were not fitted well by the absorbed LP model. However, these spectra were fitted satisfactorily using the $constant*TBabs*(bbodyrad+logpar)$ model. In the last paragraph we mentioned the previous studies in case of Mrk~421 where the authors showed that in moderate to low flux state there can be contributions from the accretion disk along with the jet, in case of BL Lac objects. 
%\sout{During this study, the source was was in a low to} \sout{moderate flux state.}
In the paper \citet{2023ApJS..266...37A}, the authors studied a very-low-activity state of Mrk~501 spanning from June 17, 2017 to July 23, 2019 (MJD 57921 to MJD 58687). In the study they showed that the 0.2--10.0~keV Swift/XRT X-ray flux was around 1~$\times~10^{-10}$ $ergs~cm^{-2}~sec^{-1}$ during that very-low-activity state. During the current study we found the X-ray flux from NICER spectra in the energy range of 0.5--10.0~keV around 3~$\times~10^{-10}$ $ergs~cm^{-2}~sec^{-1}$. From the 2--20~keV MAXI/GSC light-curve (\autoref{lc_Mrk_501}), it can be observed that during the high flux state the source flux remained around $\gtrsim0.10~photons~cm^{-2}~s^{-1}$. The source can be considered to be in a low-flux state when the 2--20~keV flux remains $\lesssim~0.05~photons~cm^{-2}~s^{-1}$. During the MJDs, 59662 and 59665 the 2--20~keV MAXI/GSC flux was around $0.05~photons~cm^{-2}~s^{-1}$. Therefore, it can be stated, during these epochs, MJD 59662 and 59665, the source was found to be in a low-flux state. In \citet{2024JHEAp..44..393T}, the authors also reported that the blazar Mrk~501 was in a low-flux state during the epochs MJD 59662 and 59665. The low-flux state of the source can justify the presence of the disk emission in the X-ray spectra. Thus we can comment that there were contribution from the disk during these epochs (MJD 59662 and 59665). We could not find any previous studies regarding Mrk~501 where the authors have claimed to observe any contribution from disk in X-ray spectral studies. Although, in this present study, %\sout{the concave nature in} 
the spectra could only be fitted and explained as a contribution from the disk only. Further studies in this regard are needed to confirm the claim of the disk emission. \\
\\
In case of PG~1553+113 the NICER+NuSTAR spectrum fitted well using the absorbed LP model. We observed a silicon edge at around 1.87~keV. The spectrum was found to be positively curved with $\alpha \sim 2.25$ and $\beta \sim 0.20$. In case of PKS~2155-304 the NICER+NuSTAR spectrum fitted perfectly using the absorbed LP model. We found the spectrum to be positively curved with $\alpha \sim 2.45$ and $\beta \sim 0.31$. These spectral curvatures can result from energy-dependent particle acceleration combined with subsequent radiative cooling \citep{2004A&A...413..489M, 2007A&A...466..521T, 2008A&A...478..395M}. 
%%%%%%%%%%%%%%%%%%%%%%%%%%%%%%%%%%%%%%%%%%%%%%%%%%%%%%%%%%%%%%%%%%%%%%%
%% Please use the acknowledgment and contribution environments. This will 
%% be anonomyized when the "anonymous" style option is used. 
\begin{acknowledgments}
This research makes use of the SciServer science platform (www.sciserver.org). This research has made use of data, software and/or web tools obtained from the High Energy Astrophysics Science Archive Research Center (HEASARC), a service of the Astrophysics Science Division at NASA/GSFC and of the Smithsonian Astrophysical Observatory's High Energy Astrophysics Division. We thank Devanand P. U. for helpful discussions and valuable suggestions during this work.
\end{acknowledgments}
%%%%%%%%%%%%%%%%%%%%%%%%%%%%%%%%%%%%%%%%%%%%%%%%%%%%%%%%%%%%%%%%%%%%%%
%\begin{contribution}
%% But authors are expected to provide more specific details, e.g. 
%RB provided the data reduction and formal analysis. RB and ACG were responsible for writing the manuscript. ACG came up with the initial research concept. ACG supervised and edited the manuscript.
%\end{contribution}
\facilities{NICER, NuSTAR, MAXI/GSC}
\software{HEASoft} 
%%%%%%%%%%%%%%%%%%%%%%%%%%%%%%%%%%%%%%%%%%%%%%%%%%%%%%%%%%%%%%%%%%%%
%%%%%%%%%%%%%%%%%%%%%%%%%%%%%%%%%%%%%%%%%%%%%%%%%%%%%%%%%%%%%%%%%%%%%%%
%\clearpage
\FloatBarrier
%\section*{appendix}
\appendix
\vspace{-.4cm}
\centering{Additional Figures}
\vspace{.4cm}
%\clearpage
\renewcommand{\thefigure}{A\arabic{figure}}
\renewcommand{\theHfigure}{A\arabic{figure}}
\setcounter{figure}{0}

\vspace{-1cm}
\begin{figure*}[h]
\centering
  \begin{minipage}[b]{0.49\textwidth}
    \includegraphics[width=\textwidth, angle=0, trim=0 1.5cm 0 0, clip]{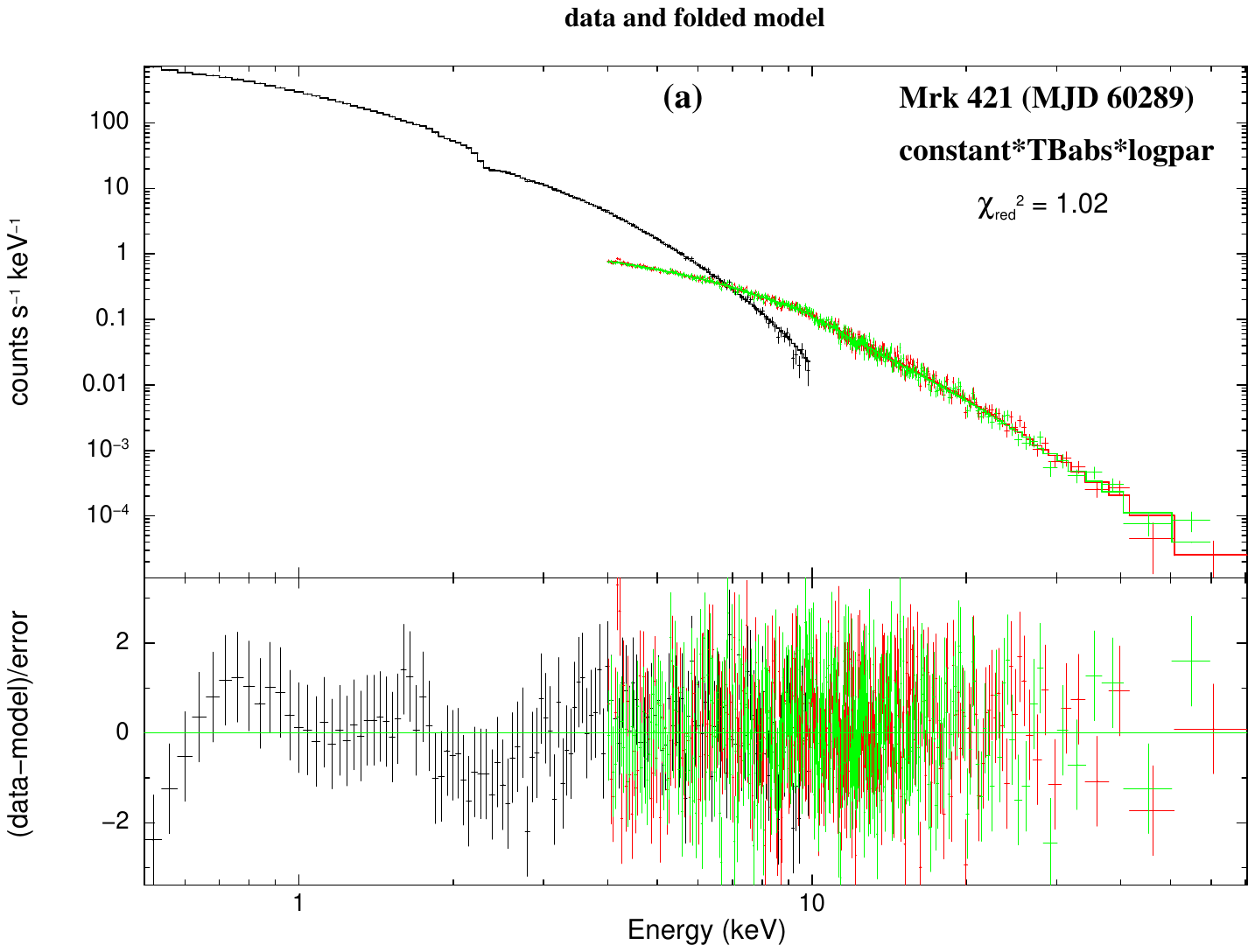}
\end{minipage}
\hfill
  \begin{minipage}[b]{0.49\textwidth}
    \includegraphics[width=\textwidth, angle=0, trim=0 1.5cm 0 0, clip]{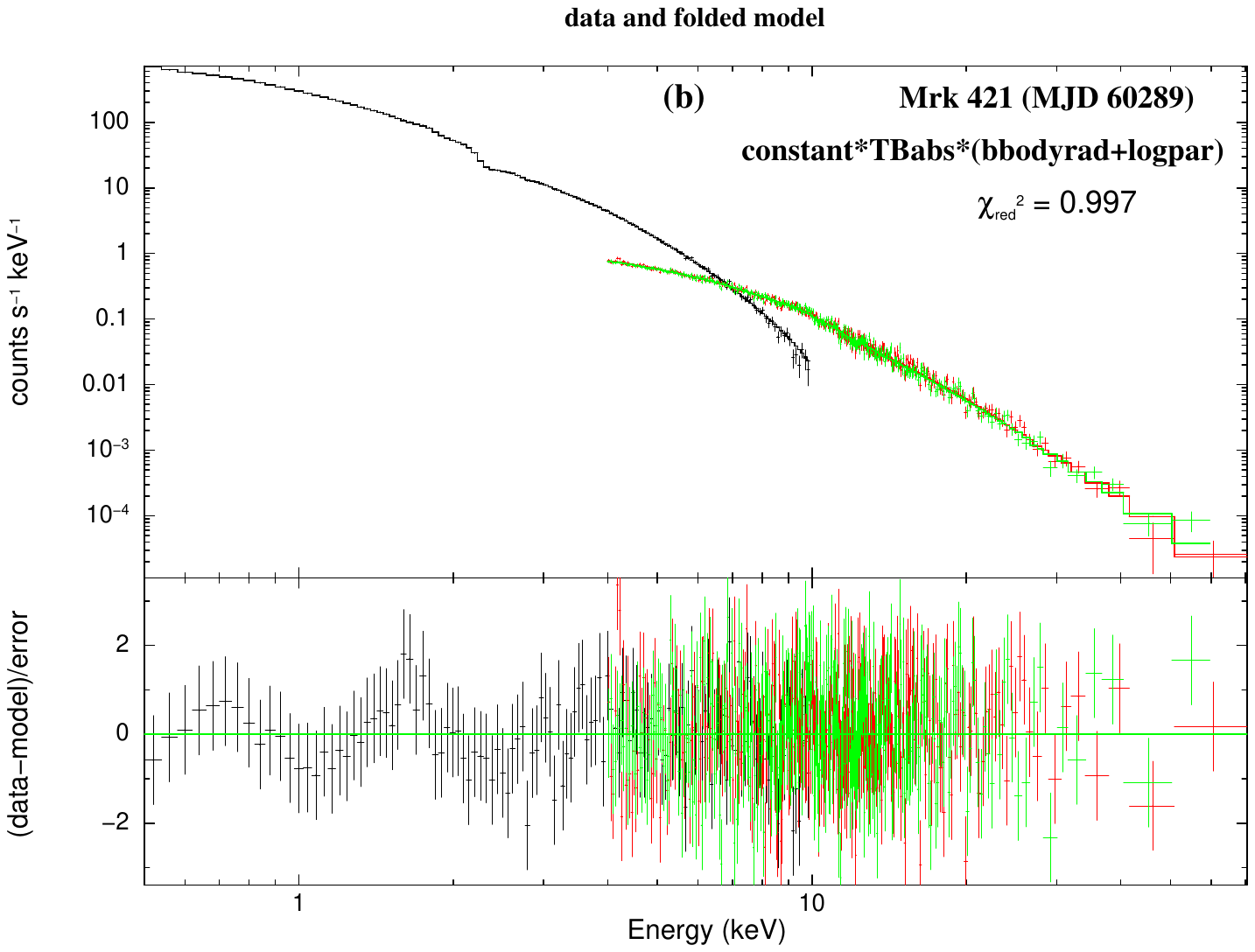}
\end{minipage}
\begin{minipage}[b]{0.51\textwidth}
    \includegraphics[width=\textwidth, angle=0, trim=0 1.5cm 0 0, clip]{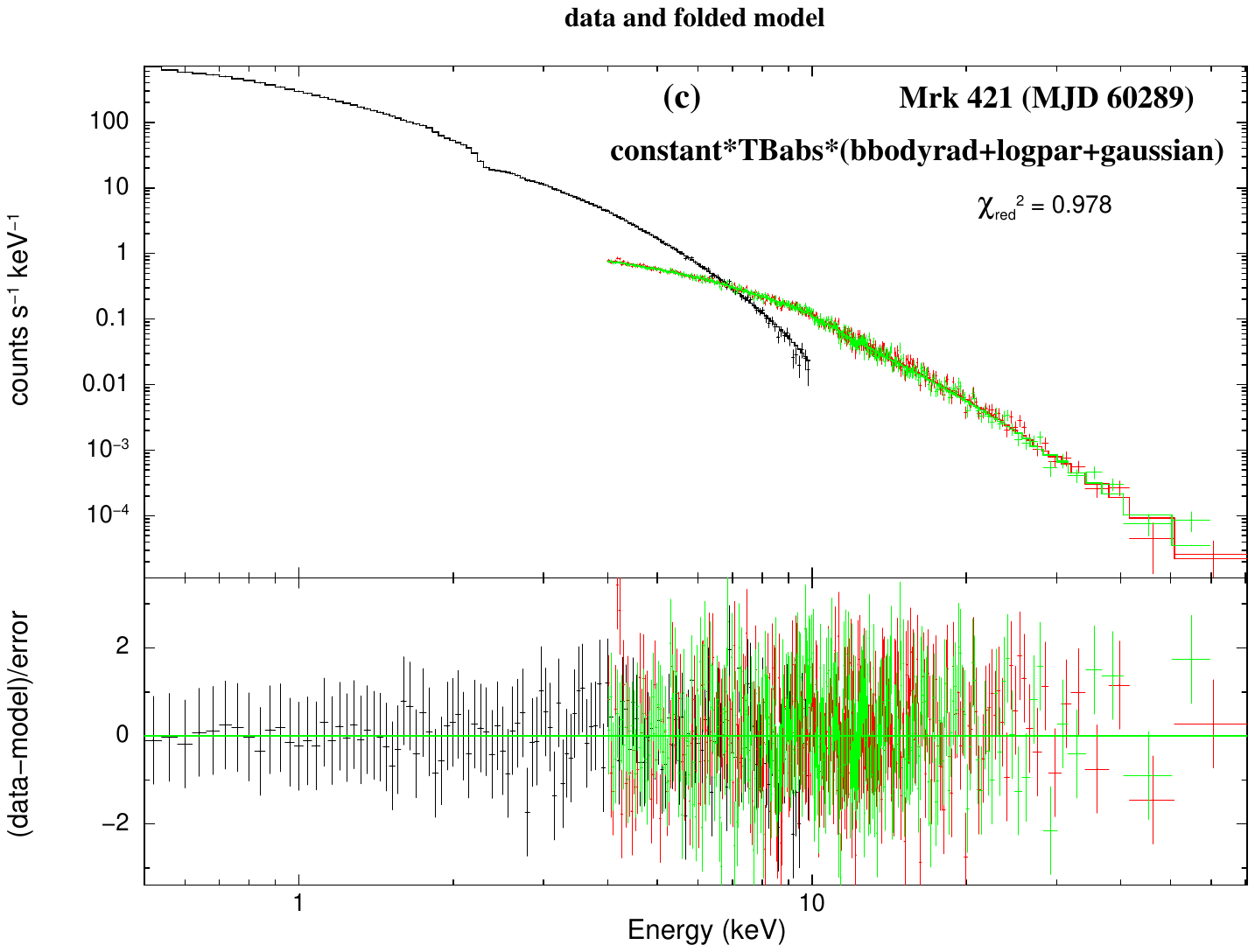}
\end{minipage}
\hfill
  \begin{minipage}[b]{0.49\textwidth}
    \includegraphics[width=\textwidth, angle=0, trim=0 1.5cm 0 0, clip]{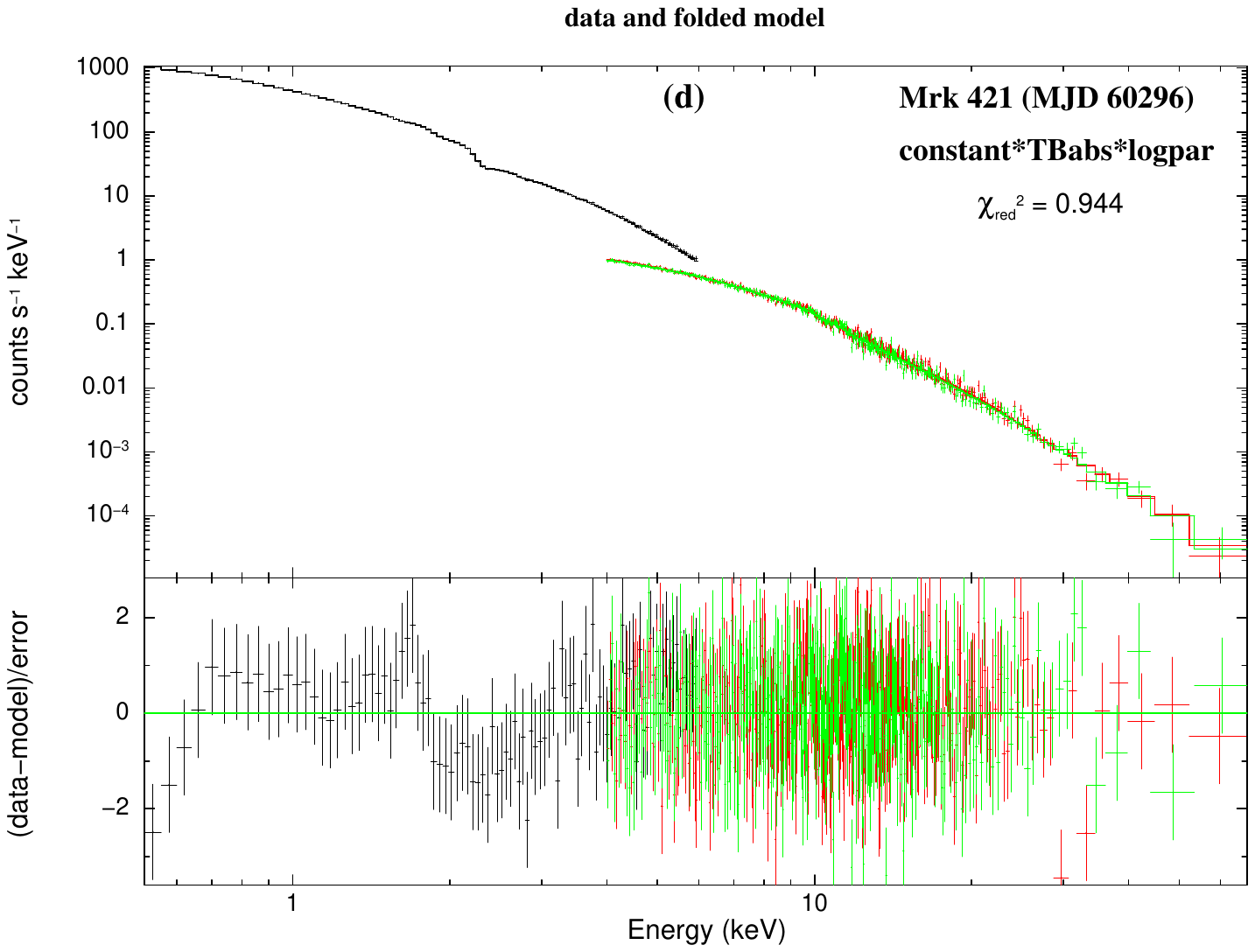}
\end{minipage}
\hfill
  \begin{minipage}[b]{0.49\textwidth}
    \includegraphics[width=\textwidth, angle=0, trim=0 1.5cm 0 0, clip]{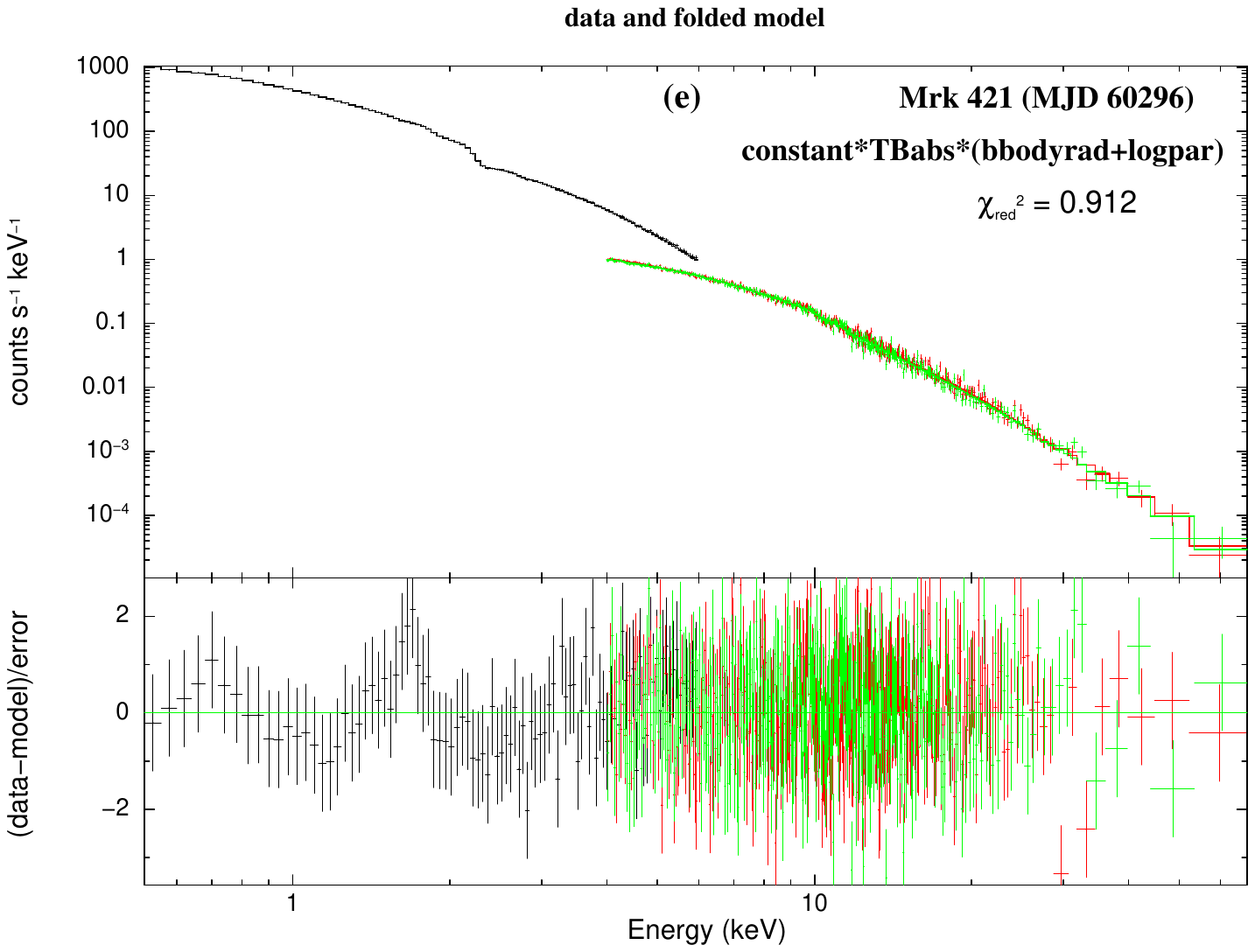}
\end{minipage}
\caption{The combined NICER (black) plus NuSTAR (red points denote FPMA and green points denote FPMB) fitted spectra of Mrk~421 during different MJDs (60289, 60296 and 60298) using different fitting models. The source name, MJD values, fitting models and $\chi_{red}^2$ are mentioned in the panels.}
\label{Mrk421_bbrad_lp_ga_rest}
\end{figure*}
\setcounter{figure}{0}
\begin{figure*}
%\continuedfloat
\centering
\begin{minipage}[b]{0.51\textwidth}
    \includegraphics[width=\textwidth, angle=0]{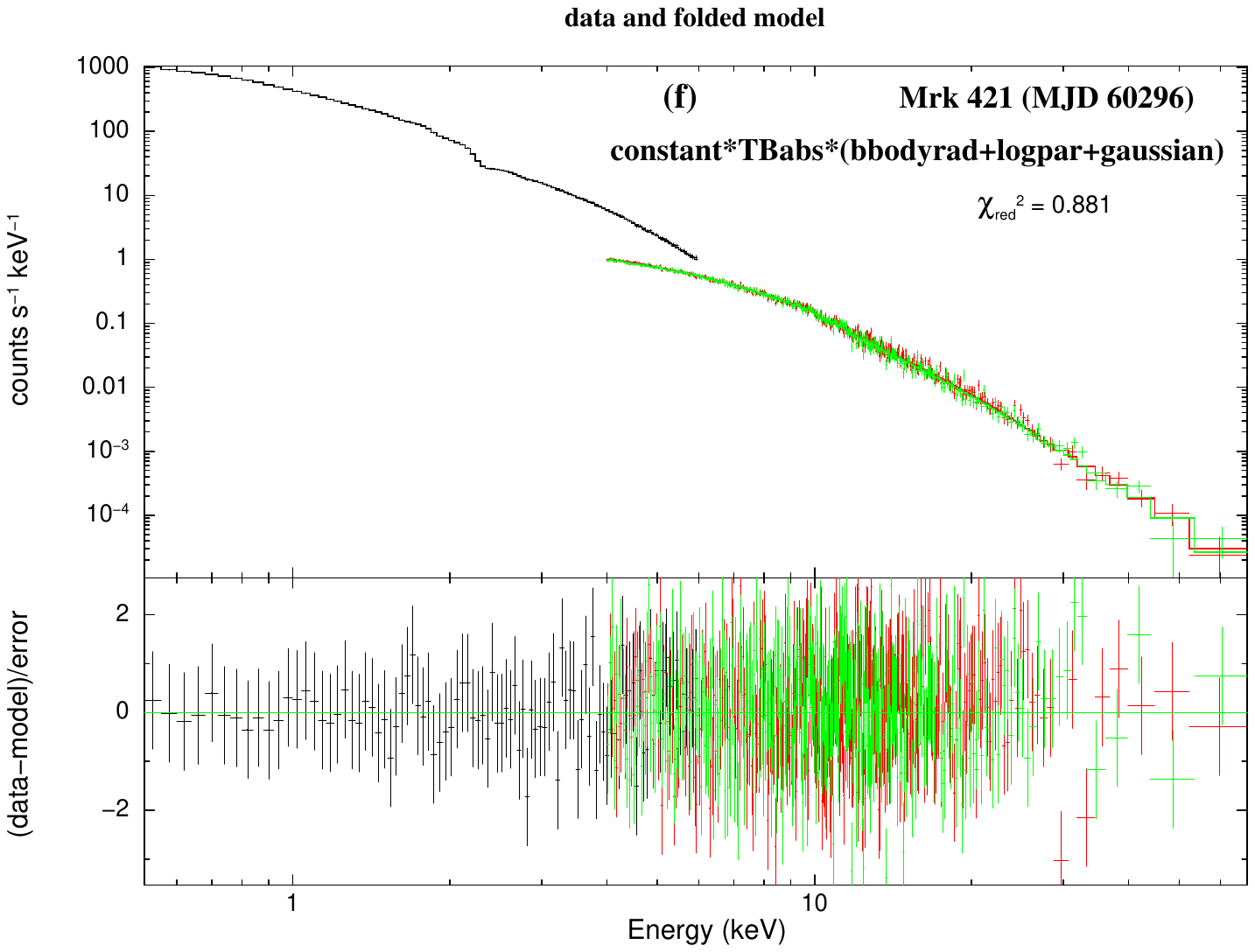}
\end{minipage}
\hfill
  \begin{minipage}[b]{0.49\textwidth}
    \includegraphics[width=\textwidth, angle=0]{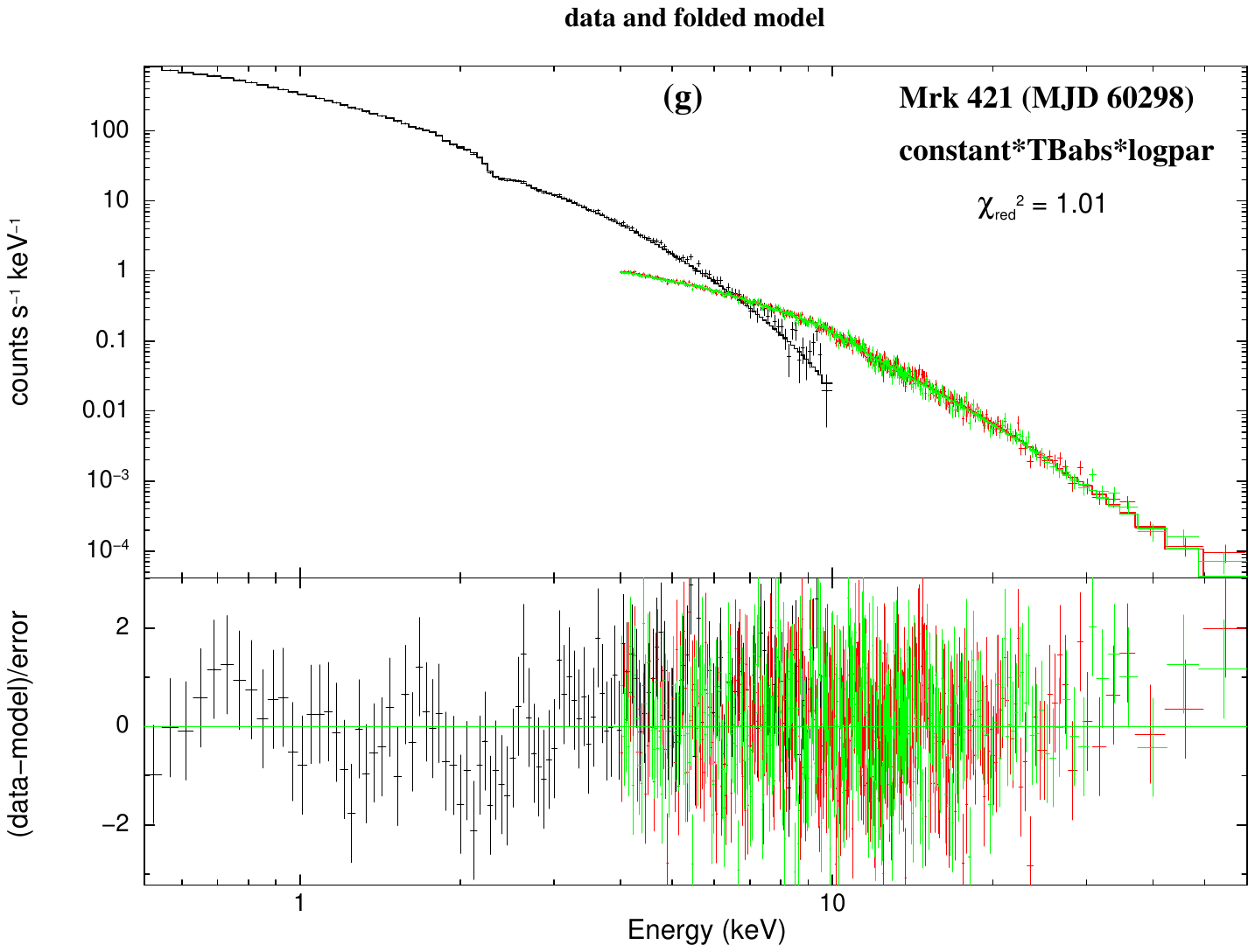}
\end{minipage}
\hfill
  \begin{minipage}[b]{0.49\textwidth}
    \includegraphics[width=\textwidth, angle=0]{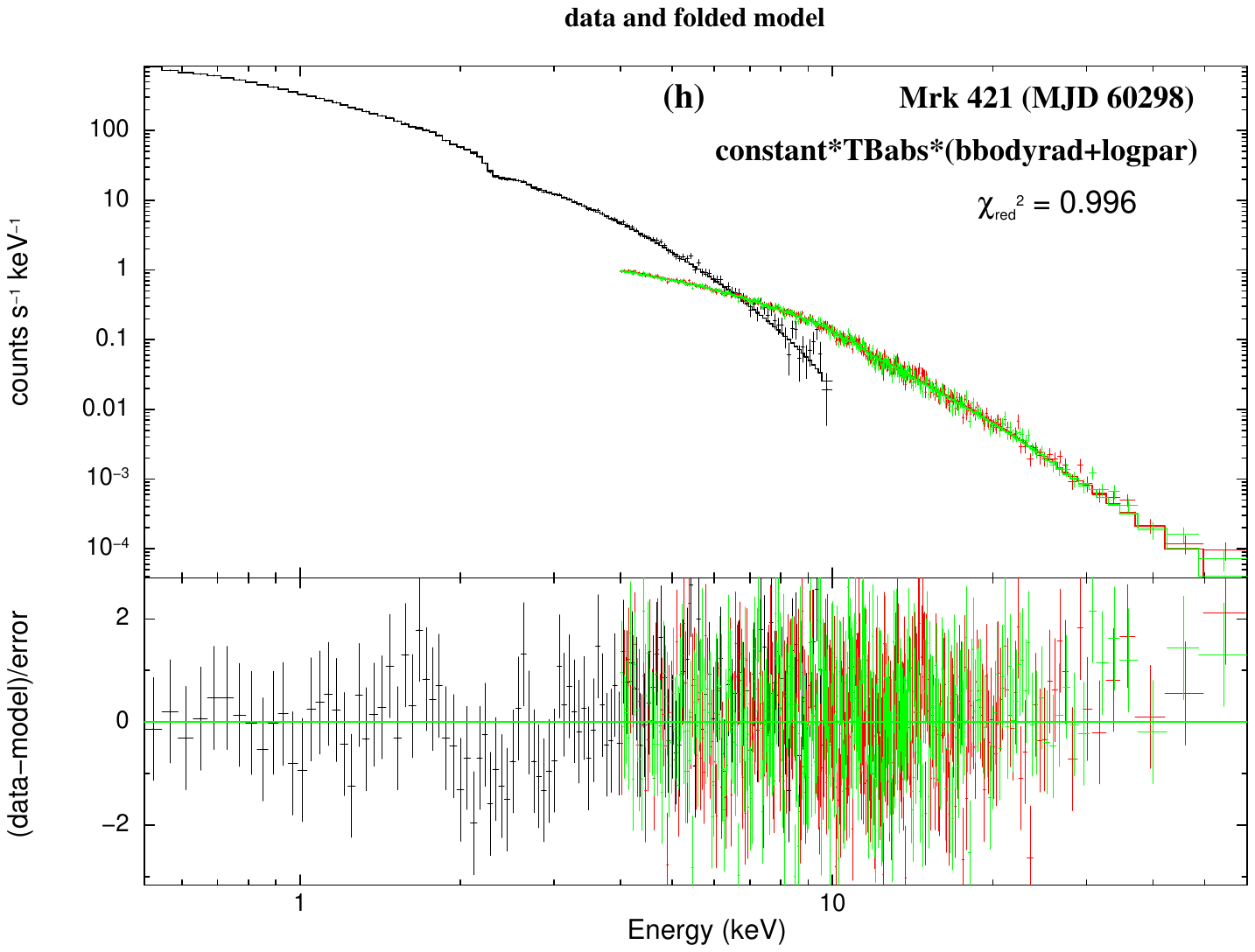}
\end{minipage}
\begin{minipage}[b]{0.51\textwidth}
    \includegraphics[width=\textwidth, angle=0]{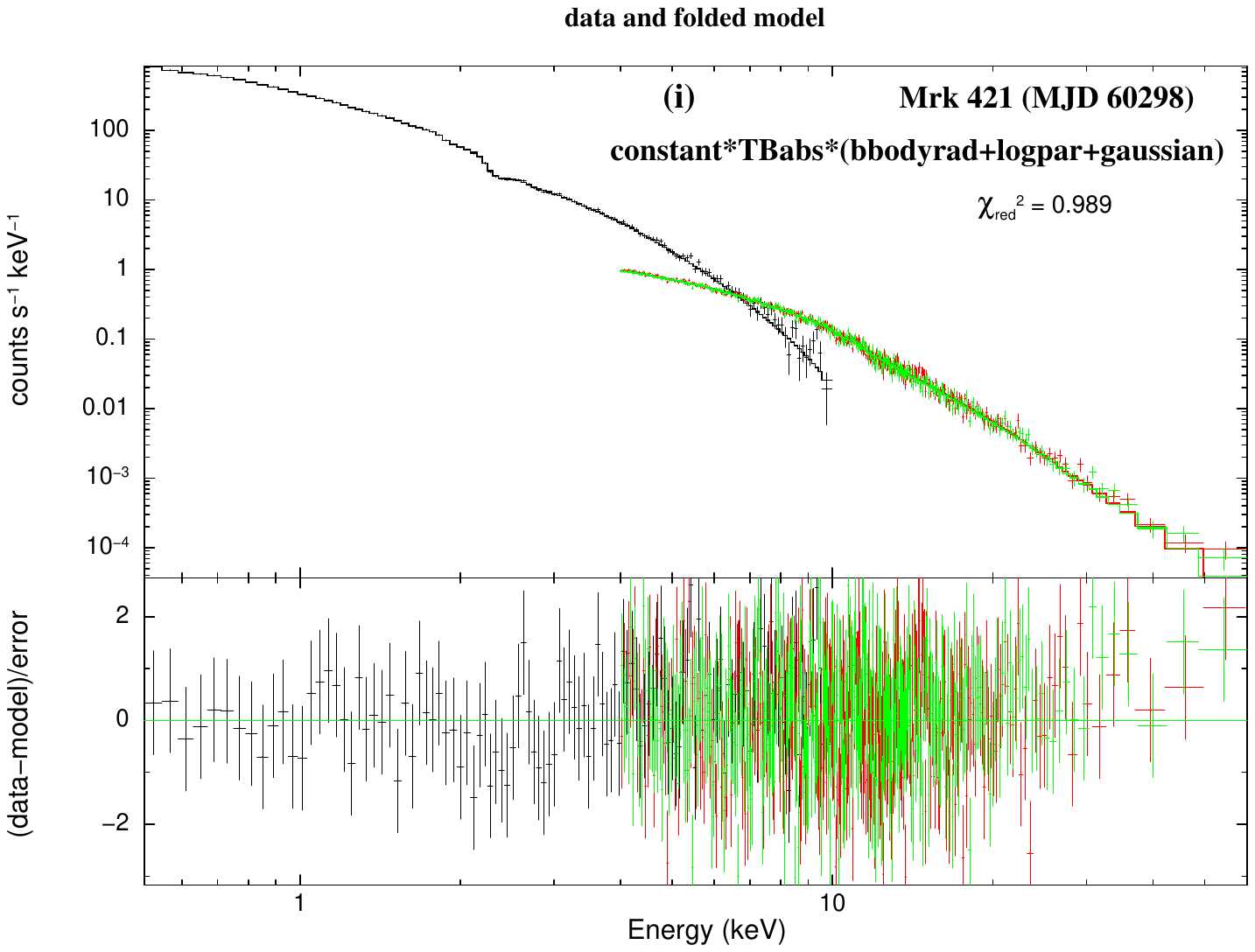}
\end{minipage}
\caption{continued: The combined NICER (black) plus NuSTAR (red points denote FPMA and green points denote FPMB) fitted spectra of Mrk~421 during different MJDs (60289, 60296 and 60298) using different fitting models. The source name, MJD values, fitting models and $\chi_{red}^2$ are mentioned in the panels.}
\end{figure*}
 \clearpage
%%%%%%%%%%%%%%%%%%%%%%%%%%%%%%%%%%%%%%%%%%%%%%%%%%%%%%%%%%%%%%%%%%%%
%%%%%%%%%%%%%%%%%%%%%%%%%%%%%%%%%%%%%%%%%%%%%%%%%%%%%%%%%%%%%%%%%%%%
\bibliography{references}{}

@ARTICLE{2013APh....50...26A,
       author = {{Abeysekara}, A.~U. and {Alfaro}, R. and {Alvarez}, C. and {{\'A}lvarez}, J.~D. and {Arceo}, R. and {Arteaga-Vel{\'a}zquez}, J.~C. and {Ayala Solares}, H.~A. and {Barber}, A.~S. and {Baughman}, B.~M. and {Bautista-Elivar}, N. and {Belmont}, E. and {BenZvi}, S.~Y. and {Berley}, D. and {Bonilla Rosales}, M. and {Braun}, J. and {Caballero-Lopez}, R.~A. and {Carrami{\~n}ana}, A. and {Castillo}, M. and {Cotti}, U. and {Cotzomi}, J. and {de la Fuente}, E. and {De Le{\'o}n}, C. and {DeYoung}, T. and {Diaz Hernandez}, R. and {Diaz-Velez}, J.~C. and {Dingus}, B.~L. and {DuVernois}, M.~A. and {Ellsworth}, R.~W. and {Fernandez}, A. and {Fiorino}, D.~W. and {Fraija}, N. and {Galindo}, A. and {Garcia-Luna}, J.~L. and {Garcia-Torales}, G. and {Garfias}, F. and {Gonz{\'a}lez}, L.~X. and {Gonz{\'a}lez}, M.~M. and {Goodman}, J.~A. and {Grabski}, V. and {Gussert}, M. and {Hampel-Arias}, Z. and {Hui}, C.~M. and {H{\"u}ntemeyer}, P. and {Imran}, A. and {Iriarte}, A. and {Karn}, P. and {Kieda}, D. and {Kunde}, G.~J. and {Lara}, A. and {Lauer}, R.~J. and {Lee}, W.~H. and {Lennarz}, D. and {Le{\'o}n Vargas}, H. and {Linares}, E.~C. and {Linnemann}, J.~T. and {Longo}, M. and {Luna-Garc{\'\i}a}, R. and {Marinelli}, A. and {Martinez}, O. and {Mart{\'\i}nez-Castro}, J. and {Matthews}, J.~A.~J. and {Miranda-Romagnoli}, P. and {Moreno}, E. and {Mostaf{\'a}}, M. and {Nava}, J. and {Nellen}, L. and {Newbold}, M. and {Noriega-Papaqui}, R. and {Oceguera-Becerra}, T. and {Patricelli}, B. and {Pelayo}, R. and {P{\'e}rez-P{\'e}rez}, E.~G. and {Pretz}, J. and {Rivi{\`e}re}, C. and {Ryan}, J. and {Rosa-Gonz{\'a}lez}, D. and {Salazar}, H. and {Salesa}, F. and {Sandoval}, A. and {Santos}, E. and {Schneider}, M. and {Silich}, S. and {Sinnis}, G. and {Smith}, A.~J. and {Sparks}, K. and {Springer}, R.~W. and {Taboada}, I. and {Toale}, P.~A. and {Tollefson}, K. and {Torres}, I. and {Ukwatta}, T.~N. and {Villase{\~n}or}, L. and {Weisgarber}, T. and {Westerhoff}, S. and {Wisher}, I.~G. and {Wood}, J. and {Yodh}, G.~B. and {Younk}, P.~W. and {Zaborov}, D. and {Zepeda}, A. and {Zhou}, H.},
        title = "{Sensitivity of the high altitude water Cherenkov detector to sources of multi-TeV gamma rays}",
      journal = {Astroparticle Physics},
         year = 2013,
        month = dec,
       volume = {50},
        pages = {26-32},
          doi = {10.1016/j.astropartphys.2013.08.002},
archivePrefix = {arXiv},
       eprint = {1306.5800},
 primaryClass = {astro-ph.HE},
       adsurl = {https://ui.adsabs.harvard.edu/abs/2013APh....50...26A}
}

@ARTICLE{2012APh....35..641A,
       author = {{Abeysekara}, A.~U. and {Aguilar}, J.~A. and {Aguilar}, S. and {Alfaro}, R. and {Almaraz}, E. and {{\'A}lvarez}, C. and {{\'A}lvarez-Romero}, J. de D. and {{\'A}lvarez}, M. and {Arceo}, R. and {Arteaga-Vel{\'a}zquez}, J.~C. and {Badillo}, C. and {Barber}, A. and {Baughman}, B.~M. and {Bautista-Elivar}, N. and {Belmont}, E. and {Ben{\'\i}tez}, E. and {BenZvi}, S.~Y. and {Berley}, D. and {Bernal}, A. and {Bonamente}, E. and {Braun}, J. and {Caballero-Lopez}, R. and {Cabrera}, I. and {Carrami{\~n}ana}, A. and {Carrasco}, L. and {Castillo}, M. and {Chambers}, L. and {Conde}, R. and {Condreay}, P. and {Cotti}, U. and {Cotzomi}, J. and {D'Olivo}, J.~C. and {de la Fuente}, E. and {De Le{\'o}n}, C. and {Delay}, S. and {Delepine}, D. and {DeYoung}, T. and {Diaz}, L. and {Diaz-Cruz}, L. and {Dingus}, B.~L. and {Duvernois}, M.~A. and {Edmunds}, D. and {Ellsworth}, R.~W. and {Fick}, B. and {Fiorino}, D.~W. and {Flandes}, A. and {Fraija}, N.~I. and {Galindo}, A. and {Garc{\'\i}a-Luna}, J.~L. and {Garc{\'\i}a-Torales}, G. and {Garfias}, F. and {Gonz{\'a}lez}, L.~X. and {Gonz{\'a}lez}, M.~M. and {Goodman}, J.~A. and {Grabski}, V. and {Gussert}, M. and {Guzm{\'a}n-Ceron}, C. and {Hampel-Arias}, Z. and {Harris}, T. and {Hays}, E. and {Hernandez-Cervantes}, L. and {H{\"u}ntemeyer}, P.~H. and {Imran}, A. and {Iriarte}, A. and {Jimenez}, J.~J. and {Karn}, P. and {Kelley-Hoskins}, N. and {Kieda}, D. and {Langarica}, R. and {Lara}, A. and {Lauer}, R. and {Lee}, W.~H. and {Linares}, E.~C. and {Linnemann}, J.~T. and {Longo}, M. and {Luna-Garc{\'\i}a}, R. and {Mart{\'\i}nez}, H. and {Mart{\'\i}nez}, J. and {Mart{\'\i}nez}, L.~A. and {Mart{\'\i}nez}, O. and {Mart{\'\i}nez-Castro}, J. and {Martos}, M. and {Matthews}, J. and {McEnery}, J.~E. and {Medina-Tanco}, G. and {Mendoza-Torres}, J.~E. and {Miranda-Romagnoli}, P.~A. and {Montaruli}, T. and {Moreno}, E. and {Mostafa}, M. and {Napsuciale}, M. and {Nava}, J. and {Nellen}, L. and {Newbold}, M. and {Noriega-Papaqui}, R. and {Oceguera-Becerra}, T. and {Olmos Tapia}, A. and {Orozco}, V. and {P{\'e}rez}, V. and {P{\'e}rez-P{\'e}rez}, E.~G. and {Perkins}, J.~S. and {Pretz}, J. and {Ramirez}, C. and {Ram{\'\i}rez}, I. and {Rebello}, D. and {Renter{\'\i}a}, A. and {Reyes}, J. and {Rosa-Gonz{\'a}lez}, D. and {Rosado}, A. and {Ryan}, J.~M. and {Sacahui}, J.~R. and {Salazar}, H. and {Salesa}, F. and {Sandoval}, A. and {Santos}, E. and {Schneider}, M. and {Shoup}, A. and {Silich}, S. and {Sinnis}, G. and {Smith}, A.~J. and {Sparks}, K. and {Springer}, W. and {Su{\'a}rez}, F. and {Suarez}, N. and {Taboada}, I. and {Tellez}, A.~F. and {Tenorio-Tagle}, G. and {Tepe}, A. and {Toale}, P.~A. and {Tollefson}, K. and {Torres}, I. and {Ukwatta}, T.~N. and {Valdes-Galicia}, J. and {Vanegas}, P. and {Vasileiou}, V. and {V{\'a}zquez}, O. and {V{\'a}zquez}, X. and {Villase{\~n}or}, L. and {Wall}, W. and {Walters}, J.~S. and {Warner}, D. and {Westerhoff}, S. and {Wisher}, I.~G. and {Wood}, J. and {Yodh}, G.~B. and {Zaborov}, D. and {Zepeda}, A.},
        title = "{On the sensitivity of the HAWC observatory to gamma-ray bursts}",
      journal = {Astroparticle Physics},
         year = 2012,
        month = may,
       volume = {35},
       number = {10},
        pages = {641-650},
          doi = {10.1016/j.astropartphys.2012.02.001},
archivePrefix = {arXiv},
       eprint = {1108.6034},
 primaryClass = {astro-ph.HE},
       adsurl = {https://ui.adsabs.harvard.edu/abs/2012APh....35..641A}
}

@ARTICLE{2018MNRAS.480.4873A,
       author = {{Aggrawal}, Vishi and {Pandey}, Ashwani and {Gupta}, Alok C. and {Zhang}, Zhongli and {Wiita}, Paul J. and {Yadav}, K.~K. and {Tiwari}, S.~N.},
        title = "{X-ray intraday variability of the TeV blazar Mrk 421 with Chandra}",
      journal = {\mnras},
         year = 2018,
        month = nov,
       volume = {480},
       number = {4},
        pages = {4873-4883},
          doi = {10.1093/mnras/sty2173},
archivePrefix = {arXiv},
       eprint = {1808.05158},
 primaryClass = {astro-ph.HE},
       adsurl = {https://ui.adsabs.harvard.edu/abs/2018MNRAS.480.4873A}
}

@ARTICLE{2002A&A...384L..23A,
       author = {{Aharonian}, F. and {Akhperjanian}, A. and {Barrio}, J. and {Beilicke}, M. and {Bernl{\"o}hr}, K. and {B{\"o}rst}, H. and {Bojahr}, H. and {Bolz}, O. and {Contreras}, J. and {Cornils}, R. and {Cortina}, J. and {Denninghoff}, S. and {Fonseca}, V. and {Girma}, M. and {Gonzalez}, J. and {G{\"o}tting}, N. and {Heinzelmann}, G. and {Hermann}, G. and {Heusler}, A. and {Hofmann}, W. and {Horns}, D. and {Jung}, I. and {Kankanyan}, R. and {Kestel}, M. and {Kettler}, J. and {Kohnle}, A. and {Konopelko}, A. and {Kornmeyer}, H. and {Kranich}, D. and {Krawczynski}, H. and {Lampeitl}, H. and {Lopez}, M. and {Lorenz}, E. and {Lucarelli}, F. and {Magnussen}, N. and {Mang}, O. and {Meyer}, H. and {Mirzoyan}, R. and {Moralejo}, A. and {Ona}, E. and {Padilla}, L. and {Panter}, M. and {Plaga}, R. and {Plyasheshnikov}, A. and {P{\"u}hlhofer}, G. and {Rauterberg}, G. and {R{\"o}hring}, A. and {Rhode}, W. and {Robrade}, J. and {Rowell}, G. and {Sahakian}, V. and {Samorski}, M. and {Schilling}, M. and {Schr{\"o}der}, F. and {Sevilla}, I. and {Siems}, M. and {Stamm}, W. and {Tluczykont}, M. and {V{\"o}lk}, H.~J. and {Wiedner}, C.~A. and {Wittek}, W.},
        title = "{TeV gamma rays from the blazar H 1426+428 and the diffuse extragalactic background radiation}",
      journal = {\aap},
         year = 2002,
        month = mar,
       volume = {384},
        pages = {L23-L26},
          doi = {10.1051/0004-6361:20020206},
       adsurl = {https://ui.adsabs.harvard.edu/abs/2002A&A...384L..23A}
}

@ARTICLE{2003A&A...406L...9A,
       author = {{Aharonian}, F. and {Akhperjanian}, A. and {Beilicke}, M. and {Bernl{\"o}hr}, K. and {B{\"o}rst}, H.-G. and {Bojahr}, H. and {Bolz}, O. and {Coarasa}, T. and {Contreras}, J.~L. and {Cortina}, J. and {Denninghoff}, S. and {Fonseca}, M.~V. and {Girma}, M. and {G{\"o}tting}, N. and {Heinzelmann}, G. and {Hermann}, G. and {Heusler}, A. and {Hofmann}, W. and {Horns}, D. and {Jung}, I. and {Kankanyan}, R. and {Kestel}, M. and {Kohnle}, A. and {Konopelko}, A. and {Kornmeyer}, H. and {Kranich}, D. and {Lampeitl}, H. and {Lopez}, M. and {Lorenz}, E. and {Lucarelli}, F. and {Mang}, O. and {Meyer}, H. and {Mirzoyan}, R. and {Moralejo}, A. and {Ona-Wilhelmi}, E. and {Panter}, M. and {Plyasheshnikov}, A. and {P{\"u}hlhofer}, G. and {de los Reyes}, R. and {Rhode}, W. and {Ripken}, J. and {Robrade}, J. and {Rowell}, G. and {Sahakian}, V. and {Samorski}, M. and {Schilling}, M. and {Siems}, M. and {Sobzynska}, D. and {Stamm}, W. and {Tluczykont}, M. and {Vitale}, V. and {V{\"o}lk}, H.~J. and {Wiedner}, C.~A. and {Wittek}, W.},
        title = "{Detection of TeV gamma-rays from the BL Lac 1ES 1959+650 in its low states and during a major outburst in 2002}",
      journal = {\aap},
         year = 2003,
        month = jul,
       volume = {406},
        pages = {L9-L13},
          doi = {10.1051/0004-6361:20030838},
       adsurl = {https://ui.adsabs.harvard.edu/abs/2003A&A...406L...9A}
}

@ARTICLE{2021ChPhC..45b5002A,
       author = {{Aharonian}, F. and {An}, Q. and {Axikegu} and {Bai}, L.~X. and {Bai}, Y.~X. and {Bao}, Y.~W. and {Bastieri}, D. and {Bi}, X.~J. and {Bi}, Y.~J. and {Cai}, H. and {Cai}, J.~T. and {Cao}, Z. and {Cao}, Z. and {Chang}, J. and {Chang}, J.~F. and {Chang}, X.~C. and {Chen}, B.~M. and {Chen}, J. and {Chen}, L. and {Chen}, L. and {Chen}, L. and {Chen}, M.~J. and {Chen}, M.~L. and {Chen}, Q.~H. and {Chen}, S.~H. and {Chen}, S.~Z. and {Chen}, T.~L. and {Chen}, X.~L. and {Chen}, Y. and {Cheng}, N. and {Cheng}, Y.~D. and {Cui}, S.~W. and {Cui}, X.~H. and {Cui}, Y.~D. and {Dai}, B.~Z. and {Dai}, H.~L. and {Dai}, Z.~G. and {Danzengluobu} and {Della Volpe}, D. and {Piazzoli}, B. D'ettorre and {Dong}, X.~J. and {Fan}, J.~H. and {Fan}, Y.~Z. and {Fan}, Z.~X. and {Fang}, J. and {Fang}, K. and {Feng}, C.~F. and {Feng}, L. and {Feng}, S.~H. and {Feng}, Y.~L. and {Gao}, B. and {Gao}, C.~D. and {Gao}, Q. and {Gao}, W. and {Ge}, M.~M. and {Geng}, L.~S. and {Gong}, G.~H. and {Gou}, Q.~B. and {Gu}, M.~H. and {Guo}, J.~G. and {Guo}, X.~L. and {Guo}, Y.~Q. and {Guo}, Y.~Y. and {Han}, Y.~A. and {He}, H.~H. and {He}, H.~N. and {He}, J.~C. and {He}, S.~L. and {He}, X.~B. and {He}, Y. and {Heller}, M. and {Hor}, Y.~K. and {Hou}, C. and {Hou}, X. and {Hu}, H.~B. and {Hu}, S. and {Hu}, S.~C. and {Hu}, X.~J. and {Huang}, D.~H. and {Huang}, Q.~L. and {Huang}, W.~H. and {Huang}, X.~T. and {Huang}, Z.~C. and {Ji}, F. and {Ji}, X.~L. and {Jia}, H.~Y. and {Jiang}, K. and {Jiang}, Z.~J. and {Jin}, C. and {Kuleshov}, D. and {Levochkin}, K. and {Li}, B.~B. and {Li}, C. and {Li}, C. and {Li}, F. and {Li}, H.~B. and {Li}, H.~C. and {Li}, H.~Y. and {Li}, J. and {Li}, K. and {Li}, W.~L. and {Li}, X. and {Li}, X. and {Li}, X.~R. and {Li}, Y. and {Li}, Y.~Z. and {Li}, Z. and {Li}, Z. and {Liang}, E.~W. and {Liang}, Y.~F. and {Lin}, S.~J. and {Liu}, B. and {Liu}, C. and {Liu}, D. and {Liu}, H. and {Liu}, H.~D. and {Liu}, J. and {Liu}, J.~L. and {Liu}, J.~S. and {Liu}, J.~Y. and {Liu}, M.~Y. and {Liu}, R.~Y. and {Liu}, S.~M. and {Liu}, W. and {Liu}, Y.~N. and {Liu}, Z.~X. and {Long}, W.~J. and {Lu}, R. and {Lv}, H.~K. and {Ma}, B.~Q. and {Ma}, L.~L. and {Ma}, X.~H. and {Mao}, J.~R. and {Masood}, A. and {Mitthumsiri}, W. and {Montaruli}, T. and {Nan}, Y.~C. and {Pang}, B.~Y. and {Pattarakijwanich}, P. and {Pei}, Z.~Y. and {Qi}, M.~Y. and {Ruffolo}, D. and {Rulev}, V. and {S{\'a}iz}, A. and {Shao}, L. and {Shchegolev}, O. and {Sheng}, X.~D. and {Shi}, J.~R. and {Song}, H.~C. and {Stenkin}, Yu. V. and {Stepanov}, V. and {Sun}, Q.~N. and {Sun}, X.~N. and {Sun}, Z.~B. and {Tam}, P.~H.~T. and {Tang}, Z.~B. and {Tian}, W.~W. and {Wang}, B.~D. and {Wang}, C. and {Wang}, H. and {Wang}, H.~G. and {Wang}, J.~C. and {Wang}, J.~S. and {Wang}, L.~P. and {Wang}, L.~Y. and {Wang}, R.~N. and {Wang}, W. and {Wang}, W. and {Wang}, X.~G. and {Wang}, X.~J. and {Wang}, X.~Y. and {Wang}, Y.~D. and {Wang}, Y.~J. and {Wang}, Y.~P. and {Wang}, Z. and {Wang}, Z. and {Wang}, Z.~H. and {Wang}, Z.~X. and {Wei}, D.~M. and {Wei}, J.~J. and {Wei}, Y.~J. and {Wen}, T. and {Wu}, C.~Y. and {Wu}, H.~R. and {Wu}, S. and {Wu}, W.~X. and {Wu}, X.~F. and {Xi}, S.~Q. and {Xia}, J. and {Xia}, J.~J. and {Xiang}, G.~M. and {Xiao}, G. and {Xiao}, H.~B. and {Xin}, G.~G. and {Xin}, Y.~L. and {Xing}, Y. and {Xu}, D.~L. and {Xu}, R.~X. and {Xue}, L. and {Yan}, D.~H.},
        title = "{Observation of the Crab Nebula with LHAASO-KM2A - a performance study}",
      journal = {Chinese Physics C},
         year = 2021,
        month = feb,
       volume = {45},
       number = {2},
          eid = {025002},
        pages = {025002},
          doi = {10.1088/1674-1137/abd01b},
archivePrefix = {arXiv},
       eprint = {2010.06205},
 primaryClass = {astro-ph.HE},
       adsurl = {https://ui.adsabs.harvard.edu/abs/2021ChPhC..45b5002A}
}

@ARTICLE{2004NIMPA.518..188B,
       author = {{Baixeras}, C. and {Bastieri}, D. and {Bigongiari}, C. and {Blanch}, O. and {Blanchot}, G. and {Bock}, R. and {Bretz}, T. and {Chilingarian}, A. and {Coarasa}, J.~A. and {Colombo}, E. and {Contreras}, J.~C. and {Corti}, D. and {Cortina}, J. and {Domingo}, C. and {Domingo}, E. and {Ferenc}, D. and {Fern{\'a}ndez}, E. and {Flix}, J. and {Fonseca}, V. and {Font}, L. and {Galante}, N. and {Gaug}, M. and {Garczarczyk}, M. and {Gebauer}, J. and {Giller}, M. and {Goebel}, F. and {Hengstebeck}, T. and {Jacone}, P. and {de Jager}, O.~C. and {Kalekin}, O. and {Kestel}, M. and {Kneiske}, T. and {Laille}, A. and {L{\'o}pez}, M. and {L{\'o}pez}, J. and {Lorenz}, E. and {Mannheim}, K. and {Mariotti}, M. and {Mart{\'\i}nez}, M. and {Mase}, K. and {Merck}, M. and {Meucci}, M. and {Miralles}, L. and {Mirzoyan}, R. and {Moralejo}, A. and {Wilhelmi}, E. O{\~n}a and {Ordu{\~n}a}, R. and {Paneque}, D. and {Paoletti}, R. and {Pascoli}, D. and {Pavel}, N. and {Pegna}, R. and {Peruzzo}, L. and {Piccioli}, A. and {Roberts}, A. and {Reyes}, R. and {Saggion}, A. and {S{\'a}nchez}, A. and {Sartori}, P. and {Scalzotto}, V. and {Schweizer}, T. and {Sillanpaa}, A. and {Sobczynska}, D. and {Stamerra}, A. and {Stepanian}, A. and {Stiehler}, R. and {Takalo}, L. and {Teshima}, M. and {Tonello}, N. and {Torres}, A. and {Turini}, N. and {Vitale}, V. and {Volkov}, S. and {Wagner}, R.~M. and {Wibig}, T. and {Wittek}, W.},
        title = "{Commissioning and first tests of the MAGIC telescope}",
      journal = {Nuclear Instruments and Methods in Physics Research A},
         year = 2004,
        month = feb,
       volume = {518},
       number = {1-2},
        pages = {188-192},
          doi = {10.1016/j.nima.2003.10.057},
       adsurl = {https://ui.adsabs.harvard.edu/abs/2004NIMPA.518..188B}
}

@ARTICLE{2014MNRAS.444.3647B,
       author = {{Bhagwan}, Jai and {Gupta}, Alok C. and {Papadakis}, I.~E. and {Wiita}, Paul J.},
        title = "{Spectral energy distributions of the BL Lac PKS 2155 - 304 from XMM-Newton}",
      journal = {\mnras},
         year = 2014,
        month = nov,
       volume = {444},
       number = {4},
        pages = {3647-3656},
          doi = {10.1093/mnras/stu1703},
archivePrefix = {arXiv},
       eprint = {1409.2963},
 primaryClass = {astro-ph.HE},
       adsurl = {https://ui.adsabs.harvard.edu/abs/2014MNRAS.444.3647B}
}

@ARTICLE{2016NewA...44...21B,
       author = {{Bhagwan}, Jai and {Gupta}, A.~C. and {Papadakis}, I.~E. and {Wiita}, Paul J.},
        title = "{Flux and spectral variability of the blazar PKS 2155 -304 with XMM-Newton: Evidence of particle acceleration and synchrotron cooling}",
      journal = {\na},
         year = 2016,
        month = apr,
       volume = {44},
        pages = {21-28},
          doi = {10.1016/j.newast.2015.08.005},
archivePrefix = {arXiv},
       eprint = {1510.02817},
 primaryClass = {astro-ph.HE},
       adsurl = {https://ui.adsabs.harvard.edu/abs/2016NewA...44...21B}
}

@ARTICLE{2024APh...15902960B,
       author = {{Borwankar}, C. and {Sharma}, M. and {Hariharan}, J. and {Venugopal}, K. and {Godambe}, S. and {Mankuzhyil}, N. and {Chandra}, P. and {Khurana}, M. and {Pathania}, A. and {Chouhan}, N. and {Dhar}, V.~K. and {Thubstan}, R. and {Norlha}, S. and {Keshavanand} and {Sarkar}, D. and {Dar}, Z.~A. and {Kotwal}, S.~V. and {Godiyal}, S. and {Kushwaha}, C.~P. and {Singh}, K.~K. and {Das}, M.~P. and {Tolamati}, A. and {Ghosal}, B. and {Chanchalani}, K. and {Pandey}, P. and {Bhatt}, N. and {Bhattcharyya}, S. and {Sahayanathan}, S. and {Koul}, M.~K. and {Dorjey}, P. and {Dorji}, N. and {Chitnis}, V.~R. and {Tickoo}, A.~K. and {Rannot}, R.~C. and {Yadav}, K.~K.},
        title = "{Observations of the Crab Nebula with MACE (Major Atmospheric Cherenkov Experiment)}",
      journal = {Astroparticle Physics},
         year = 2024,
        month = jul,
       volume = {159},
          eid = {102960},
        pages = {102960},
          doi = {10.1016/j.astropartphys.2024.102960},
archivePrefix = {arXiv},
       eprint = {2404.01649},
 primaryClass = {astro-ph.HE},
       adsurl = {https://ui.adsabs.harvard.edu/abs/2024APh...15902960B}
}

@ARTICLE{2013ApJ...768...54B,
       author = {{B{\"o}ttcher}, M. and {Reimer}, A. and {Sweeney}, K. and {Prakash}, A.},
        title = "{Leptonic and Hadronic Modeling of Fermi-detected Blazars}",
      journal = {\apj},
         year = 2013,
        month = may,
       volume = {768},
       number = {1},
          eid = {54},
        pages = {54},
          doi = {10.1088/0004-637X/768/1/54},
archivePrefix = {arXiv},
       eprint = {1304.0605},
 primaryClass = {astro-ph.HE},
       adsurl = {https://ui.adsabs.harvard.edu/abs/2013ApJ...768...54B}
}

@ARTICLE{1997A&A...320L...5B,
       author = {{Bradbury}, S.~M. and {Deckers}, T. and {Petry}, D. and {Konopelko}, A. and {Aharonian}, F. and {Akhperjanian}, A.~G. and {Barrio}, J.~A. and {Beglarian}, A.~S. and {Beteta}, J.~J.~G. and {Contreras}, J.~L. and {Cortina}, J. and {Daum}, A. and {Feigl}, E. and {Fernandez}, J. and {Fonseca}, V. and {Frass}, A. and {Funk}, B. and {Gonzalez}, J.~C. and {Haustein}, V. and {Heinzelmann}, G. and {Hemberger}, M. and {Hermann}, G. and {Hess}, M. and {Heusler}, A. and {Holl}, I. and {Hofmann}, W. and {Horns}, D. and {Kankanian}, R. and {Kirstein}, O. and {Koehler}, C. and {Kranich}, D. and {Krawczynski}, H. and {Kornmayer}, H. and {Lampeitl}, H. and {Lindner}, A. and {Lorenz}, E. and {Magnussen}, N. and {Meyer}, H. and {Mirzoyan}, R. and {Moeller}, H. and {Moralejo}, A. and {Padilla}, L. and {Panter}, M. and {Plaga}, R. and {Prahl}, J. and {Prosch}, C. and {Puehlhofer}, G. and {Rauterberg}, G. and {Rhode}, W. and {Sahakian}, V. and {Samorski}, M. and {Sanchez}, J.~A. and {Schmele}, D. and {Stamm}, W. and {Ulrich}, M. and {Voelk}, H.~J. and {Westerhoff}, S. and {Wiebel-Sooth}, B. and {Wiedner}, C.~A. and {Willmer}, M. and {Wirth}, H.},
        title = "{Detection of {\ensuremath{\gamma}}-rays above 1.5TeV from MKN 501.}",
      journal = {\aap},
         year = 1997,
        month = apr,
       volume = {320},
        pages = {L5-L8},
          doi = {10.48550/arXiv.astro-ph/9612058},
archivePrefix = {arXiv},
       eprint = {astro-ph/9612058},
 primaryClass = {astro-ph},
       adsurl = {https://ui.adsabs.harvard.edu/abs/1997A&A...320L...5B}
}

@ARTICLE{1998ApJ...501..616C,
       author = {{Catanese}, M. and {Akerlof}, C.~W. and {Badran}, H.~M. and {Biller}, S.~D. and {Bond}, I.~H. and {Boyle}, P.~J. and {Bradbury}, S.~M. and {Buckley}, J.~H. and {Burdett}, A.~M. and {Gordo}, J. Buss{\'o}ns and {Carter-Lewis}, D.~A. and {Cawley}, M.~F. and {Connaughton}, V. and {Fegan}, D.~J. and {Finley}, J.~P. and {Gaidos}, J.~A. and {Hall}, T. and {Hillas}, A.~M. and {Krennrich}, F. and {Lamb}, R.~C. and {Lessard}, R.~W. and {Masterson}, C. and {McEnery}, J.~E. and {Mohanty}, G. and {Quinn}, J. and {Rodgers}, A.~J. and {Rose}, H.~J. and {Samuelson}, F.~W. and {Schubnell}, M.~S. and {Sembroski}, G.~H. and {Srinivasan}, R. and {Weekes}, T.~C. and {Wilson}, C.~W. and {Zweerink}, J.},
        title = "{Discovery of Gamma-Ray Emission above 350 GeV from the BL Lacertae Object 1ES 2344+514}",
      journal = {\apj},
         year = 1998,
        month = jul,
       volume = {501},
       number = {2},
        pages = {616-623},
          doi = {10.1086/305857},
archivePrefix = {arXiv},
       eprint = {astro-ph/9712325},
 primaryClass = {astro-ph},
       adsurl = {https://ui.adsabs.harvard.edu/abs/1998ApJ...501..616C}
}

@ARTICLE{1999ApJ...513..161C,
       author = {{Chadwick}, P.~M. and {Lyons}, K. and {McComb}, T.~J.~L. and {Orford}, K.~J. and {Osborne}, J.~L. and {Rayner}, S.~M. and {Shaw}, S.~E. and {Turver}, K.~E. and {Wieczorek}, G.~J.},
        title = "{Very High Energy Gamma Rays from PKS 2155-304}",
      journal = {\apj},
         year = 1999,
        month = mar,
       volume = {513},
       number = {1},
        pages = {161-167},
          doi = {10.1086/306862},
archivePrefix = {arXiv},
       eprint = {astro-ph/9810209},
 primaryClass = {astro-ph},
       adsurl = {https://ui.adsabs.harvard.edu/abs/1999ApJ...513..161C}
}

@ARTICLE{2022ApJ...939...80D,
       author = {{Devanand}, P.~U. and {Gupta}, Alok C. and {Jithesh}, V. and {Wiita}, Paul J.},
        title = "{Study of X-Ray Intraday Variability of HBL Blazars Based on Observations Obtained with XMM-Newton}",
      journal = {\apj},
         year = 2022,
        month = nov,
       volume = {939},
       number = {2},
          eid = {80},
        pages = {80},
          doi = {10.3847/1538-4357/ac9064},
archivePrefix = {arXiv},
       eprint = {2209.05515},
 primaryClass = {astro-ph.HE},
       adsurl = {https://ui.adsabs.harvard.edu/abs/2022ApJ...939...80D}
}

@ARTICLE{2025ApJS..278...20D,
       author = {{Devanand}, P.~U. and {Gupta}, Alok C. and {Jithesh}, V. and {Wiita}, Paul J. and {Gupta}, Archana},
        title = "{X-Ray Spectral Variability of 13 TeV High-energy-peaked Blazars with XMM-Newton}",
      journal = {\apjs},
         year = 2025,
        month = may,
       volume = {278},
       number = {1},
          eid = {20},
        pages = {20},
          doi = {10.3847/1538-4365/adc10d},
archivePrefix = {arXiv},
       eprint = {2503.08386},
 primaryClass = {astro-ph.HE},
       adsurl = {https://ui.adsabs.harvard.edu/abs/2025ApJS..278...20D}
}

@ARTICLE{2021MNRAS.506.1198D,
       author = {{Dhiman}, Vinit and {Gupta}, Alok C. and {Gaur}, Haritma and {Wiita}, Paul J.},
        title = "{Multiband variability of the TeV blazar PG 1553+113 with XMM-Newton}",
      journal = {\mnras},
         year = 2021,
        month = sep,
       volume = {506},
       number = {1},
        pages = {1198-1208},
          doi = {10.1093/mnras/stab1743},
archivePrefix = {arXiv},
       eprint = {2106.08514},
 primaryClass = {astro-ph.HE},
       adsurl = {https://ui.adsabs.harvard.edu/abs/2021MNRAS.506.1198D}
}

@ARTICLE{1998MNRAS.299..433F,
       author = {{Fossati}, G. and {Maraschi}, L. and {Celotti}, A. and {Comastri}, A. and {Ghisellini}, G.},
        title = "{A unifying view of the spectral energy distributions of blazars}",
      journal = {\mnras},
         year = 1998,
        month = sep,
       volume = {299},
       number = {2},
        pages = {433-448},
          doi = {10.1046/j.1365-8711.1998.01828.x},
archivePrefix = {arXiv},
       eprint = {astro-ph/9804103},
 primaryClass = {astro-ph},
       adsurl = {https://ui.adsabs.harvard.edu/abs/1998MNRAS.299..433F}
}

@ARTICLE{2004APh....22..285F,
       author = {{Funk}, S. and {Hermann}, G. and {Hinton}, J. and {Berge}, D. and {Bernl{\"o}hr}, K. and {Hofmann}, W. and {Nayman}, P. and {Toussenel}, F. and {Vincent}, P.},
        title = "{The trigger system of the H.E.S.S. telescope array}",
      journal = {Astroparticle Physics},
         year = 2004,
        month = nov,
       volume = {22},
       number = {3-4},
        pages = {285-296},
          doi = {10.1016/j.astropartphys.2004.08.001},
archivePrefix = {arXiv},
       eprint = {astro-ph/0408375},
 primaryClass = {astro-ph},
       adsurl = {https://ui.adsabs.harvard.edu/abs/2004APh....22..285F}
}

@ARTICLE{2010ApJ...718..279G,
       author = {{Gaur}, Haritma and {Gupta}, Alok C. and {Lachowicz}, Pawel and {Wiita}, Paul J.},
        title = "{Detection of Intra-day Variability Timescales of Four High-energy Peaked Blazars with XMM-Newton}",
      journal = {\apj},
         year = 2010,
        month = jul,
       volume = {718},
       number = {1},
        pages = {279-291},
          doi = {10.1088/0004-637X/718/1/279},
archivePrefix = {arXiv},
       eprint = {1005.4475},
 primaryClass = {astro-ph.CO},
       adsurl = {https://ui.adsabs.harvard.edu/abs/2010ApJ...718..279G}
}

@ARTICLE{2016MNRAS.462.1508G,
       author = {{Gupta}, Alok C. and {Kalita}, Nibedita and {Gaur}, Haritma and {Duorah}, Kalpana},
        title = "{Peak of spectral energy distribution plays an important role in intra-day variability of blazars?}",
      journal = {\mnras},
         year = 2016,
        month = oct,
       volume = {462},
       number = {2},
        pages = {1508-1516},
          doi = {10.1093/mnras/stw1667},
archivePrefix = {arXiv},
       eprint = {1607.02053},
 primaryClass = {astro-ph.HE},
       adsurl = {https://ui.adsabs.harvard.edu/abs/2016MNRAS.462.1508G}
}

@ARTICLE{2006APh....25..391H,
       author = {{Holder}, J. and {Atkins}, R.~W. and {Badran}, H.~M. and {Blaylock}, G. and {Bradbury}, S.~M. and {Buckley}, J.~H. and {Byrum}, K.~L. and {Carter-Lewis}, D.~A. and {Celik}, O. and {Chow}, Y.~C.~K. and {Cogan}, P. and {Cui}, W. and {Daniel}, M.~K. and {de la Calle Perez}, I. and {Dowdall}, C. and {Dowkontt}, P. and {Duke}, C. and {Falcone}, A.~D. and {Fegan}, S.~J. and {Finley}, J.~P. and {Fortin}, P. and {Fortson}, L.~F. and {Gibbs}, K. and {Gillanders}, G. and {Glidewell}, O.~J. and {Grube}, J. and {Gutierrez}, K.~J. and {Gyuk}, G. and {Hall}, J. and {Hanna}, D. and {Hays}, E. and {Horan}, D. and {Hughes}, S.~B. and {Humensky}, T.~B. and {Imran}, A. and {Jung}, I. and {Kaaret}, P. and {Kenny}, G.~E. and {Kieda}, D. and {Kildea}, J. and {Knapp}, J. and {Krawczynski}, H. and {Krennrich}, F. and {Lang}, M.~J. and {LeBohec}, S. and {Linton}, E. and {Little}, E.~K. and {Maier}, G. and {Manseri}, H. and {Milovanovic}, A. and {Moriarty}, P. and {Mukherjee}, R. and {Ogden}, P.~A. and {Ong}, R.~A. and {Petry}, D. and {Perkins}, J.~S. and {Pizlo}, F. and {Pohl}, M. and {Quinn}, J. and {Ragan}, K. and {Reynolds}, P.~T. and {Roache}, E.~T. and {Rose}, H.~J. and {Schroedter}, M. and {Sembroski}, G.~H. and {Sleege}, G. and {Steele}, D. and {Swordy}, S.~P. and {Syson}, A. and {Toner}, J.~A. and {Valcarcel}, L. and {Vassiliev}, V.~V. and {Wakely}, S.~P. and {Weekes}, T.~C. and {White}, R.~J. and {Williams}, D.~A. and {Wagner}, R.},
        title = "{The first VERITAS telescope}",
      journal = {Astroparticle Physics},
         year = 2006,
        month = jul,
       volume = {25},
       number = {6},
        pages = {391-401},
          doi = {10.1016/j.astropartphys.2006.04.002},
archivePrefix = {arXiv},
       eprint = {astro-ph/0604119},
 primaryClass = {astro-ph},
       adsurl = {https://ui.adsabs.harvard.edu/abs/2006APh....25..391H}
}

@ARTICLE{2003ApJ...583L...9H,
       author = {{Holder}, J. and {Bond}, I.~H. and {Boyle}, P.~J. and {Bradbury}, S.~M. and {Buckley}, J.~H. and {Carter-Lewis}, D.~A. and {Cui}, W. and {Dowdall}, C. and {Duke}, C. and {de la Calle Perez}, I. and {Falcone}, A. and {Fegan}, D.~J. and {Fegan}, S.~J. and {Finley}, J.~P. and {Fortson}, L. and {Gaidos}, J.~A. and {Gibbs}, K. and {Gammell}, S. and {Hall}, J. and {Hall}, T.~A. and {Hillas}, A.~M. and {Horan}, D. and {Jordan}, M. and {Kertzman}, M. and {Kieda}, D. and {Kildea}, J. and {Knapp}, J. and {Kosack}, K. and {Krawczynski}, H. and {Krennrich}, F. and {LeBohec}, S. and {Linton}, E.~T. and {Lloyd-Evans}, J. and {Moriarty}, P. and {M{\"u}ller}, D. and {Nagai}, T.~N. and {Ong}, R. and {Page}, M. and {Pallassini}, R. and {Petry}, D. and {Power-Mooney}, B. and {Quinn}, J. and {Rebillot}, P. and {Reynolds}, P.~T. and {Rose}, H.~J. and {Schroedter}, M. and {Sembroski}, G.~H. and {Swordy}, S.~P. and {Vassiliev}, V.~V. and {Wakely}, S.~P. and {Walker}, G. and {Weekes}, T.~C.},
        title = "{Detection of TeV Gamma Rays from the BL Lacertae Object 1ES 1959+650 with the Whipple 10 Meter Telescope}",
      journal = {\apjl},
         year = 2003,
        month = jan,
       volume = {583},
       number = {1},
        pages = {L9-L12},
          doi = {10.1086/367816},
archivePrefix = {arXiv},
       eprint = {astro-ph/0212170},
 primaryClass = {astro-ph},
       adsurl = {https://ui.adsabs.harvard.edu/abs/2003ApJ...583L...9H}
}

@ARTICLE{2002ApJ...571..753H,
       author = {{Horan}, D. and {Badran}, H.~M. and {Bond}, I.~H. and {Bradbury}, S.~M. and {Buckley}, J.~H. and {Carson}, M.~J. and {Carter-Lewis}, D.~A. and {Catanese}, M. and {Cui}, W. and {Dunlea}, S. and {Das}, D. and {de la Calle Perez}, I. and {D'Vali}, M. and {Fegan}, D.~J. and {Fegan}, S.~J. and {Finley}, J.~P. and {Gaidos}, J.~A. and {Gibbs}, K. and {Gillanders}, G.~H. and {Hall}, T.~A. and {Hillas}, A.~M. and {Holder}, J. and {Jordan}, M. and {Kertzman}, M. and {Kieda}, D. and {Kildea}, J. and {Knapp}, J. and {Kosack}, K. and {Krennrich}, F. and {Lang}, M.~J. and {LeBohec}, S. and {Lessard}, R. and {Lloyd-Evans}, J. and {McKernan}, B. and {Moriarty}, P. and {Muller}, D. and {Ong}, R. and {Pallassini}, R. and {Petry}, D. and {Quinn}, J. and {Reay}, N.~W. and {Reynolds}, P.~T. and {Rose}, H.~J. and {Sembroski}, G.~H. and {Sidwell}, R. and {Stanton}, N. and {Swordy}, S.~P. and {Vassiliev}, V.~V. and {Wakely}, S.~P. and {Weekes}, T.~C.},
        title = "{Detection of the BL Lacertae Object H1426+428 at TeV Gamma-Ray Energies}",
      journal = {\apj},
         year = 2002,
        month = jun,
       volume = {571},
       number = {2},
        pages = {753-762},
          doi = {10.1086/340019},
archivePrefix = {arXiv},
       eprint = {astro-ph/0202185},
 primaryClass = {astro-ph},
       adsurl = {https://ui.adsabs.harvard.edu/abs/2002ApJ...571..753H}
}

@ARTICLE{2015MNRAS.451.1356K,
       author = {{Kalita}, Nibedita and {Gupta}, Alok C. and {Wiita}, Paul J. and {Bhagwan}, Jai and {Duorah}, Kalpana},
        title = "{Multiband variability in the blazar 3C 273 with XMM-Newton}",
      journal = {\mnras},
         year = 2015,
        month = aug,
       volume = {451},
       number = {2},
        pages = {1356-1365},
          doi = {10.1093/mnras/stv1027},
archivePrefix = {arXiv},
       eprint = {1506.00586},
 primaryClass = {astro-ph.GA},
       adsurl = {https://ui.adsabs.harvard.edu/abs/2015MNRAS.451.1356K}
}

@ARTICLE{2017MNRAS.469.3824K,
       author = {{Kalita}, Nibedita and {Gupta}, Alok C. and {Wiita}, Paul J. and {Dewangan}, Gulab C. and {Duorah}, Kalpana},
        title = "{Origin of X-rays in the low state of the FSRQ 3C 273: evidence of inverse Compton emission}",
      journal = {\mnras},
         year = 2017,
        month = aug,
       volume = {469},
       number = {4},
        pages = {3824-3839},
          doi = {10.1093/mnras/stx1108},
archivePrefix = {arXiv},
       eprint = {1705.02721},
 primaryClass = {astro-ph.HE},
       adsurl = {https://ui.adsabs.harvard.edu/abs/2017MNRAS.469.3824K}
}

@ARTICLE{2019ApJ...880...19K,
       author = {{Kalita}, Nibedita and {Sawangwit}, Utane and {Gupta}, Alok C. and {Wiita}, Paul J.},
        title = "{Signature of Stochastic Acceleration and Cooling Processes in an Outburst Phase of the TeV Blazar ON 231}",
      journal = {\apj},
         year = 2019,
        month = jul,
       volume = {880},
       number = {1},
          eid = {19},
        pages = {19},
          doi = {10.3847/1538-4357/ab2765},
archivePrefix = {arXiv},
       eprint = {1907.03408},
 primaryClass = {astro-ph.HE},
       adsurl = {https://ui.adsabs.harvard.edu/abs/2019ApJ...880...19K}
}

@ARTICLE{2004NewAR..48..367K,
       author = {{Krawczynski}, Henric},
        title = "{TeV blazars - observations and models}",
      journal = {\nar},
         year = 2004,
        month = apr,
       volume = {48},
       number = {5-6},
        pages = {367-373},
          doi = {10.1016/j.newar.2003.12.008},
archivePrefix = {arXiv},
       eprint = {astro-ph/0309443},
 primaryClass = {astro-ph},
       adsurl = {https://ui.adsabs.harvard.edu/abs/2004NewAR..48..367K}
}

@ARTICLE{2009A&A...506L..17L,
       author = {{Lachowicz}, P. and {Gupta}, A.~C. and {Gaur}, H. and {Wiita}, P.~J.},
        title = "{A \raisebox{-0.5ex}\textasciitilde4.6 h quasi-periodic oscillation in the BL Lacertae PKS 2155-304?}",
      journal = {\aap},
         year = 2009,
        month = nov,
       volume = {506},
       number = {2},
        pages = {L17-L20},
          doi = {10.1051/0004-6361/200913161},
archivePrefix = {arXiv},
       eprint = {0909.2113},
 primaryClass = {astro-ph.HE},
       adsurl = {https://ui.adsabs.harvard.edu/abs/2009A&A...506L..17L}
}

@ARTICLE{2022ApJS..262....4N,
       author = {{Noel}, A. Priyana and {Gaur}, Haritma and {Gupta}, Alok C. and {Wierzcholska}, Alicja and {Ostrowski}, Micha{\l} and {Dhiman}, Vinit and {Bhatta}, Gopal},
        title = "{X-Ray Intraday Variability of the TeV Blazar Markarian 421 with XMM-Newton}",
      journal = {\apjs},
         year = 2022,
        month = sep,
       volume = {262},
       number = {1},
          eid = {4},
        pages = {4},
          doi = {10.3847/1538-4365/ac7799},
archivePrefix = {arXiv},
       eprint = {2206.02159},
 primaryClass = {astro-ph.HE},
       adsurl = {https://ui.adsabs.harvard.edu/abs/2022ApJS..262....4N}
}

@ARTICLE{2017ApJ...841..123P,
       author = {{Pandey}, Ashwani and {Gupta}, Alok C. and {Wiita}, Paul J.},
        title = "{X-Ray Intraday Variability of Five TeV Blazars with NuSTAR}",
      journal = {\apj},
         year = 2017,
        month = jun,
       volume = {841},
       number = {2},
          eid = {123},
        pages = {123},
          doi = {10.3847/1538-4357/aa705e},
archivePrefix = {arXiv},
       eprint = {1705.02719},
 primaryClass = {astro-ph.HE},
       adsurl = {https://ui.adsabs.harvard.edu/abs/2017ApJ...841..123P}
}

@ARTICLE{2018ApJ...859...49P,
       author = {{Pandey}, Ashwani and {Gupta}, Alok C. and {Wiita}, Paul J.},
        title = "{X-Ray Flux and Spectral Variability of Six TeV Blazars with NuSTAR}",
      journal = {\apj},
         year = 2018,
        month = may,
       volume = {859},
       number = {1},
          eid = {49},
        pages = {49},
          doi = {10.3847/1538-4357/aabc5b},
archivePrefix = {arXiv},
       eprint = {1804.10126},
 primaryClass = {astro-ph.HE},
       adsurl = {https://ui.adsabs.harvard.edu/abs/2018ApJ...859...49P}
}

@ARTICLE{2022MNRAS.511.3101P,
       author = {{Pavana Gowtami}, G.~S. and {Gaur}, Haritma and {Gupta}, Alok C. and {Wiita}, Paul J. and {Liao}, Mai and {Ward}, Martin},
        title = "{X-ray intraday variability and power spectral density profiles of the blazar 3C 273 with XMM-Newton during 2000-2021}",
      journal = {\mnras},
         year = 2022,
        month = apr,
       volume = {511},
       number = {2},
        pages = {3101-3112},
          doi = {10.1093/mnras/stac286},
archivePrefix = {arXiv},
       eprint = {2109.13118},
 primaryClass = {astro-ph.HE},
       adsurl = {https://ui.adsabs.harvard.edu/abs/2022MNRAS.511.3101P}
}

@ARTICLE{1992Natur.358..477P,
       author = {{Punch}, M. and {Akerlof}, C.~W. and {Cawley}, M.~F. and {Chantell}, M. and {Fegan}, D.~J. and {Fennell}, S. and {Gaidos}, J.~A. and {Hagan}, J. and {Hillas}, A.~M. and {Jiang}, Y. and {Kerrick}, A.~D. and {Lamb}, R.~C. and {Lawrence}, M.~A. and {Lewis}, D.~A. and {Meyer}, D.~I. and {Mohanty}, G. and {O'Flaherty}, K.~S. and {Reynolds}, P.~T. and {Rovero}, A.~C. and {Schubnell}, M.~S. and {Sembroski}, G. and {Weekes}, T.~C. and {Whitaker}, T. and {Wilson}, C.},
        title = "{Detection of TeV photons from the active galaxy Markarian 421}",
      journal = {\nat},
         year = 1992,
        month = aug,
       volume = {358},
       number = {6386},
        pages = {477-478},
          doi = {10.1038/358477a0},
       adsurl = {https://ui.adsabs.harvard.edu/abs/1992Natur.358..477P}
}

@ARTICLE{1996ApJ...456L..83Q,
       author = {{Quinn}, J. and {Akerlof}, C.~W. and {Biller}, S. and {Buckley}, J. and {Carter-Lewis}, D.~A. and {Cawley}, M.~F. and {Catanese}, M. and {Connaughton}, V. and {Fegan}, D.~J. and {Finley}, J.~P. and {Gaidos}, J. and {Hillas}, A.~M. and {Lamb}, R.~C. and {Krennrich}, F. and {Lessard}, R. and {McEnery}, J.~E. and {Meyer}, D.~I. and {Mohanty}, G. and {Rodgers}, A.~J. and {Rose}, H.~J. and {Sembroski}, G. and {Schubnell}, M.~S. and {Weekes}, T.~C. and {Wilson}, C. and {Zweerink}, J.},
        title = "{Detection of Gamma Rays with E > 300 GeV from Markarian 501}",
      journal = {\apjl},
         year = 1996,
        month = jan,
       volume = {456},
        pages = {L83},
          doi = {10.1086/309878},
       adsurl = {https://ui.adsabs.harvard.edu/abs/1996ApJ...456L..83Q}
}

@ARTICLE{2025A&ARv..33....8R,
       author = {{Raiteri}, Claudia M.},
        title = "{The variability of blazars throughout the electromagnetic spectrum}",
      journal = {\aapr},
         year = 2025,
        month = nov,
       volume = {33},
       number = {1},
          eid = {8},
        pages = {8},
          doi = {10.1007/s00159-025-00165-4},
archivePrefix = {arXiv},
       eprint = {2511.18975},
 primaryClass = {astro-ph.HE},
       adsurl = {https://ui.adsabs.harvard.edu/abs/2025A&ARv..33....8R}
}

@ARTICLE{2009ApJ...696.2170R,
       author = {{Rani}, Bindu and {Wiita}, Paul J. and {Gupta}, Alok C.},
        title = "{Nearly Periodic Fluctuations in the Long-term X-Ray Light Curves of the Blazars AO 0235+164 and 1ES 2321+419}",
      journal = {\apj},
         year = 2009,
        month = may,
       volume = {696},
       number = {2},
        pages = {2170-2178},
          doi = {10.1088/0004-637X/696/2/2170},
archivePrefix = {arXiv},
       eprint = {0903.2606},
 primaryClass = {astro-ph.CO},
       adsurl = {https://ui.adsabs.harvard.edu/abs/2009ApJ...696.2170R}
}

@ARTICLE{1995PASP..107..803U,
       author = {{Urry}, C. Megan and {Padovani}, Paolo},
        title = "{Unified Schemes for Radio-Loud Active Galactic Nuclei}",
      journal = {\pasp},
         year = 1995,
        month = sep,
       volume = {107},
        pages = {803},
          doi = {10.1086/133630},
archivePrefix = {arXiv},
       eprint = {astro-ph/9506063},
 primaryClass = {astro-ph},
       adsurl = {https://ui.adsabs.harvard.edu/abs/1995PASP..107..803U}
}

@ARTICLE{2019ApJ...884..125Z,
       author = {{Zhang}, Zhongli and {Gupta}, Alok C. and {Gaur}, Haritma and {Wiita}, Paul J. and {An}, Tao and {Gu}, Minfeng and {Hu}, Dan and {Xu}, Haiguang},
        title = "{X-Ray Intraday Variability of the TeV Blazar Mrk 421 with Suzaku}",
      journal = {\apj},
         year = 2019,
        month = oct,
       volume = {884},
       number = {2},
          eid = {125},
        pages = {125},
          doi = {10.3847/1538-4357/ab3f3a},
archivePrefix = {arXiv},
       eprint = {1908.08149},
 primaryClass = {astro-ph.HE},
       adsurl = {https://ui.adsabs.harvard.edu/abs/2019ApJ...884..125Z}
}

@ARTICLE{2021ApJ...909..103Z,
       author = {{Zhang}, Zhongli and {Gupta}, Alok C. and {Gaur}, Haritma and {Wiita}, Paul J. and {An}, Tao and {Lu}, Yang and {Fan}, Shida and {Xu}, Haiguang},
        title = "{X-Ray Intraday Variability of the TeV Blazar PKS 2155-304 with Suzaku during 2005-2014}",
      journal = {\apj},
         year = 2021,
        month = mar,
       volume = {909},
       number = {2},
          eid = {103},
        pages = {103},
          doi = {10.3847/1538-4357/abdd38},
archivePrefix = {arXiv},
       eprint = {2101.05977},
 primaryClass = {astro-ph.HE},
       adsurl = {https://ui.adsabs.harvard.edu/abs/2021ApJ...909..103Z}
}

@ARTICLE{2024MNRAS.532.3285Z,
       author = {{Zhou}, Dongtao and {Zhang}, Zhongli and {Gupta}, Alok C. and {Kushwaha}, Pankaj and {Wiita}, Paul J. and {Gu}, Minfeng and {Xu}, Haiguang},
        title = "{X-ray flux and spectral variability of the blazar OJ 287 with Suzaku}",
      journal = {\mnras},
         year = 2024,
        month = aug,
       volume = {532},
       number = {3},
        pages = {3285-3298},
          doi = {10.1093/mnras/stae1722},
archivePrefix = {arXiv},
       eprint = {2408.02371},
 primaryClass = {astro-ph.HE},
       adsurl = {https://ui.adsabs.harvard.edu/abs/2024MNRAS.532.3285Z}
}

@INPROCEEDINGS{1996ASPC..101...17A,
       author = {{Arnaud}, K.~A.},
        title = "{XSPEC: The First Ten Years}",
    booktitle = {Astronomical Data Analysis Software and Systems V},
         year = 1996,
       editor = {{Jacoby}, George H. and {Barnes}, Jeannette},
       series = {Astronomical Society of the Pacific Conference Series},
       volume = {101},
        month = jan,
        pages = {17},
       adsurl = {https://ui.adsabs.harvard.edu/abs/1996ASPC..101...17A}
}

@ARTICLE{2016A&A...594A.116H,
       author = {{HI4PI Collaboration} and {Ben Bekhti}, N. and {Fl{\"o}er}, L. and {Keller}, R. and {Kerp}, J. and {Lenz}, D. and {Winkel}, B. and {Bailin}, J. and {Calabretta}, M.~R. and {Dedes}, L. and {Ford}, H.~A. and {Gibson}, B.~K. and {Haud}, U. and {Janowiecki}, S. and {Kalberla}, P.~M.~W. and {Lockman}, F.~J. and {McClure-Griffiths}, N.~M. and {Murphy}, T. and {Nakanishi}, H. and {Pisano}, D.~J. and {Staveley-Smith}, L.},
        title = "{HI4PI: A full-sky H I survey based on EBHIS and GASS}",
      journal = {\aap},
         year = 2016,
        month = oct,
       volume = {594},
          eid = {A116},
        pages = {A116},
          doi = {10.1051/0004-6361/201629178},
archivePrefix = {arXiv},
       eprint = {1610.06175},
 primaryClass = {astro-ph.GA},
       adsurl = {https://ui.adsabs.harvard.edu/abs/2016A&A...594A.116H}
}

@ARTICLE{2013ApJ...770..103H,
       author = {{Harrison}, Fiona A. and {Craig}, William W. and {Christensen}, Finn E. and {Hailey}, Charles J. and {Zhang}, William W. and {Boggs}, Steven E. and {Stern}, Daniel and {Cook}, W. Rick and {Forster}, Karl and {Giommi}, Paolo and {Grefenstette}, Brian W. and {Kim}, Yunjin and {Kitaguchi}, Takao and {Koglin}, Jason E. and {Madsen}, Kristin K. and {Mao}, Peter H. and {Miyasaka}, Hiromasa and {Mori}, Kaya and {Perri}, Matteo and {Pivovaroff}, Michael J. and {Puccetti}, Simonetta and {Rana}, Vikram R. and {Westergaard}, Niels J. and {Willis}, Jason and {Zoglauer}, Andreas and {An}, Hongjun and {Bachetti}, Matteo and {Barri{\`e}re}, Nicolas M. and {Bellm}, Eric C. and {Bhalerao}, Varun and {Brejnholt}, Nicolai F. and {Fuerst}, Felix and {Liebe}, Carl C. and {Markwardt}, Craig B. and {Nynka}, Melania and {Vogel}, Julia K. and {Walton}, Dominic J. and {Wik}, Daniel R. and {Alexander}, David M. and {Cominsky}, Lynn R. and {Hornschemeier}, Ann E. and {Hornstrup}, Allan and {Kaspi}, Victoria M. and {Madejski}, Greg M. and {Matt}, Giorgio and {Molendi}, Silvano and {Smith}, David M. and {Tomsick}, John A. and {Ajello}, Marco and {Ballantyne}, David R. and {Balokovi{\'c}}, Mislav and {Barret}, Didier and {Bauer}, Franz E. and {Blandford}, Roger D. and {Brandt}, W. Niel and {Brenneman}, Laura W. and {Chiang}, James and {Chakrabarty}, Deepto and {Chenevez}, Jerome and {Comastri}, Andrea and {Dufour}, Francois and {Elvis}, Martin and {Fabian}, Andrew C. and {Farrah}, Duncan and {Fryer}, Chris L. and {Gotthelf}, Eric V. and {Grindlay}, Jonathan E. and {Helfand}, David J. and {Krivonos}, Roman and {Meier}, David L. and {Miller}, Jon M. and {Natalucci}, Lorenzo and {Ogle}, Patrick and {Ofek}, Eran O. and {Ptak}, Andrew and {Reynolds}, Stephen P. and {Rigby}, Jane R. and {Tagliaferri}, Gianpiero and {Thorsett}, Stephen E. and {Treister}, Ezequiel and {Urry}, C. Megan},
        title = "{The Nuclear Spectroscopic Telescope Array (NuSTAR) High-energy X-Ray Mission}",
      journal = {\apj},
         year = 2013,
        month = jun,
       volume = {770},
       number = {2},
          eid = {103},
        pages = {103},
          doi = {10.1088/0004-637X/770/2/103},
archivePrefix = {arXiv},
       eprint = {1301.7307},
 primaryClass = {astro-ph.IM},
       adsurl = {https://ui.adsabs.harvard.edu/abs/2013ApJ...770..103H}
}

@ARTICLE{2004A&A...413..489M,
       author = {{Massaro}, E. and {Perri}, M. and {Giommi}, P. and {Nesci}, R.},
        title = "{Log-parabolic spectra and particle acceleration in the BL Lac object Mkn 421: Spectral analysis of the complete BeppoSAX wide band X-ray data set}",
      journal = {\aap},
         year = 2004,
        month = jan,
       volume = {413},
        pages = {489-503},
          doi = {10.1051/0004-6361:20031558},
archivePrefix = {arXiv},
       eprint = {astro-ph/0312260},
 primaryClass = {astro-ph},
       adsurl = {https://ui.adsabs.harvard.edu/abs/2004A&A...413..489M}
}

@ARTICLE{2022A&A...663A.178M,
       author = {{Mondal}, S. and {Rani}, P. and {Stalin}, C.~S. and {Chakrabarti}, S.~K. and {Rakshit}, S.},
        title = "{Flux and spectral variability of Mrk 421 during its moderate activity state using NuSTAR: Possible accretion disc contribution?}",
      journal = {\aap},
         year = 2022,
        month = jul,
       volume = {663},
          eid = {A178},
        pages = {A178},
          doi = {10.1051/0004-6361/202141990},
archivePrefix = {arXiv},
       eprint = {2204.02132},
 primaryClass = {astro-ph.HE},
       adsurl = {https://ui.adsabs.harvard.edu/abs/2022A&A...663A.178M}
}

@ARTICLE{2000ApJ...542..914W,
       author = {{Wilms}, J. and {Allen}, A. and {McCray}, R.},
        title = "{On the Absorption of X-Rays in the Interstellar Medium}",
      journal = {\apj},
         year = 2000,
        month = oct,
       volume = {542},
       number = {2},
        pages = {914-924},
          doi = {10.1086/317016},
archivePrefix = {arXiv},
       eprint = {astro-ph/0008425},
 primaryClass = {astro-ph},
       adsurl = {https://ui.adsabs.harvard.edu/abs/2000ApJ...542..914W}
}

@ARTICLE{2020A&C....3300412T,
       author = {{Taghizadeh-Popp}, M. and {Kim}, J.~W. and {Lemson}, G. and {Medvedev}, D. and {Raddick}, M.~J. and {Szalay}, A.~S. and {Thakar}, A.~R. and {Booker}, J. and {Chhetri}, C. and {Dobos}, L. and {Rippin}, M.},
        title = "{SciServer: A science platform for astronomy and beyond}",
      journal = {Astronomy and Computing},
         year = 2020,
        month = oct,
       volume = {33},
          eid = {100412},
        pages = {100412},
          doi = {10.1016/j.ascom.2020.100412},
archivePrefix = {arXiv},
       eprint = {2001.08619},
 primaryClass = {astro-ph.IM},
       adsurl = {https://ui.adsabs.harvard.edu/abs/2020A&C....3300412T}
}

@ARTICLE{2016A&A...587A.151K,
       author = {{Kaastra}, J.~S. and {Bleeker}, J.~A.~M.},
        title = "{Optimal binning of X-ray spectra and response matrix design}",
      journal = {\aap},
         year = 2016,
        month = mar,
       volume = {587},
          eid = {A151},
        pages = {A151},
          doi = {10.1051/0004-6361/201527395},
archivePrefix = {arXiv},
       eprint = {1601.05309},
 primaryClass = {astro-ph.IM},
       adsurl = {https://ui.adsabs.harvard.edu/abs/2016A&A...587A.151K}
}

@ARTICLE{2025arXiv251208531K,
       author = {{Kizhakkekalam}, Sangeetha and {Bhatta}, Gopal and {K}, Navaneeth P and {Adhikari}, Tek P.},
        title = "{NICER Perspective on TeV Blazar Mrk\raisebox{-0.5ex}\textasciitilde421: X-ray Variability and Particle Acceleration}",
      journal = {arXiv e-prints},
         year = 2025,
        month = dec,
          eid = {arXiv:2512.08531},
        pages = {arXiv:2512.08531},
          doi = {10.48550/arXiv.2512.08531},
archivePrefix = {arXiv},
       eprint = {2512.08531},
 primaryClass = {astro-ph.HE},
       adsurl = {https://ui.adsabs.harvard.edu/abs/2025arXiv251208531K}
}

@ARTICLE{2025JHEAp..4800417A,
       author = {{Ata}, S.~A. and {Ahmed}, Nasser M. and {Beheary}, M.~M. and {Kamal}, F.~Y.},
        title = "{X-ray flux variability of Blazar Mrk 501 observed using NuSTAR}",
      journal = {Journal of High Energy Astrophysics},
         year = 2025,
        month = aug,
       volume = {48},
          eid = {100417},
        pages = {100417},
          doi = {10.1016/j.jheap.2025.100417},
       adsurl = {https://ui.adsabs.harvard.edu/abs/2025JHEAp..4800417A}
}

@ARTICLE{2008A&A...478..395M,
       author = {{Massaro}, F. and {Tramacere}, A. and {Cavaliere}, A. and {Perri}, M. and {Giommi}, P.},
        title = "{X-ray spectral evolution of TeV BL Lacertae objects: eleven years of observations with BeppoSAX, XMM-Newton and Swift satellites}",
      journal = {\aap},
         year = 2008,
        month = feb,
       volume = {478},
       number = {2},
        pages = {395-401},
          doi = {10.1051/0004-6361:20078639},
archivePrefix = {arXiv},
       eprint = {0712.2116},
 primaryClass = {astro-ph},
       adsurl = {https://ui.adsabs.harvard.edu/abs/2008A&A...478..395M}
}

@ARTICLE{2007A&A...466..521T,
       author = {{Tramacere}, A. and {Massaro}, F. and {Cavaliere}, A.},
        title = "{Signatures of synchrotron emission and of electron acceleration in the X-ray spectra of Mrk 421}",
      journal = {\aap},
         year = 2007,
        month = may,
       volume = {466},
       number = {2},
        pages = {521-529},
          doi = {10.1051/0004-6361:20066723},
archivePrefix = {arXiv},
       eprint = {astro-ph/0702151},
 primaryClass = {astro-ph},
       adsurl = {https://ui.adsabs.harvard.edu/abs/2007A&A...466..521T}
}

@ARTICLE{2016ApJ...819..156B,
       author = {{Balokovi{\'c}}, M. and {Paneque}, D. and {Madejski}, G. and {Furniss}, A. and {Chiang}, J. and {Ajello}, M. and {Alexander}, D.~M. and {Barret}, D. and {Blandford}, R.~D. and {Boggs}, S.~E. and {Christensen}, F.~E. and {Craig}, W.~W. and {Forster}, K. and {Giommi}, P. and {Grefenstette}, B. and {Hailey}, C. and {Harrison}, F.~A. and {Hornstrup}, A. and {Kitaguchi}, T. and {Koglin}, J.~E. and {Madsen}, K.~K. and {Mao}, P.~H. and {Miyasaka}, H. and {Mori}, K. and {Perri}, M. and {Pivovaroff}, M.~J. and {Puccetti}, S. and {Rana}, V. and {Stern}, D. and {Tagliaferri}, G. and {Urry}, C.~M. and {Westergaard}, N.~J. and {Zhang}, W.~W. and {Zoglauer}, A. and {NuSTAR Team} and {Archambault}, S. and {Archer}, A. and {Barnacka}, A. and {Benbow}, W. and {Bird}, R. and {Buckley}, J.~H. and {Bugaev}, V. and {Cerruti}, M. and {Chen}, X. and {Ciupik}, L. and {Connolly}, M.~P. and {Cui}, W. and {Dickinson}, H.~J. and {Dumm}, J. and {Eisch}, J.~D. and {Falcone}, A. and {Feng}, Q. and {Finley}, J.~P. and {Fleischhack}, H. and {Fortson}, L. and {Griffin}, S. and {Griffiths}, S.~T. and {Grube}, J. and {Gyuk}, G. and {Huetten}, M. and {H{\r{a}}kansson}, N. and {Holder}, J. and {Humensky}, T.~B. and {Johnson}, C.~A. and {Kaaret}, P. and {Kertzman}, M. and {Khassen}, Y. and {Kieda}, D. and {Krause}, M. and {Krennrich}, F. and {Lang}, M.~J. and {Maier}, G. and {McArthur}, S. and {Meagher}, K. and {Moriarty}, P. and {Nelson}, T. and {Nieto}, D. and {Ong}, R.~A. and {Park}, N. and {Pohl}, M. and {Popkow}, A. and {Pueschel}, E. and {Reynolds}, P.~T. and {Richards}, G.~T. and {Roache}, E. and {Santander}, M. and {Sembroski}, G.~H. and {Shahinyan}, K. and {Smith}, A.~W. and {Staszak}, D. and {Telezhinsky}, I. and {Todd}, N.~W. and {Tucci}, J.~V. and {Tyler}, J. and {Vincent}, S. and {Weinstein}, A. and {Wilhelm}, A. and {Williams}, D.~A. and {Zitzer}, B. and {VERITAS Collaboration} and {Ahnen}, M.~L. and {Ansoldi}, S. and {Antonelli}, L.~A. and {Antoranz}, P. and {Babic}, A. and {Banerjee}, B. and {Bangale}, P. and {Barres de Almeida}, U. and {Barrio}, J.~A. and {Becerra Gonz{\'a}lez}, J. and {Bednarek}, W. and {Bernardini}, E. and {Biasuzzi}, B. and {Biland}, A. and {Blanch}, O. and {Bonnefoy}, S. and {Bonnoli}, G. and {Borracci}, F. and {Bretz}, T. and {Carmona}, E. and {Carosi}, A. and {Chatterjee}, A. and {Clavero}, R. and {Colin}, P. and {Colombo}, E. and {Contreras}, J.~L. and {Cortina}, J. and {Covino}, S. and {Da Vela}, P. and {Dazzi}, F. and {De Angelis}, A. and {De Lotto}, B. and {de O{\~n}a Wilhelmi}, E. and {Delgado Mendez}, C. and {Di Pierro}, F. and {Dominis Prester}, D. and {Dorner}, D. and {Doro}, M. and {Einecke}, S. and {Elsaesser}, D. and {Fern{\'a}ndez-Barral}, A. and {Fidalgo}, D. and {Fonseca}, M.~V. and {Font}, L. and {Frantzen}, K. and {Fruck}, C. and {Galindo}, D. and {Garc{\'\i}a L{\'o}pez}, R.~J. and {Garczarczyk}, M. and {Garrido Terrats}, D. and {Gaug}, M. and {Giammaria}, P. and {Glawion (Eisenacher}, D. and {Godinovi{\'c}}, N. and {Gonz{\'a}lez Mu{\~n}oz}, A. and {Guberman}, D. and {Hahn}, A. and {Hanabata}, Y. and {Hayashida}, M. and {Herrera}, J. and {Hose}, J. and {Hrupec}, D. and {Hughes}, G. and {Idec}, W. and {Kodani}, K. and {Konno}, Y. and {Kubo}, H. and {Kushida}, J. and {La Barbera}, A. and {Lelas}, D. and {Lindfors}, E. and {Lombardi}, S. and {Longo}, F. and {L{\'o}pez}, M. and {L{\'o}pez-Coto}, R. and {L{\'o}pez-Oramas}, A. and {Lorenz}, E. and {Majumdar}, P. and {Makariev}, M. and {Mallot}, K. and {Maneva}, G. and {Manganaro}, M. and {Mannheim}, K. and {Maraschi}, L. and {Marcote}, B. and {Mariotti}, M. and {Mart{\'\i}nez}, M. and {Mazin}, D. and {Menzel}, U. and {Miranda}, J.~M. and {Mirzoyan}, R. and {Moralejo}, A. and {Moretti}, E. and {Nakajima}, D. and {Neustroev}, V. and {Niedzwiecki}, A. and {Nievas Rosillo}, M. and {Nilsson}, K. and {Nishijima}, K. and {Noda}, K.},
        title = "{Multiwavelength Study of Quiescent States of Mrk 421 with Unprecedented Hard X-Ray Coverage Provided by NuSTAR in 2013}",
      journal = {\apj},
         year = 2016,
        month = mar,
       volume = {819},
       number = {2},
          eid = {156},
        pages = {156},
          doi = {10.3847/0004-637X/819/2/156},
archivePrefix = {arXiv},
       eprint = {1512.02235},
 primaryClass = {astro-ph.HE},
       adsurl = {https://ui.adsabs.harvard.edu/abs/2016ApJ...819..156B}
}

@ARTICLE{2015A&A...576A.126A,
       author = {{Aleksi{\'c}}, J. and {Ansoldi}, S. and {Antonelli}, L.~A. and {Antoranz}, P. and {Babic}, A. and {Bangale}, P. and {Barres de Almeida}, U. and {Barrio}, J.~A. and {Becerra Gonz{\'a}lez}, J. and {Bednarek}, W. and {Berger}, K. and {Bernardini}, E. and {Biland}, A. and {Blanch}, O. and {Bock}, R.~K. and {Bonnefoy}, S. and {Bonnoli}, G. and {Borracci}, F. and {Bretz}, T. and {Carmona}, E. and {Carosi}, A. and {Carreto Fidalgo}, D. and {Colin}, P. and {Colombo}, E. and {Contreras}, J.~L. and {Cortina}, J. and {Covino}, S. and {Da Vela}, P. and {Dazzi}, F. and {De Angelis}, A. and {De Caneva}, G. and {De Lotto}, B. and {Delgado Mendez}, C. and {Doert}, M. and {Dom{\'\i}nguez}, A. and {Dominis Prester}, D. and {Dorner}, D. and {Doro}, M. and {Einecke}, S. and {Eisenacher}, D. and {Elsaesser}, D. and {Farina}, E. and {Ferenc}, D. and {Fonseca}, M.~V. and {Font}, L. and {Frantzen}, K. and {Fruck}, C. and {Garc{\'\i}a L{\'o}pez}, R.~J. and {Garczarczyk}, M. and {Garrido Terrats}, D. and {Gaug}, M. and {Giavitto}, G. and {Godinovi{\'c}}, N. and {Gonz{\'a}lez Mu{\~n}oz}, A. and {Gozzini}, S.~R. and {Hadamek}, A. and {Hadasch}, D. and {Herrero}, A. and {Hildebrand}, D. and {Hose}, J. and {Hrupec}, D. and {Idec}, W. and {Kadenius}, V. and {Kellermann}, H. and {Knoetig}, M.~L. and {Krause}, J. and {Kushida}, J. and {La Barbera}, A. and {Lelas}, D. and {Lewandowska}, N. and {Lindfors}, E. and {Longo}, F. and {Lombardi}, S. and {L{\'o}pez}, M. and {L{\'o}pez-Coto}, R. and {L{\'o}pez-Oramas}, A. and {Lorenz}, E. and {Lozano}, I. and {Makariev}, M. and {Mallot}, K. and {Maneva}, G. and {Mankuzhiyil}, N. and {Mannheim}, K. and {Maraschi}, L. and {Marcote}, B. and {Mariotti}, M. and {Mart{\'\i}nez}, M. and {Mazin}, D. and {Menzel}, U. and {Meucci}, M. and {Miranda}, J.~M. and {Mirzoyan}, R. and {Moralejo}, A. and {Munar-Adrover}, P. and {Nakajima}, D. and {Niedzwiecki}, A. and {Nilsson}, K. and {Nowak}, N. and {Orito}, R. and {Overkemping}, A. and {Paiano}, S. and {Palatiello}, M. and {Paneque}, D. and {Paoletti}, R. and {Paredes}, J.~M. and {Paredes-Fortuny}, X. and {Partini}, S. and {Persic}, M. and {Prada}, F. and {Prada Moroni}, P.~G. and {Prandini}, E. and {Preziuso}, S. and {Puljak}, I. and {Reinthal}, R. and {Rhode}, W. and {Rib{\'o}}, M. and {Rico}, J. and {RodriguezGarcia}, J. and {R{\"u}gamer}, S. and {Saggion}, A. and {Saito}, K. and {Salvati}, M. and {Satalecka}, K. and {Scalzotto}, V. and {Scapin}, V. and {Schultz}, C. and {Schweizer}, T. and {Shore}, S.~N. and {Sillanp{\"a}{\"a}}, A. and {Sitarek}, J. and {Snidaric}, I. and {Sobczynska}, D. and {Spanier}, F. and {Stamatescu}, V. and {Stamerra}, A. and {Steinbring}, T. and {Storz}, J. and {Sun}, S. and {Suri{\'c}}, T. and {Takalo}, L. and {Tavecchio}, F. and {Temnikov}, P. and {Terzi{\'c}}, T. and {Tescaro}, D. and {Teshima}, M. and {Thaele}, J. and {Tibolla}, O. and {Torres}, D.~F. and {Toyama}, T. and {Treves}, A. and {Uellenbeck}, M. and {Vogler}, P. and {Wagner}, R.~M. and {Zandanel}, F. and {Zanin}, R. and {MAGIC Collaboration} and {Archambault}, S. and {Behera}, B. and {Beilicke}, M. and {Benbow}, W. and {Bird}, R. and {Buckley}, J.~H. and {Bugaev}, V. and {Cerruti}, M. and {Chen}, X. and {Ciupik}, L. and {Collins-Hughes}, E. and {Cui}, W. and {Dumm}, J. and {Eisch}, J.~D. and {Falcone}, A. and {Federici}, S. and {Feng}, Q. and {Finley}, J.~P. and {Fleischhack}, H. and {Fortin}, P. and {Fortson}, L. and {Furniss}, A. and {Griffin}, S. and {Griffiths}, S.~T. and {Grube}, J. and {Gyuk}, G. and {Hanna}, D. and {Holder}, J. and {Hughes}, G. and {Humensky}, T.~B. and {Johnson}, C.~A. and {Kaaret}, P. and {Kertzman}, M. and {Khassen}, Y. and {Kieda}, D. and {Krawczynski}, H. and {Krennrich}, F. and {Kumar}, S. and {Lang}, M.~J. and {Maier}, G. and {McArthur}, S. and {Meagher}, K. and {Moriarty}, P. and {Mukherjee}, R.},
        title = "{The 2009 multiwavelength campaign on Mrk 421: Variability and correlation studies}",
      journal = {\aap},
         year = 2015,
        month = apr,
       volume = {576},
          eid = {A126},
        pages = {A126},
          doi = {10.1051/0004-6361/201424216},
archivePrefix = {arXiv},
       eprint = {1502.02650},
 primaryClass = {astro-ph.HE},
       adsurl = {https://ui.adsabs.harvard.edu/abs/2015A&A...576A.126A}
}

@ARTICLE{2011ApJ...738...25A,
       author = {{Acciari}, V.~A. and {Aliu}, E. and {Arlen}, T. and {Aune}, T. and {Beilicke}, M. and {Benbow}, W. and {Boltuch}, D. and {Bradbury}, S.~M. and {Buckley}, J.~H. and {Bugaev}, V. and {Byrum}, K. and {Cannon}, A. and {Cesarini}, A. and {Ciupik}, L. and {Cui}, W. and {Dickherber}, R. and {Duke}, C. and {Falcone}, A. and {Finley}, J.~P. and {Finnegan}, G. and {Fortson}, L. and {Furniss}, A. and {Galante}, N. and {Gall}, D. and {Gillanders}, G.~H. and {Godambe}, S. and {Grube}, J. and {Guenette}, R. and {Gyuk}, G. and {Hanna}, D. and {Holder}, J. and {Hui}, C.~M. and {Humensky}, T.~B. and {Imran}, A. and {Kaaret}, P. and {Karlsson}, N. and {Kertzman}, M. and {Kieda}, D. and {Konopelko}, A. and {Krawczynski}, H. and {Krennrich}, F. and {Lang}, M.~J. and {Maier}, G. and {McArthur}, S. and {McCutcheon}, M. and {Moriarty}, P. and {Ong}, R.~A. and {Otte}, A.~N. and {Ouellette}, M. and {Pandel}, D. and {Perkins}, J.~S. and {Pichel}, A. and {Pohl}, M. and {Quinn}, J. and {Ragan}, K. and {Reyes}, L.~C. and {Reynolds}, P.~T. and {Roache}, E. and {Rose}, H.~J. and {Rovero}, A.~C. and {Schroedter}, M. and {Sembroski}, G.~H. and {Senturk}, G. Demet and {Steele}, D. and {Swordy}, S.~P. and {Theiling}, M. and {Thibadeau}, S. and {Varlotta}, A. and {Vassiliev}, V.~V. and {Vincent}, S. and {Wagner}, R.~G. and {Wakely}, S.~P. and {Ward}, J.~E. and {Weekes}, T.~C. and {Weinstein}, A. and {Weisgarber}, T. and {Williams}, D.~A. and {Wissel}, S. and {Wood}, M. and {Zitzer}, B. and {Garson}, III, A. and {Lee}, K. and {Sadun}, A.~C. and {Carini}, M. and {Barnaby}, D. and {Cook}, K. and {Maune}, J. and {Pease}, A. and {Smith}, S. and {Walters}, R. and {Berdyugin}, A. and {Lindfors}, E. and {Nilsson}, K. and {Pasanen}, M. and {Sainio}, J. and {Sillanpaa}, A. and {Takalo}, L.~O. and {Villforth}, C. and {Montaruli}, T. and {Baker}, M. and {Lahteenmaki}, A. and {Tornikoski}, M. and {Hovatta}, T. and {Nieppola}, E. and {Aller}, H.~D. and {Aller}, M.~F.},
        title = "{TeV and Multi-wavelength Observations of Mrk 421 in 2006-2008}",
      journal = {\apj},
         year = 2011,
        month = sep,
       volume = {738},
       number = {1},
          eid = {25},
        pages = {25},
          doi = {10.1088/0004-637X/738/1/25},
archivePrefix = {arXiv},
       eprint = {1106.1210},
 primaryClass = {astro-ph.HE},
       adsurl = {https://ui.adsabs.harvard.edu/abs/2011ApJ...738...25A}
}

@ARTICLE{2009ApJ...695..596H,
       author = {{Horan}, D. and {Acciari}, V.~A. and {Bradbury}, S.~M. and {Buckley}, J.~H. and {Bugaev}, V. and {Byrum}, K.~L. and {Cannon}, A. and {Celik}, O. and {Cesarini}, A. and {Chow}, Y.~C.~K. and {Ciupik}, L. and {Cogan}, P. and {Falcone}, A.~D. and {Fegan}, S.~J. and {Finley}, J.~P. and {Fortin}, P. and {Fortson}, L.~F. and {Gall}, D. and {Gillanders}, G.~H. and {Grube}, J. and {Gyuk}, G. and {Hanna}, D. and {Hays}, E. and {Kertzman}, M. and {Kildea}, J. and {Konopelko}, A. and {Krawczynski}, H. and {Krennrich}, F. and {Lang}, M.~J. and {Lee}, K. and {Moriarty}, P. and {Nagai}, T. and {Niemiec}, J. and {Ong}, R.~A. and {Perkins}, J.~S. and {Pohl}, M. and {Quinn}, J. and {Reynolds}, P.~T. and {Rose}, H.~J. and {Sembroski}, G.~H. and {Smith}, A.~W. and {Steele}, D. and {Swordy}, S.~P. and {Toner}, J.~A. and {Vassiliev}, V.~V. and {Wakely}, S.~P. and {Weekes}, T.~C. and {White}, R.~J. and {Williams}, D.~A. and {Wood}, M.~D. and {Zitzer}, B. and {Aller}, H.~D. and {Aller}, M.~F. and {Baker}, M. and {Barnaby}, D. and {Carini}, M.~T. and {Charlot}, P. and {Dumm}, J.~P. and {Fields}, N.~E. and {Hovatta}, T. and {Jordan}, B. and {Kovalev}, Y.~A. and {Kovalev}, Y.~Y. and {Krimm}, H.~A. and {Kurtanidze}, O.~M. and {L{\"a}hteenm{\"a}ki}, A. and {LeCampion}, J.~F. and {Maune}, J. and {Montaruli}, T. and {Sadun}, A.~C. and {Smith}, S. and {Tornikoski}, M. and {Turunen}, M. and {Walters}, R.},
        title = "{Multiwavelength Observations of Markarian 421 in 2005-2006}",
      journal = {\apj},
         year = 2009,
        month = apr,
       volume = {695},
       number = {1},
        pages = {596-618},
          doi = {10.1088/0004-637X/695/1/596},
archivePrefix = {arXiv},
       eprint = {0901.1225},
 primaryClass = {astro-ph.HE},
       adsurl = {https://ui.adsabs.harvard.edu/abs/2009ApJ...695..596H}
}

@ARTICLE{2008ApJ...677..906F,
       author = {{Fossati}, G. and {Buckley}, J.~H. and {Bond}, I.~H. and {Bradbury}, S.~M. and {Carter-Lewis}, D.~A. and {Chow}, Y.~C.~K. and {Cui}, W. and {Falcone}, A.~D. and {Finley}, J.~P. and {Gaidos}, J.~A. and {Grube}, J. and {Holder}, J. and {Horan}, D. and {Horns}, D. and {Jordan}, M.~M. and {Kieda}, D.~B. and {Kildea}, J. and {Krawczynski}, H. and {Krennrich}, F. and {Lang}, M.~J. and {LeBohec}, S. and {Lee}, K. and {Moriarty}, P. and {Ong}, R.~A. and {Petry}, D. and {Quinn}, J. and {Sembroski}, G.~H. and {Wakely}, S.~P. and {Weekes}, T.~C.},
        title = "{Multiwavelength Observations of Markarian 421 in 2001 March: An Unprecedented View on the X-Ray/TeV Correlated Variability}",
      journal = {\apj},
         year = 2008,
        month = apr,
       volume = {677},
       number = {2},
        pages = {906-925},
          doi = {10.1086/527311},
archivePrefix = {arXiv},
       eprint = {0710.4138},
 primaryClass = {astro-ph},
       adsurl = {https://ui.adsabs.harvard.edu/abs/2008ApJ...677..906F}
}

@ARTICLE{2005ApJ...630..130B,
       author = {{B{\l}a{\.z}ejowski}, M. and {Blaylock}, G. and {Bond}, I.~H. and {Bradbury}, S.~M. and {Buckley}, J.~H. and {Carter-Lewis}, D.~A. and {Celik}, O. and {Cogan}, P. and {Cui}, W. and {Daniel}, M. and {Duke}, C. and {Falcone}, A. and {Fegan}, D.~J. and {Fegan}, S.~J. and {Finley}, J.~P. and {Fortson}, L. and {Gammell}, S. and {Gibbs}, K. and {Gillanders}, G.~G. and {Grube}, J. and {Gutierrez}, K. and {Hall}, J. and {Hanna}, D. and {Holder}, J. and {Horan}, D. and {Humensky}, B. and {Kenny}, G. and {Kertzman}, M. and {Kieda}, D. and {Kildea}, J. and {Knapp}, J. and {Kosack}, K. and {Krawczynski}, H. and {Krennrich}, F. and {Lang}, M. and {LeBohec}, S. and {Linton}, E. and {Lloyd-Evans}, J. and {Maier}, G. and {Mendoza}, D. and {Milovanovic}, A. and {Moriarty}, P. and {Nagai}, T.~N. and {Ong}, R.~A. and {Power-Mooney}, B. and {Quinn}, J. and {Quinn}, M. and {Ragan}, K. and {Reynolds}, P.~T. and {Rebillot}, P. and {Rose}, H.~J. and {Schroedter}, M. and {Sembroski}, G.~H. and {Swordy}, S.~P. and {Syson}, A. and {Valcarel}, L. and {Vassiliev}, V.~V. and {Wakely}, S.~P. and {Walker}, G. and {Weekes}, T.~C. and {White}, R. and {Zweerink}, J. and {VERITAS Collaboration} and {Mochejska}, B. and {Smith}, B. and {Aller}, M. and {Aller}, H. and {Ter{\"a}sranta}, H. and {Boltwood}, P. and {Sadun}, A. and {Stanek}, K. and {Adams}, E. and {Foster}, J. and {Hartman}, J. and {Lai}, K. and {B{\"o}ttcher}, M. and {Reimer}, A. and {Jung}, I.},
        title = "{A Multiwavelength View of the TeV Blazar Markarian 421: Correlated Variability, Flaring, and Spectral Evolution}",
      journal = {\apj},
         year = 2005,
        month = sep,
       volume = {630},
       number = {1},
        pages = {130-141},
          doi = {10.1086/431925},
archivePrefix = {arXiv},
       eprint = {astro-ph/0505325},
 primaryClass = {astro-ph},
       adsurl = {https://ui.adsabs.harvard.edu/abs/2005ApJ...630..130B}
}

@ARTICLE{2000ApJ...542L.105T,
       author = {{Takahashi}, T. and {Kataoka}, J. and {Madejski}, G. and {Mattox}, J. and {Urry}, C.~M. and {Wagner}, S. and {Aharonian}, F. and {Catanese}, M. and {Chiappetti}, L. and {Coppi}, P. and {Degrange}, B. and {Fossati}, G. and {Kubo}, H. and {Krawczynski}, H. and {Makino}, F. and {Marshall}, H. and {Maraschi}, L. and {Piron}, F. and {Remillard}, R. and {Takahara}, F. and {Tashiro}, M. and {Terasranta}, H. and {Weekes}, T.},
        title = "{Complex Spectral Variability from Intensive Multiwavelength Monitoring of Markarian 421 in 1998}",
      journal = {\apjl},
         year = 2000,
        month = oct,
       volume = {542},
       number = {2},
        pages = {L105-L109},
          doi = {10.1086/312929},
archivePrefix = {arXiv},
       eprint = {astro-ph/0008505},
 primaryClass = {astro-ph},
       adsurl = {https://ui.adsabs.harvard.edu/abs/2000ApJ...542L.105T}
}

@ARTICLE{2002A&A...389..742W,
       author = {{Wu}, Xue-Bing and {Liu}, F.~K. and {Zhang}, T.~Z.},
        title = "{Supermassive black hole masses of AGNs with elliptical hosts}",
      journal = {\aap},
         year = 2002,
        month = jul,
       volume = {389},
        pages = {742-751},
          doi = {10.1051/0004-6361:20020577},
archivePrefix = {arXiv},
       eprint = {astro-ph/0203158},
 primaryClass = {astro-ph},
       adsurl = {https://ui.adsabs.harvard.edu/abs/2002A&A...389..742W}
}

@ARTICLE{2002ApJ...569L..35F,
       author = {{Falomo}, R. and {Kotilainen}, J.~K. and {Treves}, A.},
        title = "{The Black Hole Mass of BL Lacertae Objects from the Stellar Velocity Dispersion of the Host Galaxy}",
      journal = {\apjl},
         year = 2002,
        month = apr,
       volume = {569},
       number = {1},
        pages = {L35-L38},
          doi = {10.1086/340642},
archivePrefix = {arXiv},
       eprint = {astro-ph/0203199},
 primaryClass = {astro-ph},
       adsurl = {https://ui.adsabs.harvard.edu/abs/2002ApJ...569L..35F}
}

@ARTICLE{2003ApJ...583..134B,
       author = {{Barth}, Aaron J. and {Ho}, Luis C. and {Sargent}, Wallace L.~W.},
        title = "{The Black Hole Masses and Host Galaxies of BL Lacertae Objects}",
      journal = {\apj},
         year = 2003,
        month = jan,
       volume = {583},
       number = {1},
        pages = {134-144},
          doi = {10.1086/345083},
archivePrefix = {arXiv},
       eprint = {astro-ph/0209562},
 primaryClass = {astro-ph},
       adsurl = {https://ui.adsabs.harvard.edu/abs/2003ApJ...583..134B}
}

@ARTICLE{2026ApJ...998..107H,
       author = {{Huang}, Tao and {Gupta}, Alok C. and {Cui}, Lang and {Tripathi}, Ashutosh and {Huang}, Yongfeng and {Devanand}, P.~U. and {Liu}, Xiang},
        title = "{X-Ray Intraday Variability of the Blazar OJ 287 Observed with XMM-Newton}",
      journal = {\apj},
         year = 2026,
        month = feb,
       volume = {998},
       number = {1},
          eid = {107},
        pages = {107},
          doi = {10.3847/1538-4357/ae38b9},
archivePrefix = {arXiv},
       eprint = {2603.06097},
 primaryClass = {astro-ph.HE},
       adsurl = {https://ui.adsabs.harvard.edu/abs/2026ApJ...998..107H}
}

@inproceedings{10.1117/12.2231304,
author = {Keith C. Gendreau and Zaven Arzoumanian and Phillip W. Adkins and Cheryl L. Albert and John F. Anders and Andrew T. Aylward and Charles L. Baker and Erin R. Balsamo and William A. Bamford and Suyog S. Benegalrao and Daniel L. Berry and Shiraz Bhalwani and J. Kevin Black and Carl Blaurock and Ginger M. Bronke and Gary L. Brown and Jason G. Budinoff and Jeffrey D. Cantwell and Thoniel Cazeau and Philip T. Chen and Thomas G. Clement and Andrew T. Colangelo and Jerry S. Coleman and Jonathan D. Coopersmith and William E. Dehaven and John P. Doty and Mark D. Egan and Teruaki Enoto and Terry W.-M. Fan and Deneen M. Ferro and Richard Foster and Nicholas M. Galassi and Luis D. Gallo and Chris M. Green and Dave Grosh and Kong Q. Ha and Monther A. Hasouneh and Kristofer B. Heefner and Phyllis Hestnes and Lisa J. Hoge and Tawanda M. Jacobs and John L. J{\o}rgensen and Michael A. Kaiser and James W. Kellogg and Steven J. Kenyon and Richard G. Koenecke and Robert P. Kozon and Beverly LaMarr and Mike D. Lambertson and Anne M. Larson and Steven Lentine and Jesse H. Lewis and Michael G. Lilly and Kuochia Alice Liu and Andrew Malonis and Sridhar S. Manthripragada and Craig B. Markwardt and Bryan D. Matonak and Isaac E. Mcginnis and Roger L. Miller and Alissa L. Mitchell and Jason W. Mitchell and Jelila S. Mohammed and Charles A. Monroe and Kristina M. Montt de Garcia and Peter D. Mul{\'e} and Louis T. Nagao and Son N. Ngo and Eric D. Norris and Dwight A. Norwood and Joseph Novotka and Takashi Okajima and Lawrence G. Olsen and Chimaobi O. Onyeachu and Henry Y. Orosco and Jacqualine R. Peterson and Kristina N. Pevear and Karen K. Pham and Sue E. Pollard and John S. Pope and Daniel F. Powers and Charles E. Powers and Samuel R. Price and Gregory Y. Prigozhin and Julian B. Ramirez and Winston J. Reid and Ronald A. Remillard and Eric M. Rogstad and Glenn P. Rosecrans and John N. Rowe and Jennifer A. Sager and Claude A. Sanders and Bruce Savadkin and Maxine R. Saylor and Alexander F. Schaeffer and Nancy S. Schweiss and Sean R. Semper and Peter J. Serlemitsos and Larry V. Shackelford and Yang Soong and Jonathan Struebel and Michael L. Vezie and Joel S. Villasenor and Luke B. Winternitz and George I. Wofford and Michael R. Wright and Mike Y. Yang and Wayne H. Yu},
title = {{The  Neutron star Interior Composition Explorer (NICER): design and development}},
volume = {9905},
booktitle = {Space Telescopes and Instrumentation 2016: Ultraviolet to Gamma Ray},
editor = {Jan-Willem A. den Herder and Tadayuki Takahashi and Marshall Bautz},
organization = {International Society for Optics and Photonics},
publisher = {SPIE},
pages = {99051H},
year = {2016},
doi = {10.1117/12.2231304},
URL = {https://doi.org/10.1117/12.2231304}
}

@ARTICLE{2019MmSAI..90..154G,
       author = {{Ghisellini}, G.},
        title = "{Extra-galactic jets: a hard X-ray view}",
      journal = {\memsai},
         year = 2019,
        month = jan,
       volume = {90},
        pages = {154},
          doi = {10.48550/arXiv.1911.11777},
archivePrefix = {arXiv},
       eprint = {1911.11777},
 primaryClass = {astro-ph.HE},
       adsurl = {https://ui.adsabs.harvard.edu/abs/2019MmSAI..90..154G}
}

@ARTICLE{1997ApJ...478L..79N,
       author = {{Narayan}, Ramesh and {Garcia}, Michael R. and {McClintock}, Jeffrey E.},
        title = "{Advection-dominated Accretion and Black Hole Event Horizons}",
      journal = {\apjl},
         year = 1997,
        month = apr,
       volume = {478},
       number = {2},
        pages = {L79-L82},
          doi = {10.1086/310554},
archivePrefix = {arXiv},
       eprint = {astro-ph/9701139},
 primaryClass = {astro-ph},
       adsurl = {https://ui.adsabs.harvard.edu/abs/1997ApJ...478L..79N}
}

@ARTICLE{2000ApJ...539..798N,
       author = {{Narayan}, Ramesh and {Igumenshchev}, Igor V. and {Abramowicz}, Marek A.},
        title = "{Self-similar Accretion Flows with Convection}",
      journal = {\apj},
         year = 2000,
        month = aug,
       volume = {539},
       number = {2},
        pages = {798-808},
          doi = {10.1086/309268},
archivePrefix = {arXiv},
       eprint = {astro-ph/9912449},
 primaryClass = {astro-ph},
       adsurl = {https://ui.adsabs.harvard.edu/abs/2000ApJ...539..798N}
}

@ARTICLE{2009MNRAS.396L.105G,
       author = {{Ghisellini}, G. and {Maraschi}, L. and {Tavecchio}, F.},
        title = "{The Fermi blazars' divide}",
      journal = {\mnras},
         year = 2009,
        month = jun,
       volume = {396},
       number = {1},
        pages = {L105-L109},
          doi = {10.1111/j.1745-3933.2009.00673.x},
archivePrefix = {arXiv},
       eprint = {0903.2043},
 primaryClass = {astro-ph.CO},
       adsurl = {https://ui.adsabs.harvard.edu/abs/2009MNRAS.396L.105G}
}
\bibliographystyle{aasjournalv7}
%%%%%%%%%%%%%%%%%%%%%%%%%%%%%%%%%%%%%%%%%%%%%%%%%%%%%%%%%%%%%%%%%%%%
\end{document}